\documentclass[twocolumn]{svjour3-b}          % twocolumn
\usepackage{color}
\usepackage{graphicx}
\usepackage{amsmath}
\usepackage{amssymb}
\usepackage{cite}
\usepackage{amsfonts}
\newcommand{\bm}[1]{\mbox{\boldmath$#1$}}
\newcommand{\Td}{T$_{\rm d}$\ }

\journalname{Brazilian Journal of Physics}
\begin{document}

\title{Tetrahedral Order in Liquid Crystals%\thanks{Grants or other notes
}

\titlerunning{Tetrahedral Liquid Crystals}       

\author{Harald Pleiner       \and
        Helmut R. Brand 
}

\institute{Harald Pleiner \at
              Max Planck Institute for Polymer Research,
              Mainz, Germany\\
             Tel.: +49-6131-379246\\
             Fax: +49-6131-379340\\
              \email{pleiner@mpip-mainz.mpg.de}          
           \and
           Helmut R. Brand \at
              Theoretische Physik III, Universit\"at Bayreuth, 95440 Bayreuth, 
Germany\\ 
\email{brand@uni-bayreuth.de}
}

\date{Received: 8 June 2016 / Published online: 15 August 2016 \\ 
\copyright\ The Authors 2016. This article is published with open access at Springerlink.com}
%\date{\today}

\maketitle

\begin{abstract}
We review the impact of tetrahedral order on the macroscopic dynamics of bent-core liquid crystals. We discuss tetrahedral order comparing with other types of orientational order, like nematic, polar nematic, polar smectic, and active polar order. In particular, we present hydrodynamic equations for phases, where only tetrahedral order exists or tetrahedral order is combined with nematic order. Among the latter we discriminate between
three cases, where the nematic director (a) orients along a 4-fold, (b) along a 3-fold symmetry axis of the tetrahedral structure, or (c) is homogeneously uncorrelated with the tetrahedron. For the optically isotropic \Td phase, which only has tetrahedral order, we focus on the coupling of flow with {\it e.g.} temperature gradients and on the specific orientation behavior in external electric fields. For the transition to the nematic phase, electric fields lead to a temperature shift that is linear in the field strength. Electric fields induce nematic order, again linear in the field strength. If strong enough, electric fields can change the tetrahedral structure and symmetry leading to a polar phase. We briefly deal with the T phase that arises when tetrahedral order occurs in a system of chiral molecules. To case (a), defined above, belong (i) the non-polar, achiral, optically uniaxial D2d phase with ambidextrous helicity (due to a linear gradient free energy contribution) and with orientational frustration in external fields, (ii) the non-polar tetragonal S4 phase, (iii) the non-polar, orthorhombic D2 phase that is structurally chiral featuring ambidextrous chirality, (iv) the polar orthorhombic C2v phase, and (v) the polar, structurally chiral, monoclinic C2 phase. Case (b) results in a trigonal C3v phase that behaves like a biaxial polar nematic phase. An example for case (c) is a splay bend phase, where the ground state is inhomogeneous due to a linear gradient free energy contribution. Finally we discuss some experiments that show typical effects related to the existence of tetrahedral order. A summary and perspective is given.

\keywords{Bent-core liquid crystals \and Hydrodynamics \and
Phase Behavior \and Symmetries \and Macroscopic Properties \and
Electric Octupolar Order}
\end{abstract}

%%%%%%%%%%%%%%%%%%%%%%%%%%%%%%%%%%%%%%%%%%%%%%%
%%%%%%%%%%%%%%%%%%%%%%%%%%%%%%%%%%%%%%%%%%%%%%%

\section{Introduction} \label{intro}

The quantitative macroscopic description of liquid crystals (LC) in terms of partial differential dynamic equations, free energy functionals, Ginzburg-Landau energies and the like, has been developed over the last 50 years \cite{MPP} - \cite{mbuch}. For a long time, it has been sufficient to consider nematic-like phases with preferred (non-polar) directions or smectic (and columnar) phases with one-dimensional layer (or two-dimensional lattice) structures. With the advent of bent-core (or banana) LC \cite{CM92,Takezoe96} this has changed considerably. It turned out that, first, polar (vector) order plays an important role, in particular in smectic-like structures, and second, tetrahedral order is an additional type of order, necessary to describe the phases found in these new materials. We will only briefly discuss the role of polar order, but concentrate on the various aspects of tetrahedral order. To do that we first recall the traditional types of orientational order, give some motivations why these are insufficient in the case of bent-core materials, and then discuss tetrahedral order and and its interplay with other types of orientational order. In the subsequent Sections   \ref{Td} - \ref{TNnotcorr} we give detailed accounts of the hydrodynamics of the various phases where tetrahedral order is involved. The experimental situation is discussed in Section \ref{exper}, followed by a Summary and 
Perspective, Section \ref{Sum}.

Liquid crystals generally are anisotropic fluids. This is due to the existence of one (or more) preferred directions in the fluid, either due to rotational order ({\it e.g.} nematic LCs), 
or translational order ({\it e.g.} smectic and columnar LCs), or both ({\it e.g.} 
smectic C LCs). As a result, at least some of their (macroscopic) material properties, like {\it e.g.} heat conduction, viscosity, or sound velocity, depend on the orientation. Thereby, the rotational and/or translational symmetry of isotropic liquids is spontaneously broken by the occurrence of ordered structures at an equilibrium phase transition. The latter can be obtained by changing some control parameters, like temperature, pressure or concentration (in a mixture) leading to the distinction of thermotropic, barotropic, and lyotropic LCs, respectively.

%%%%%%%%%%%%%%%%%%%%%%%%%%%%%%%%%%%%%%%%%%%%%%%%%

\subsection{Nematic order \label{nemaorder}}

Well-known are (uniaxial thermotropic) nematic LC, consisting of rod-like or plate-like molecules that are disordered in the isotropic state. When cooling down into the nematic phase they spontaneously align their preferred molecular axes in the mean forming a macroscopic preferred direction, the director $\bm{n}$, with $\bm{n}^2=1$ rendering the phase anisotropic. In nematic LC one cannot discriminate between the director and its opposite direction (no "head" or "tail"), even if the molecular axes do have this distinction. Thus, $\bm{n}$ is not a true vector. The common procedure to circumvent this shortcoming is to use use $\bm{n}$ as a true vector with the additional requirement that all macroscopic equations are invariant under the exchange of $\bm{n}$ with $-\bm{n}$.

The (uniaxial) nematic order is described by the quadrupolar, traceless symmetric order parameter tensor, correctly reflecting the $\bm{n}$ to $-\bm{n}$ invariance,
\begin{equation} \label{Qij}
Q_{ij} = \frac12 S (n_i n_j - \frac13 \delta_{ij})
\end{equation}
Due to its quadrupolar structure no vector can be extracted from it. 
Here, $S$, the strength of the order, is the second moment of the microscopic orientations of all molecules with respect to the preferred direction $\bm{n}$. There is $S=0$ in the isotropic phase, and the extremum values $S=1$ and $S=-1/2$ for perfect order in the rod-like and plate-like case, respectively. A realistic value for many rod-like nematic LC is $S\approx 0.8$ within the nematic phase, which decreases approaching the nematic to isotropic phase transition, with a jump of $\Delta S \approx 0.4$ at the transition. It can be measured {\it e.g.} through the dielectric anisotropy \cite{PGdG}. 

The order is spontaneous and the orientation of $\bm{n}$ is arbitrary, as long as there are no orienting external fields or boundaries. Therefore, the two rotations of the director, $\delta\bm{n}$ (with $\bm{n}\cdot \delta\bm{n}=0$), are the slow additional hydrodynamic variables ("symmetry variables") due to the nematic order \cite{MPP}. Deviations of the order parameter $S$ from its equilibrium value relax in a finite time (except near phase transitions) and are often neglected. 

In case the molecules order themselves with respect to two different molecular directions, a biaxial order results with the order parameter tensor
\begin{equation} \label{Qijbiax}
Q_{ij} = \frac12 S (n_i n_j - \frac13 \delta_{ij}) + \eta (m_i m_j - l_i l_j)
\end{equation}
where $\eta$ is a measure for biaxiality. The directions $\bm{m}$ and $\bm{l}$ are directors with a $\bm{m}$ to $-\bm{m}$ and $\bm{l}$ to $-\bm{l}$ invariance. Depending on the symmetries of the phase considered there can be additional relations between the directors, {\it e.g.} for orthorhombic phases the three directors have to be mutually orthogonal. The symmetry variables are the three rotations of the rigid tripod $\{\bm{n}, \bm{m}, \bm{l}\}$, {\it e.g.} described by $\bm{m}\cdot \delta \bm{n} +\bm{n}\cdot \delta \bm{m}$, $\bm{m}\cdot \delta \bm{l} +\bm{l}\cdot \delta \bm{m}$, and $\bm{l}\cdot \delta \bm{n} +\bm{n}\cdot \delta \bm{l}$. For finite rotations, that is in a nonlinear description, these variables do not commute \cite{Nbiax,biaxnemliu}. 

%%%%%%%%%%%%%%%%%%%%%%%%%%%%%%%%%%%%%%%%%%%%%%%%%%

\subsection{Polar order \label{polarorder}}

A different type of orientational order is polar order. In polar nematic LC the preferred direction is a true vector, with "head" and "tail" distinguishable. The order parameter is the polarization vector \cite{polnema}
\begin{equation} \label{Pi}
P_{i} =P p_i
\end{equation}
with $P$, the value of the spontaneous polarization that characterizes the strength of the polar order, and $\bm{p}$ the unit vector that denotes the (arbitrary) orientation of the polar direction. For rod-like systems, thermotropic polar nematic LC are rare in nature. One reason might be that a finite sample of homogeneous ($\bm{p} = \textrm{const.}$) polar nematic LC exhibit opposite surface charges in the planes perpendicular to $\bm{p}$, which give rise to destabilizing electrostatic forces. Second, the homogeneous state is not the energetic ground state, since the existence of $\bm{P}$ allows for spontaneous structures with a constant splay texture \cite{splay}. However, the latter cannot 
be space filling
and is necessarily connected to defects.\footnote{In chiral nematics the existence of a pseudoscalar allows for the existence of spontaneous twist, which can fill space without defects.} 

The hydrodynamic variables are the two rotations of the polarization direction, $\delta\bm{p}$ (with $\bm{p}\cdot \delta\bm{p}=0$). The absolute value of the polarization $P$ is linearly susceptible to an external electric field and is therefore often kept as (slowly relaxing) variable in a macroscopic description including, {\it e.g.}, pyroelectricity.

%%%%
%%%%
\begin{figure}      
\includegraphics[width=5cm]{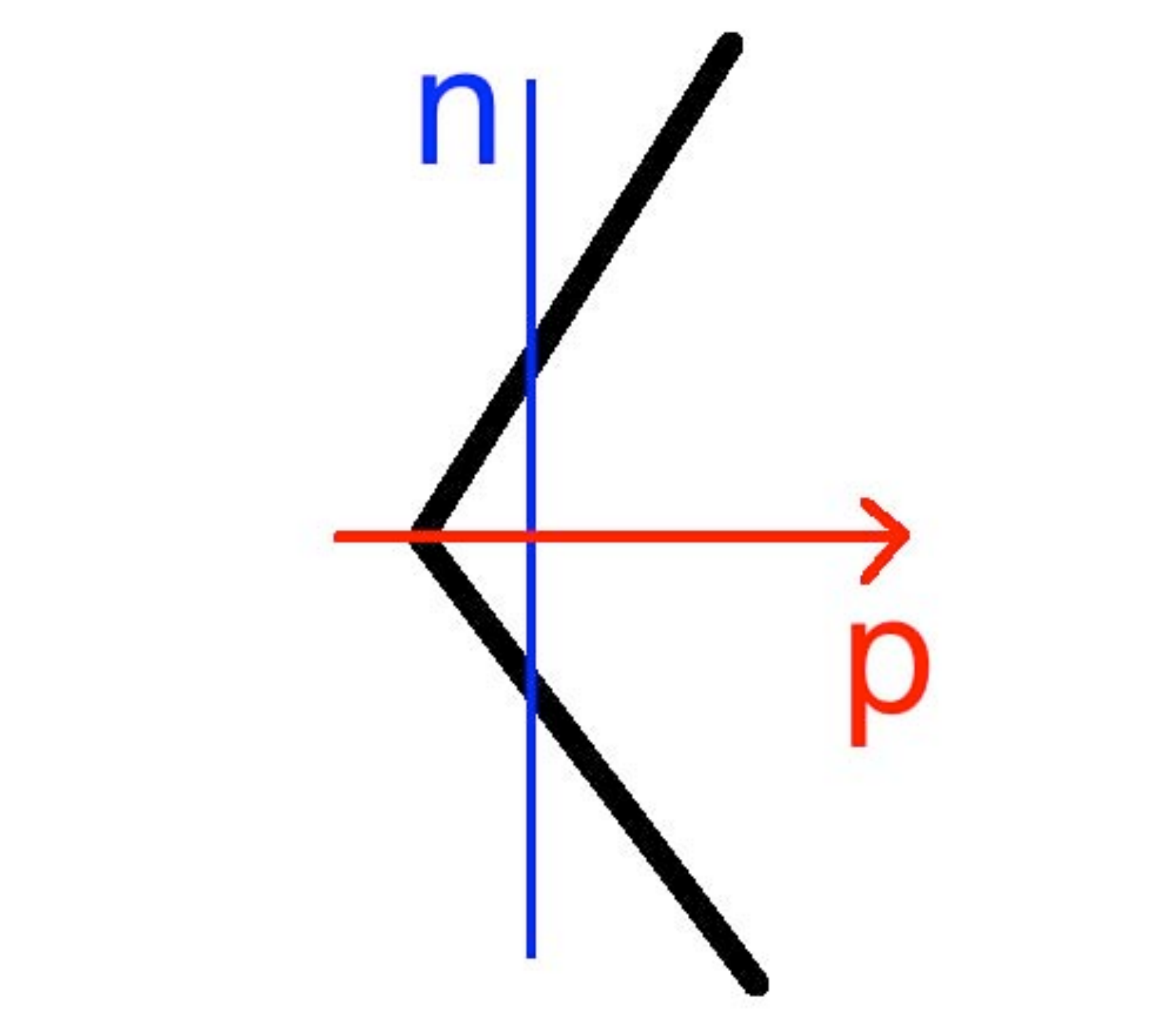}
\caption{Model bent-core molecule with nematic ($\bm{n}$) and polar ($\bm{p}$) directions \label{fig:1}}
\end{figure}
%%%%
%%%%
%
%%%%
%%%%
\begin{figure}       
\includegraphics[width=4cm]{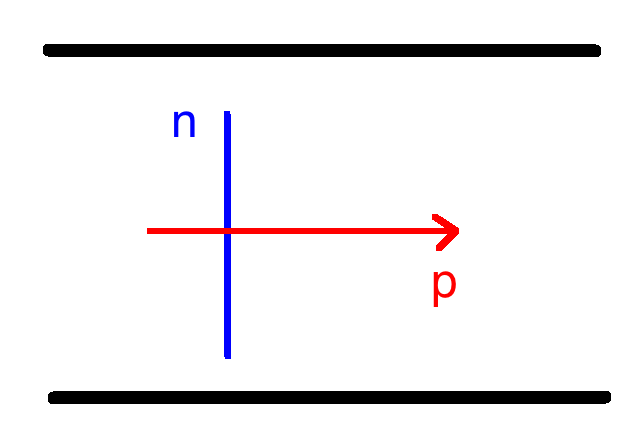}\hspace{0.1cm}
\includegraphics[width=4cm]{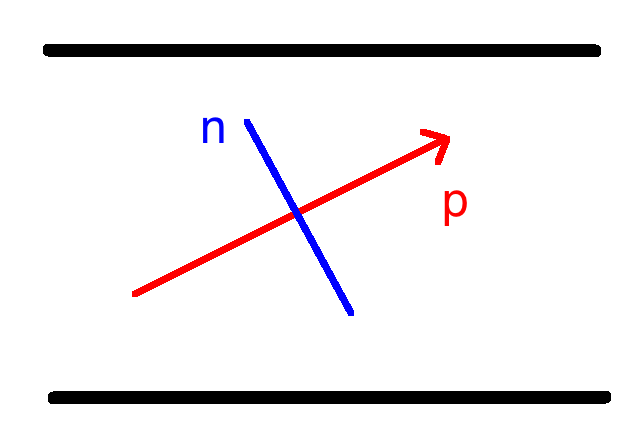}\\[0.1cm]
\includegraphics[width=4cm]{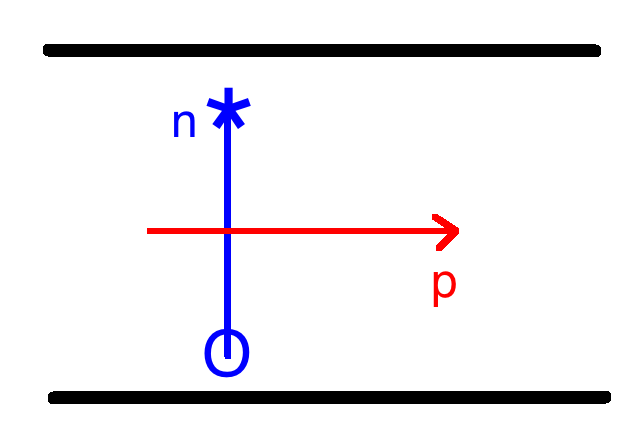}\hspace{0.1cm}
\includegraphics[width=4cm]{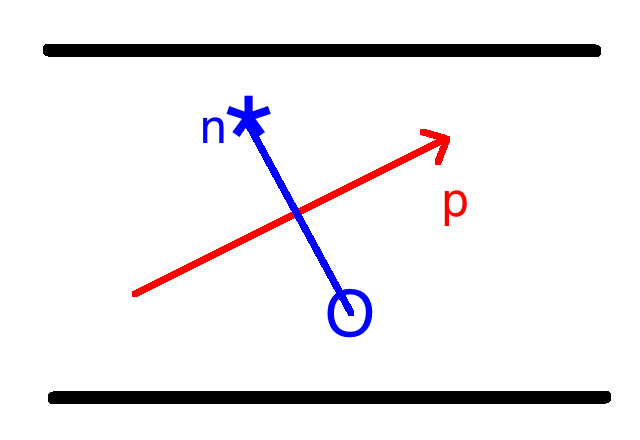}
\caption{Orientations in polar smectic LC: (upper left) untilted C$_P$ phase, (upper right) p-tilted C$_{B1}$ phase, (lower left) n-tilted C$_{B2}$ phase and (lower right) twice tilted C$_G$ phase;  
stars and rings mean a vector points out of, or into, the drawing plane, respectively.
\label{fig:2}}
\end{figure}
%%%%
%%%%
For bent-core molecules the situation is rather different. Typically they have a non-polar (long) axis, $\bm{n}_{mol}$ and perpendicularly a (short) polar axis $\bm{p}_{mol}$, cf. Fig. \ref{fig:1}. They can align in layered structures to form various smectic phases with polar properties (Fig.    \ref{fig:2}). There are two untilted phases, C$_{\textrm{P}}$ with the nematic axis $\bm{n}$ along the layer normal and the polar axis $\bm{p}$ within the layer 
\cite{CM92}, and C$_{\textrm{P'}}$, vice versa. Both are of $C_{2v}$ symmetry. One can tilt the C$_{\textrm{P}}$ structure in two ways. Rotation about the polar axis results in the C$_{\textrm{B2}}$ phase of $C_2$ symmetry and rotating about the $\bm{p \times n}$ direction gives the C$_{\textrm{B1}}$ phase of $C_{1h}$ symmetry. If both ways of tilting are combined, the C$_G$  phase arises, where no symmetry element is left ($C_1$) \cite{D}. The C$_{\textrm{B2}}$ and C$_{\textrm{G}}$ phases are chiral due to their structure, even when the molecules are achiral. This allows for the occurrence  of energetically equivalent left- and right-handed helices \cite{G1,G2} (called ambidextrous chirality). 
If layers of different polarization direction and/or different tilt direction are stacked, various overall structures with different properties can be obtained, {\it e.g.} ferro-, ferri-, or antiferroelectric, and helical or non-helical\cite{E1,E2}.  
There is also the possibility of polar columnar phases \cite{I}. If bent-core molecules align both their axes in a biaxial nematic way, the result is a polar biaxial nematic phase (N$_{\textrm{I}}$) of $C_{2V}$ symmetry \cite{E2,F}.

A different type of polar order can occur in active systems,\footnote{There is also the possibility of an axial, non-polar {\it active} order \cite{AG}.} like schools of fish, 
bird flocks, insect swarms, growing bacteria, and biological motors. If the active part of those systems moves relative to a passive background, the relative velocity denotes a preferred polar direction \cite{AF}. This type of order is dynamic, since it is provided by the motion of the active part and vanishes, when the motion stops. It does not occur spontaneously by an equilibrium phase transition from an disordered to an ordered state, but is due to the (internal) driving typically via chemical reactions (food consumption, metabolism). The active state is a non-equilibrium one, constantly dissipating the supplied energy. If the driving stops, the system falls back into a passive, disordered state \cite{AH}. The order parameter in this case is the relative velocity
\begin{equation} \label{Fi}
F_{i} =F f_i
\end{equation}
where the amount of the velocity, $F$, is a measure of the strength of the order.  Its non-zero value even in a stationary state, is due to the driving and indicates the non-equilibrium nature of the system. The unit velocity, $\bm{f}$, is a polar vector and characterizes the direction of the polar order. This direction is not prescribed by the driving and is therefore arbitrary. Its two rotations are the symmetry variables.
In that respect $\bm{f}$ is similar to the polar nematic case, $\bm{p}$. However,
a velocity changes sign under time reversal, while a polarization does not. Therefore, the hydrodynamics of active polar systems is quite different from that of (passive) polar nematic LC. In particular, the former systems exhibit new non-trivial couplings among various hydrodynamic variables, linear advection properties, active stresses, second sound, and asymmetries between forward and backward traveling sound excitations, which is a clear 
indication of non-equilibrium  \cite{AF,AI}. 

%%%%%%%%%%%%%%%%%%%%%%%%%%%%%%%%%%%%%%%%%%%%%%%%%%

\subsection{Tetrahedral order \label{tetraorder}}

During the development of more sophisticated bent-core materials, it became apparent that an important ingredient in the macroscopic description was still missing. Compounds were found that showed a phase transition between two different (optically) isotropic phases. Being one of them the true isotropic phase without any order, the other must have a type of order that is not detectable in a microscope, meaning the dielectric tensor has to be isotropic. That immediately rules out polar or nematic order. Obviously, tetrahedral order 
\cite{Fel} - \cite{BPC02} does qualify, 
since it is described by a third rank tensor $T_{ijk}$, that cannot influence lower rank material tensors, like the rank-2 dielectric tensor. In analogy to the nematic order that has the  quadrupolar structure of a second moment of an orientational distribution, tetrahedral order can be called octupolar, since it is related to the third moment.

\begin{figure}      
\includegraphics[width=8.3cm]{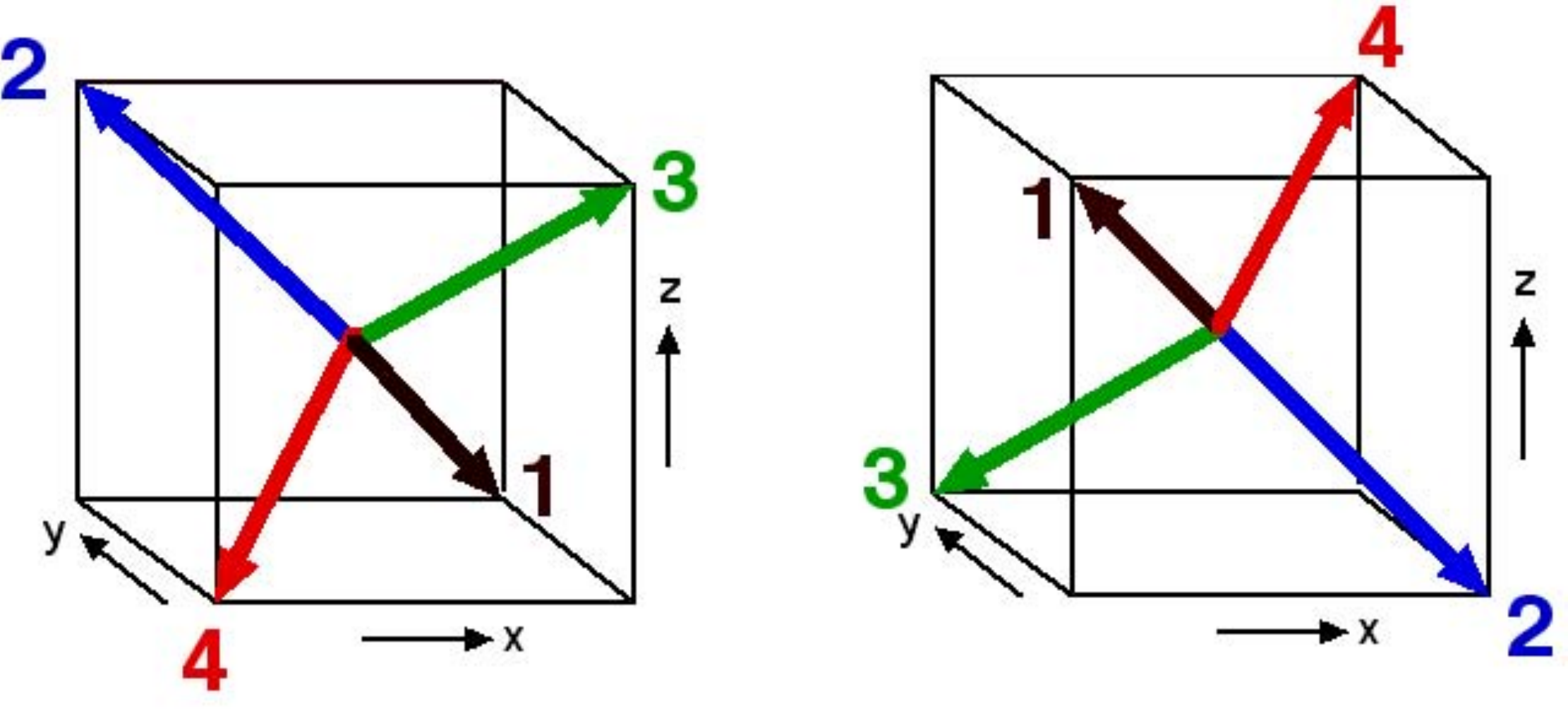}
\caption{The four tetrahedral vectors pointing to the edge of a tetrahedron and (right) the mirror image - from \cite{BP-D2d} \label{fig:3}}
\end{figure}

\begin{figure}      
\includegraphics[width=3.5cm]{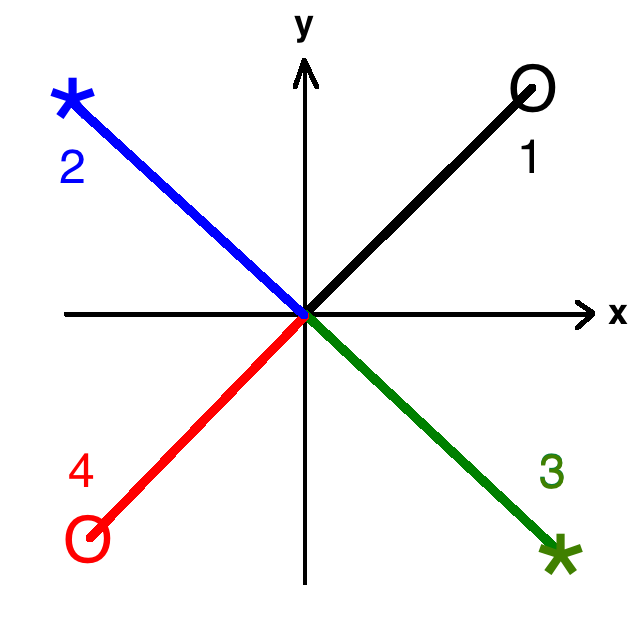}\hspace{0.4cm}
\includegraphics[width=3.5cm]{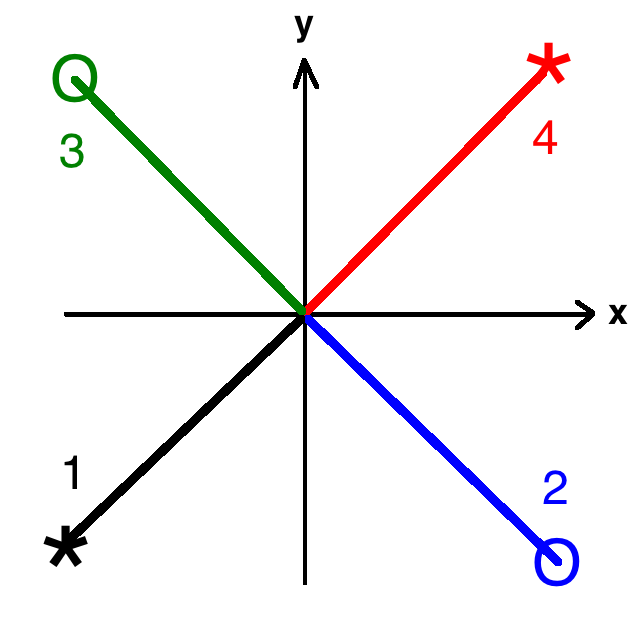}
\caption{A projection of the tetrahedral structure and its mirror image of Fig. \ref{fig:3} onto the x/y plane; stars and rings mean a vector points out of, or into, the drawing plane, respectively. The two structures differ by a 90$^0$ rotation about the z-axis -- note that the actual numbering of the tetrahedral vectors is irrelevant, since they are all equivalent.\label{fig:4}}
\end{figure}

The tetrahedral (octupolar) order parameter 
\begin{equation} \label{Tijk}
T_{ijk}= N \sum_{\beta=1}^{4} n_{i}^{\beta} n_{j}^{\beta} n_{k}^{\beta}
\end{equation}
is a fully symmetric rank-3 tensor expressed by an amplitude $N$ and by the 4 tetrahedral unit vectors,
$\mathbf{n^{\beta}}$, with $\beta \in \{1,2,3,4\}$ defining a tetrahedron, cf. Fig. \ref{fig:3}. 
For actual calculations one can use {\it e.g.} the representation of $T_{ijk}$ \cite{Fel,RadLub02}
\begin{equation}\label{unit}
T_{ijk} = \frac{N}{\sqrt{3}}
\begin{pmatrix}
1&1&-1&-1\\
1&-1&1&-1\\
-1&1&1&-1
\end{pmatrix}
\end{equation}
that is used in the left hand sides of Figs. \ref{fig:3} and \ref{fig:4}. The tetrahedral structure has four 3-fold symmetry axes, the tetrahedral vectors $n_i^\beta$, and three 2-fold (proper) and 4-fold improper ($\bar 4$), symmetry axis, the Cartesian directions x,y,z in Eq. (\ref{unit}).
The latter means that a 90$^0$ rotation about such an axis has to be followed by a spatial inversion, in order to arrive at the initial structure, cf. Fig. \ref{fig:4}. Inversion is an operation, where a structure is either reflected through a point in space, or mirrored by three mutually orthogonal planes. If the resulting structure is different from the original one, inversion symmetry is broken. The existence of improper rotation axes is a sign of broken
inversion symmetry. 

Another useful representation of the tetrahedral structure is
\begin{equation} \label{unitE}
T_{ijk}=\frac{N}{3}
\begin{pmatrix}
0 & -\sqrt{2} & -\sqrt{2} & 2\sqrt{2}\\
0 & -\sqrt{6} & \sqrt{6} & 0\\
3 & -1 & -1 & -1
\end{pmatrix}
\end{equation}
where one tetrahedral vector is along the z-axis, and one of the remaining three has a lateral projection along the x-axis, only. Note that both structures, Eqs. (\ref{unit}) and (\ref{unitE}), have an inverted counter part (all signs within the braces changed, shown on the right hand sides of Figs. \ref{fig:3} and \ref{fig:4}); those are different from the original ones, but can be used equivalently to describe the structures.

A phase that has only tetrahedral order is of $T_d$ symmetry, a subgroup of cubic symmetry. It breaks inversion symmetry, since $T_{ijk}$ changes sign under inversion due to the odd number of (tetrahedral) vectors involved. It is non-polar, {\it i.e.} one cannot extract a vector from $T_{ijk}$, because of $T_{ikk}\equiv0$. It does not imply any nematic order, since $T_{ikl} T_{jkl} \sim\delta_{ij}$ is isotropic. Because of the two mirror planes, defined by two non-adjacent tetrahedral vectors ({\it e.g.} 1/4 or 2/3 in Fig. \ref{fig:4}), chirality is also excluded. Only if the molecules themselves are chiral, a phase of (chiral) $T$ symmetry arises, which will be dealt with below, separately. 
%%%%
%%%%
%%%%
%%%%
%
%%%%
%%%%
%%%%
%%%%

A trivial example for broken inversion symmetry is polar order, since $\bm{p}$ becomes $-\bm{p}$ after inversion, which is different from the original structure.
On the other hand, nematic order $Q_{ij}$ is invariant under inversion. Inversion symmetry is also broken in chiral systems, where a pseudoscalar exists, typically called $q_0$, that changes sign under inversion and reflects the two possibilities of left- and right-handed structures. However, the broken inversion symmetry in tetrahedral order is quite different, since it does not show polarity, nor chirality.

Among the peculiar features, which we will discuss in detail in Section  \ref{Td}, is the possibility that an applied external electric field leads to (i) a temperature shift of the (optically isotropic) tetrahedral to nematic phase transition and (ii) induces nematic order in the \Td phase. Both, the transition shift and the induced order are proportional to the field strength \cite{N}, rather than to the square of the field strength, as it is common when only ordinary isotropic phases are involved. The experimental detection of such linear-field strength transition shifts \cite{Weissflog} accompanied by flow phenomena in some bent-core materials is another indication that tetrahedral order plays an important role for these systems. 
Another consequence of the broken inversion symmetry in tetrahedral phases is the reversible coupling of flow to (vectorial) generalized forces, like electric fields or gradients of temperature or concentration. This piezo-like dynamic coupling 
constitutes ambi-polarity: although polar fluxes are induced, the inverted $T_{ijk}$ structure gives fluxes in the opposite direction and the overall phase is non-polar.

An external electric field (but not a magnetic one) not only orients the tetrahedral structure such that one of the tetrahedral vectors, $n_i^1$, is parallel or antiparallel to the field \cite{Fel3}, it also imposes a torque on the other three \cite{L}. If the tetrahedral structure is soft enough and the field $E$ high enough, a deformed structure is obtained, where the angle between $n_i^1$ and the others is reduced, asymptotically ($E \to \infty$) to $90^o$ giving a pyramidal structure, cf. Section \ref{Tdstrong}. In addition,
there is an overall rotation about $n_i^1$ (or $E_i$), whose rotation sense is reversed for the inverted tetrahedral structure or if the electric field is reversed.

If the molecules are chiral, the chiral tetrahedral phase T can be obtained 
\cite{Fel,PB14}. 
Its $T$ symmetry\footnote{We keep the standard notation for this type of symmetry, confident that confusion with the temperature $T$ does not occur.} lacks any mirror plane and the tetrahedral $\bar 4$ axes of the \Td phase are reduced to (proper) 2-fold axes. There is a pseudoscalar $q_0$ of molecular origin. Similar to the case of cholesteric LC a helical structure of a given handedness reduces the free energy, but only if the helical axis is one of the tetrahedral vectors (the 3-fold symmetry axes). There are appropriate static as well as dissipative Lehmann-type couplings \cite{Lehmann} among rotations of the tetrahedral structure and {\it e.g.} the thermal degree of freedom. In contrast to the \Td phase, there is flow alignment in the chiral T phase, in particular, for simple shear flow (with the vorticity direction along one of the tetrahedral vectors) the tetrahedral structure is rotated about this direction by an angle that depends on $q_0$, but not on the shear rate, \cite{PB14} and Section \ref{chiralT}.

%%%%%%%%%%%%%%%%%%%%%%%%%%%%%%%%%%%%%%%%%%%%%%%%%%

\subsection{Combined tetrahedral and nematic order \label{tetranemaorder}}

In a system where both, nematic ($Q_{ij}$) and tetrahedral order ($T_{ijk}$) exist simultaneously, the overall symmetry of the possible phases depend on the relative orientation of the two structures. This problem is investigated using a Landau description setting up a free energy in terms of the two order parameter tensors
\begin{equation} \label{Genergy}
E_L = E_L^Q + E_L^T + E_L^{(mix)}
\end{equation}
where $E_L^Q$ and $ E_L^T $ are the well-known Landau energy expressions for a (pure) nematic \cite{PGdG} and a (pure) tetrahedral phase \cite{Fel3,RadLub02}, respectively. They contain terms quadratic, cubic and quartic in the order parameter in the first case, but only quadratic and quartic ones in the second case, due to the broken inversion symmetry of $T_{ijk}$. Since the orientational symmetry is broken {\it spontaneously} in a nematic phase as well as in a tetrahedral phase, $E_L^Q$ and $ E_L^T$ cannot depend on the relative orientation of the two structures. Thus, the minimum of the mixed Landau energy 
\begin{eqnarray} \label{Gmixenergy}
E_L^{(mix)} &=& d_1 Q_{il} Q_{jm} T_{ilk} T_{jmk} \nonumber \\ &+& \frac{d_2}{2}   (Q_{im} Q_{jl} T_{ilk} T_{jmk} + Q_{ij} Q_{lm} T_{ilk} T_{jmk})
\end{eqnarray}
gives the ground state of the combined system. It is of fourth order in the order parameters and might therefore be small, in particular close to the phase transition.  We will therefore consider two cases, first the strong coupling limit, where $E_L^{(mix)}$ is large and leads to a rigid relative orientation of the two structures ("correlated order") in Section  \ref{TNcorr}, and the weak coupling limit, where $E_L^{(mix)}$ is neglected ("uncorrelated order") in Section  \ref{TNnotcorr}, where other energies become important. 

For the case of correlated order there are two well-defined distinct geometries (restricting ourselves for the moment to uniaxial nematic order): Either the nematic director is along of one of the 2-fold (or $\bar 4$) symmetry axes of the tetrahedral structure, or it is along one of the 3-fold axes, the tetrahedral vectors. To show that these two possibilities indeed lead to energetic minima, one can take, without loss of generality, the director along the z-axis, with the tetrahedral structure given in Eq. (\ref{unit}) and Eq. (\ref{unitE}), for the first and second case respectively. Indeed, the first case is the ground state, if $d_1 + d_2 >0$, while $d_1 + d_2 <0$ gives the second one.

In the first case (the director along a $\bar 4$ axis) a tetragonal biaxial structure is obtained that lacks inversion symmetry. It is of $D_{2d}$ symmetry and its hydrodynamics is discussed in detail in Section  \ref{D2d}. 
The hallmark of this phase is the existence of ambidextrous helicity \cite{PB14}. Although this phase is not chiral, the formation of a non-homogeneous ground state in the form of a helix is possible. The combined tetragonal biaxial structure rotates about one of the $\bar 4$ axes that are perpendicular to the director, when going along this helical axis. The physical origin is the broken inversion symmetry that allows for a linear gradient energy, $\varepsilon_l = \xi T_{ijk} n_i \nabla_j n_k$,
involving linearly $\nabla_x n_y + \nabla_y n_y$ (with the helical axis along the z-direction). One can discriminate left- and right-handed helices because the inverted tetragonal biaxial structure is different from the original one, in particular, what can be called right-handed for the original structure is left-handed for the inverted one and {\it vice versa}. Obviously, the energetic gain due to the helix does not depend on the handedness and is therefore coined "ambidextrous". In that respect it is similar to the ambidextrous chirality in the smectic C$_{B2}$ and C$_G$ phases, which are however (structurally) chiral, in contrast to the D2d phase, where we call this phenomenon ambidextrous helicity, cf. Section \ref{ambi}.

Another peculiar feature of this phase is the orientational frustration in an external electric field. A nematic director is oriented in an electric field due to the dielectric anisotropy either along the field or perpendicular to it. The orientation energy is proportional to the square of the field strength. In the \Td phase the electric orientation effect is cubic in the field strength and orients one of the tetrahedral vectors parallel or antiparallel to the field direction \cite{Fel3}. In the D2d phase, where the director and the tetrahedral vectors are at an oblique angle $\phi_{D}$ with $\cos \phi_{D} = \pm 1/\sqrt{3}$, the nematic and tetrahedral orientation cannot be achieved simultaneously. As a result, depending on the values of the coupling constants one can obtain {\it e.g.} for small fields the nematic-type of orientation that deviates, however, for higher fields towards the tetrahedral orientation, cf. Section \ref{External}. Similar frustration effects occur for the orientation by boundaries, but not for magnetic fields, since the latter do not orient the tetrahedral structure. 

We also discuss relative rotations in the D2d phase, where the director and the tetrahedral structure deviate from their equilibrium orientation. If the relaxation of this variable is slow enough, it can influence the dynamics of the D2d phase, cf. Section \ref{AddRem}.

If the uniaxial nematic direction $\bm{n}$ along one of the $\bar 4$ axes of the tetrahedral structure is accompanied by (orthogonal) biaxial nematic directions $\bm{m}$ and $\bm{l}$, phases of even lower symmetry occur. Depending on whether the nematic structure is tetragonal or orthorhombic, and on how $\bm{m}$ and $\bm{l}$ are oriented with respect to the tetrahedral vectors, non-polar phases of $S_4$ (S4) and $D_2$ (D2) symmetry occur, cf. Section \ref{D2S4}, as well as polar ones with $C_{2v}$ (C2v) and $C_2$ (C2) symmetry, cf. Section \ref{C32v}. Phase transitions among various tetrahedral phases are described in \cite{RadLub02}. 

In the tetragonal S4 phase the nematic directors $\bm{m}$ and $\bm{l}$ are equivalent, but oriented obliquely within the plane perpendicular to $\bm{n}$, Section \ref{S4}. The $\bar 4$ symmetry axis (along $\bm{n}$) is the only symmetry element left. Its hydrodynamics is rather similar to that of the D2d phase. 

The D2 phase is orthorhombic with all nematic directions along one of the $\bar 4$ tetrahedral directions, Section \ref{D2}. Thereby, the three preferred directions are reduced to 2-fold symmetry axes, which are the only symmetry elements left. For the hydrodynamic description the most important additional feature of the D2 phase (compared to D2d) is its chirality, since it only contains proper rotation axes and no mirror planes anymore. Chirality is due to the structure (and not due to the chirality of molecules) described by the pseudoscalar $q_0 = n_i n_j m_k m_p l_q l_r \epsilon_{ikq} T_{jpr}$ with the orthorhombic nematic directors $\bm{n}$, $\bm{m}$, $\bm{l}$. This gives rise to ambidextrous chirality, since the inverted structure (with the opposite chirality) is energetically equivalent and comes in addition to the ambidextrous helicity that is already present in the D2d phase. To make things even more complicated, both effects favor helices about all three 2-fold axes generating strong frustration due to the steric incompatibility of helices about different axes (similar to the case of biaxial cholesteric LC \cite{PB90}). 

If one orients the orthorhombic directors $\bm{m}$ and $\bm{l}$ within the tetrahedral planes $n^1/n^4$ and $n^2/n^3$ (instead of along the $\bar 4$ axes as in the D2 phase) one gets the C2v phase, Section \ref{C2v}, which is polar, but achiral, since the planes $n^1/n^4$ and $n^2/n^3$ are mirror planes. The polar axis is the (former) $\bar 4$ axis along $\bm{n}$, since a flip of that axis can no longer be compensated by a $\pi/4$ rotation, since $\bm{m}$ and $\bm{l}$ are not equivalent. If $\bm{m}$ and $\bm{l}$ are oriented obliquely within the plane perpendicular to $\bm{n}$, the polar C2 phase occurs, Section \ref{C2}. It is chiral, since the mirror planes are removed. It could also be obtained by replacing the tetragonal biaxial nematic structure of a S4 phase by an orthorhombic one. The hydrodynamics of these polar phases will be discussed together with that of the C3v phase, introduced next.

In the second case of correlated nematic and tetrahedral structure (with the director along one of the {\it tetrahedral} axes, instead of a $\bar 4$ axis) the resulting trigonal biaxial structure is polar. The preferred polar direction, $p_i \sim T_{ijk} Q_{jk}$, is given by the tetrahedral vector along the director.  The hydrodynamics of this C3v phase is rather similar to that of the (uniaxial) polar nematic phase \cite{polnema}, but has one additional hydrodynamic degree of freedom, the rotation about the polar axis, and a reduced symmetry, $C_{3v}$ (compared to $C_{\infty v}$ in polar nematics). This gives rise to a more complicated structure of all rank-3 (and higher) material tensors. A brief discussion of the hydrodynamics is given in Section \ref{C3v}.

In the case $E_L^{(mix)} =0$ there is no energy that locks a homogeneous nematic and a homogeneous tetrahedral structure ({\it "homogeneously uncorrelated"}). Therefore we amend the Landau energy (\ref{Genergy}) by Ginzburg-type gradient terms
\begin{eqnarray} \label{GLenergy}
E_{GL} &=& \gamma (\nabla_k Q_{ij})^2 + \delta (\nabla_k T_{ijl})^2 + {\cal D}\, T_{ijk} \nabla_k Q_{ij}.
\end{eqnarray}
The linear gradient term exists because of the broken inversion symmetry of $T_{ijk}$ and is not related to chirality. It allows for inhomogeneous phases having a lower energy than the homogeneous ones. In Section \ref{TNnotcorr} we discuss as an example, splay-bend textures of the nematic director accompanied by those of the tetrahedral structure \cite{O}. 
In the nematic splay-bend texture the orientation of the director periodically oscillates along an axis in the plane. Taking this axis as one of the $\bar 4$ axes (and a second one to define the splay-bend plane) the tetrahedral splay-bend structure (with the same periodicity) is rotated independently about the second $\bar 4$ axes by a constant angle $\phi$. The total energy of the combined system is negative due to linear gradient term, despite the energy density not being constant in space. The state with a constant energy density is energetically slightly less favorable. This also holds for a generalization of the nematic splay-bend texture with a constant tilt into the third dimension.

In closing this section we briefly address the work 
performed in the area of microscopic and molecular modeling.
Macroscopic and molecular symmetries of unconventional nematic phases
have been studied in detail in Ref. \cite{mettout}.
The analysis presented in \cite{mettout} also includes a polar order 
parameter as well as third rank tensors of tetrahedral/octupolar type
along with a discussion of possible isotropic - nematic phase transitions.
Microscopic models leading to phase diagrams for liquid crystalline
phases formed by bent-core molecules using a generalized Lebwohl-Lasher 
lattice model with quadrupolar and octupolar anisotropic interactions  
were studied in \cite{longa1,longa2,longa3}.
The techniques used include mean field theory and Monte Carlo
simulations. Among the liquid crystalline phases found are 
the tetrahedral phase (\Td in our notation), a tetrahedral nematic phase (D2d) with $D_{2d}$ symmetry, as well as a chiral tetrahedral nematic phase (D2) with
$D_2$ symmetry. In addition, the classical phases expected for rod-like 
molecules, namely the uniaxial nematic phase and the orthorhombic  biaxial one (with $D_{2h}$ symmetry) were found.
In Ref. \cite{longa2} an estimate for the pitch in the chiral tetrahedral nematic 
phase has been presented and the possibility to find ambidextrous chirality 
for the nonchiral nematic phase formed by bent-core molecules was
elucidated. In Ref. \cite{longa3}  
the spontaneous formation of macroscopic domains of opposite optical activity 
has been investigated in the context of bent-core systems and
ferrocenomesogens for two spatial dimensions.

%%%%%%%%%%%%%%%%%%%%%%%%%%%%%%%%%%%%%%%%%%%%%%%%%%%
%%%%%%%%%%%%%%%%%%%%%%%%%%%%%%%%%%%%%%%%%%%%%%%%%%

\section{Tetrahedral Hydrodynamics} \label{Td}

\subsection{Hydrodynamics of the \Td phase without external fields} \label{Tdhydro}

Hydrodynamics is a powerful systematic tool to describe the dynamics of macroscopic systems.
It is applicable to situations where most of the many degrees of freedom have locally relaxed to their equilibrium values, and only a few ones are slow enough to be dealt with explicitly by partial differential equations. The latter class comprises the conserved quantities that cannot relax locally, but can only be transported, like mass, momentum, and energy. In an Eulerian description their local densities, $\rho(\bm{r},t)$, $g_i(\bm{r},t)$, and $\varepsilon(\bm{r},t)$, respectively, are space-time fields obeying the conservation laws 
\begin{eqnarray} 
 \frac{\partial}{\partial t} \rho + \nabla_i g_i &=& 0 \label{rhodyn} \\
\frac{\partial}{\partial t} g_i + \nabla_j \sigma_{ij} &=& 0 \label{gdyn} \\
\frac{\partial}{\partial t} \varepsilon + \nabla_i j_i^\varepsilon &=& 0 \label{epsdyn}
\end{eqnarray}
where the nabla operator denotes partial spatial derivation $\bm{\nabla}= {\partial}/{\partial \bm{r}}$. The mass current $g_i$ in Eq.  (\ref{rhodyn}) is the momentum density, while the stress tensor $\sigma_{ij}$ and the heat current $j_i^\varepsilon$ are still to be determined. Angular momentum conservation does not give rise to an additional dynamic equation, but to some restrictions for the stress tensor. For an extended discussion cf. \cite{MPP,mbuch}.

For isotropic liquids the conserved quantities are the only hydrodynamic variables and Eqs. (\ref{rhodyn})-(\ref{epsdyn}) are the basis for the universal set of Navier-Stokes equations. For fluids with internal structures, like LC, the symmetry or Goldstone variables have to be taken into account, additionally. In the \Td phase the tetrahedral structure breaks 3-dimensional rotational symmetry spontaneously, {\it i.e.}, the orientation of the structure is arbitrary. Any rigid rotation of the structure leads to a different state, which however, has the same internal energy as any of the others. Therefore, there is no restoring force and an $\omega=0$ excitation (Goldstone mode) results. For an inhomogeneous distortion with a characteristic, macroscopic wavelength $k$, the dynamics $\omega(k)$ is slow and vanishes in the hydrodynamic limit $\omega(k)\to 0$ for $k \to 0$. The same behavior is found for the conserved variables as is obvious from Eqs. (\ref{rhodyn})-(\ref{epsdyn}).

From the general changes of the order parameter from its
equilibrium value, $\delta T_{ijk}=T_{ijk}-T_{ijk}^{eq}$, the projection
\begin{equation}
\delta\Gamma_{i}\equiv\frac{1}{4\alpha}\epsilon_{ipq}T_{pkl}\,\delta T_{qkl}
\label{omega}%
\end{equation}
with the conventional normalization $27 \alpha= 32\, N^2$, describes the rotations of the tetrahedral structure according to the broken rotational symmetry. Note that $\delta \Gamma_i$ is even under spatial inversion.
Since $\delta\Gamma_{i}$ is not conserved, its dynamic equation has the form of a balance equation
\begin{equation}
\frac{\partial}{\partial t}\Gamma_{i}+Y_{i}=0 \label{omegadyn}
\end{equation}
with a yet undetermined quasi-current $Y_i$.

Since (finite) rotations in three dimension do not commute, $\delta\Gamma_{i}$ is not a vector, nor are its components rotation angles (except in linear approximation). Indeed,
\begin{align}
(\delta_{1}\delta_{2}-\delta_{2}\delta_{1})\Gamma_{i}  &  =\frac{1}{2\alpha
}\epsilon_{ipq}(\delta_{1}T_{pjk})(\delta_{2}T_{qjk})\nonumber\label{noncommute}\\
&  =2\epsilon_{ipq}(\delta_{1}\Gamma_{p})(\delta_{2}\Gamma_{q})
\end{align}
two subsequent changes cannot be interchanged.
This is similar to {\it e.g.} rotations in biaxial nematics
\cite{Nbiax,biaxnemliu} or of the preferred direction in superfluid $^{3}$He-A \cite{ho}.
Equation  
(\ref{omega}) can be inverted to give $\delta T_{qkl}=2\,\epsilon_{ipq} T_{pkl}\,\delta\Gamma_{i}$.
It is easy to check that this special $\delta T_{ijk}$ fulfills the requirements for the absence of polar order, $\delta T_{ijj}=0$, and of nematic order, $\delta T_{ikl}T_{jkl}+T_{ikl}\delta T_{jkl}=0$. 
 
All the relations above are given in terms of Cartesian coordinates, which however only serve as a proxy for rotationally invariant descriptions using vectors and tensors and their appropriate products. Therefore, we can use for actual calculations any representation of $T_{ijk}$ that is suitable. Most of the time we use the orientations given in Eqs. (\ref{unit}) and (\ref{unitE}). 

In addition to the hydrodynamic variables discussed above, there are systems or situations, where a few mesoscopic, fast-relaxing degrees of freedom become slow enough to be relevant for a macroscopic description. Examples are elastic strains in viscoelastic fluids or the scalar nematic order parameter $S$ in nematic LC close to phase transitions and in the vicinity of defect cores.
In the \Td phase the strength of tetrahedral order, $N$, is of that character. Most of the time we will assume that $N$ has already relaxed to its equilibrium value $N_0$, which one can then take as unity. When deviations $\delta N = N - N_0$ are relevant for the macroscopic dynamics, the balance equation
\begin{equation}
\frac{\partial}{\partial t}N+X=0 \label{Tdyn}
\end{equation}
with the quasi-current $X$ will be used. In contrast to the symmetry variables, however, even homogeneous changes $\delta N$ cost energy and lead to a finite restoring force. As a consequence, appropriate excitations are non-hydrodynamic $\omega(k \to 0) = \omega_0 $
with a finite gap $\omega_0\neq 0$. Some aspects of the order dynamics will be discussed at the end of this section.

In order to set up the complete dynamics of the \Td phase we apply thermodynamics locally to the relevant variables. The first law of thermodynamics (Gibbs relation), describing energy conservation including heat, 
\begin{equation} \label{Gibbs}
d \varepsilon = \mu d\rho + v_i d g_i + h_{i}^{\Gamma\prime} d \Gamma_i + \Psi^\Gamma_{ij} d\nabla_j \Gamma_i + T d\sigma
\end{equation}
relates changes of all variables  to changes of the entropy density $d \sigma$. The prefactors are the conjugate quantities, chemical potential $\mu$, velocity $v_i$, molecular tetrahedral fields $h_{i}^{\Gamma\prime}$, $\Psi^\Gamma_{ij}$, and temperature $T$. Like in nematic LC we have added gradients of the symmetry variable, since without external fields $h_{i}^{\Gamma\prime}$ has to vanish. The two contributions can be combined to $h_i^\Gamma d \Gamma_i$ with 
\begin{equation}\label{hGamma}
h_i^\Gamma = h_i^{\Gamma\prime} - \nabla_j \Psi^\Gamma_{ij} - 2 \epsilon_{ikl}\Psi^\Gamma_{kj}  \nabla_j \Gamma_l
\end{equation}
The last (nonlinear) contributions is due to Eq. (\ref{noncommute}).

Setting up a phenomenological expression for the energy density, 
\begin{equation} \label{vareps}
 \varepsilon = \varepsilon_0 + \frac{1}{2\rho} \bm{g}^2  +  \frac12 K_{ijkl}^\Gamma (\nabla_j \Gamma_i)(\nabla_l \Gamma_k)
 \end{equation}
the conjugate quantities follow by partial derivation, $\mu= \partial \varepsilon/\partial \rho$, $v_i= \partial \varepsilon/\partial g_i$, $h_{i}^{\Gamma\prime}= \partial \varepsilon/\partial \Gamma_i$, $\Psi^\Gamma_{ij}= \partial \varepsilon/\partial \nabla_j \Gamma_i$, and $T= \partial \varepsilon/\partial \sigma$. The first contribution to Eq. (\ref{vareps}) is the free energy density of an isotropic liquid expressed by well-known susceptibilities, like compressibility, specific heat, and thermal expansion. The second one is the kinetic energy and the last one is the gradient energy for inhomogeneous rotations. It contains three (Frank-like) susceptibilities
\begin{align}
K_{ijkl}^{\Gamma}  &  =K_{1}^{\Gamma}(\delta_{ij}\delta_{kl}+\delta_{il}%
\delta_{jk})+K_{2}^{\Gamma}\delta_{ik}\delta_{jl} \nonumber
\\
&  +K_{3}^{\Gamma}T_{jlp}T_{ikp}\text{\,.} \label{kgamma}%
\end{align}
completing the statics of the \Td phase.

The Gibbs relation allows to interchange the entropy density with the energy density as a dynamic variable 
\begin{equation} \label{sigmadyn}
\frac{\partial}{\partial t} \sigma + \nabla_i j_i^\sigma = R /T
\end{equation}
making Eq. (\ref{epsdyn}) redundant. The source term, the entropy production, is written in terms of the dissipation function $R$. With the help of the dynamic Eqs. (\ref{rhodyn}) - (\ref{epsdyn}), (\ref{omegadyn}) and (\ref{sigmadyn}) the Gibbs relation (\ref{Gibbs}) can be written as
\begin{equation}\label{dissfunction}
R= -\sigma_{ij} \nabla_j v_i - j_i^\sigma \nabla_i T + Y_i  \,h_i^\Gamma + g_i  \nabla_i \mu \geq 0
\end{equation}
suppressing an irrelevant surface term ($ \textrm{div}  \bm{j}$). 

According to the second law of thermodynamics, entropy is conserved, {\it i.e.} $R=0$, only for reversible processes, while irreversible ones are dissipative with $R>0$. It is possible to write any current or quasi-current as a sum of a reversible (superscript R) and a dissipative (superscript D) part, $\sigma_{ij} = \sigma_{ij}^R + \sigma_{ij}^D$, $j_i^\sigma =j_i^{ \sigma, R} + j_i^{\sigma, D}$, and $Y_i = Y_i^R + Y_i^D$, while $g_i$, the momentum density is reversible and has no irreversible part. The reversible (dissipative) parts have the same (opposite) time reversal behavior as the time derivative of the appropriate variable. All variables (and conjugates) have a definite time reversal behavior, {\it e.g.}, $\sigma$, $\rho$, $T$, $\mu$, $\nabla_j\Gamma_i$, $\Psi^\Gamma_{ij}$, and $\varepsilon$ are invariant, while $g_i$ and $v_i$ change sign implying that also $\sigma_{ij}^R$, $j_i^{\sigma,D}$, and $Y_i^D$ are invariant and $\sigma_{ij}^D$, $j_i^{\sigma,R}$, and $Y_i^R$ change sign.

The framework of linear irreversible thermodynamics is used to derive the irreversible parts of the currents and quasi-currents. It has a solid microscopic basis in linear response theory guaranteeing compatibility with general physical principles. Descriptions based on linear irreversible thermodynamics have successfully applied even to systems driven far from equilibrium. 
It is based on a linear relationship between the irreversible currents and the thermodynamic forces that drive the system out of equilibrium. In equilibrium, $T^{(eq)}$, $\mu^{(eq)}$, and $v_i^{(eq)}$ are constant (the latter is typically put to zero due to Galilean invariance), while $h_{i}^{\Gamma(eq)}$ is zero. Thus, $\nabla_i T$, $\nabla_i \mu$, $\nabla_j v_i$, and $h_{i}^\Gamma$ are candidates for thermodynamic forces. However, $\nabla_i \mu$ must not enter the dissipation function, since $g_i$ is reversible, and only symmetrized gradients $A_{ij} = (1/2)(\nabla_j v_i + \nabla_i v_j)$ are allowed, since solid body rotations must not change the entropy of the system\footnote{A solid body rotation is equivalent to a change of the point, from which a system is viewed.}.

Taking into account spatial inversion behavior additionally, we find the symmetry-allowed contributions
\begin{equation}  \label{diss}
 R= \frac{\kappa}{2} (\bm{\nabla} T)^2 + \frac{1}{2\gamma^\Gamma} (\bm{h}^\Gamma)^2
+ \frac12 \nu_{ijkl} A_{ij} A_{kl} 
 \end{equation}
with the (dissipative) transport parameters, heat conduction $T \kappa$, and tetrahedral rotational viscosity $\gamma^\Gamma$, and with the viscosity tensor $\nu_{ijkl}$. The first two terms are isotropic, while the rank-4 material tensor has a form different from the isotropic case \cite{Fel2}
\begin{eqnarray}\label{viscos}
\nu_{ijkl}&=&\eta_{1}(\delta_{jl}\delta_{ik} + \delta_{il} \delta_{jk}
- \frac {2}{3} \delta_{ij} \delta_{kl})+  \zeta \delta_{ij}\delta_{lk}  \nonumber \\ &&+ \eta_{2} T_{ijp} T_{klp}
\end{eqnarray}
with an additional deformational viscosity, $\eta_2$, due to the tetrahedral order. It leads to additional stresses in  symmetric shear flows, but not in elongational ones. Positivity of $R$ requires some positivity conditions on the transport parameters, in particular $\gamma^\Gamma >0$, $\eta_1>0$, $\zeta - (2/3)\eta_1 >0$ and $\eta_1 + \eta_2 >0$. 

The dissipative currents follow from Eq. (\ref{diss}) by partial derivation
\begin{eqnarray}
 j_i^{\sigma, D} &=& - \kappa \nabla_i T  \label{jsigmadiss} \\
Y_i^D &=& \frac{1}{\gamma^\Gamma} h_i^\Gamma  \label{Ydiss} \\
\sigma_{ij}^D &=& \nu_{ijkl} A_{kl} \label{sigmaijdiss}
\end{eqnarray}

The reversible part of the dynamics has two origins. Either there are phenomenological reversible contributions to the currents, or the contributions are due to symmetry and/or other general requirements. The first kind comes with phenomenological reversible material coefficients, similar to the case of dissipative ones. However, in the reversible case there is no potential from which such contributions could be derived, since $R=0$. Instead, one looks for couplings between reversible currents and forces that are allowed by time reversal symmetry and inversion symmetry and choose the phenomenological parameters such that $R=0$. In isotropic liquids no such phenomenological reversible couplings exist. On the other hand, in nematic LC the reversible couplings between director rotations and simple shear flow (flow alignment and back flow) are a well-known example. Its phenomenological parameter, generally called $\lambda$, determines the director alignment angle under shear flow. In the \Td phase there is no flow alignment effect, since there is no preferred direction that could be aligned, but there is a (reversible) coupling between flow and the thermal degree of freedom
\begin{eqnarray}\label{sigmarev}
\sigma_{ij}^{ph} &=& -  \Gamma_2 T_{ijk} \nabla_k T 
\\ \label{heatrev} 
j_{i}^{\sigma, ph}&=& 
 \Gamma_2 T_{ijk} A_{jk} 
\end{eqnarray}

\begin{figure}      
\begin{center}
\includegraphics[width=7.5cm]{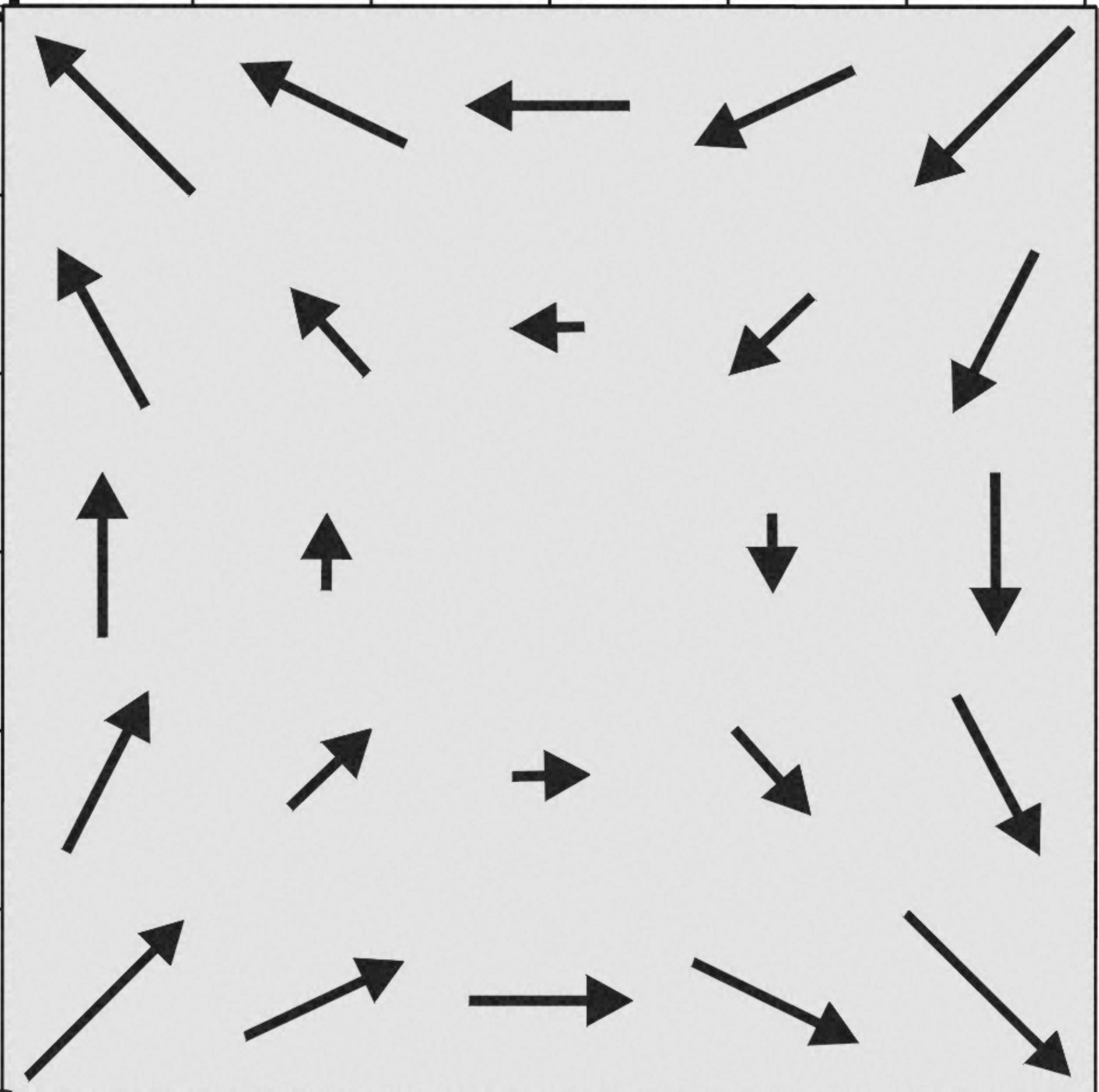}
\caption{Planar symmetric shear flow induced by a temperature gradient perpendicular to the shear plane - from \cite{BPC02}. \label{fig:5}}
\end{center}
\end{figure}

In particular, a temperature gradient generates (symmetric) shear stresses, the geometry of which depends on the orientation of the temperature gradient with respect to the tetrahedral orientation. Assuming the temperature gradient (taken as z-axis) is along one of the $\bar 4$ axis, the stresses induced by Eq. (\ref{heatrev}) are $\sigma_{xy} = \sigma_{yx}$ according to the structure Eq. (\ref{unit}). Due to the viscous stress - strain rate coupling, there is a stationary planar flow pattern perpendicular to the temperature gradient shown in Fig. \ref{fig:5}.

{\it Vice versa}, a shear flow, say, in the x/y plane that defines the perpendicular directions ($\bm{z}$ or $\bm{-z}$) produces via Eq. (\ref{sigmarev}) a heat flux in a definite polar direction ($\bm{z}$ for $\Gamma_2<0$ or $\bm{-z}$ for $\Gamma_2>0$). Of course, 
this does nor imply that the T$_d$ phase is polar.  
If the tetrahedral structure is inverted, $T_{ijk}$ changes sign and the induced currents will point in the opposite direction, what could be viewed as induced ambi-polarity. If both variants are present in different parts of the same sample, this ambi-polarity shows up directly.
%%%%
%%%%
%%%%
%%%%

The reversible transport parameter $\Gamma_2$ is 
without a rigorous upper bound in magnitude
and can have either sign. 
It is easy to check that these contributions cancel each other in the entropy production, Eq. (\ref{dissfunction}), since $\sigma_{ij}^{ph} A_{ij}= - j_i^{\sigma, ph} \nabla_i T $.

The non-phenomenological contributions to the reversible currents are mainly due to transport. In the Eulerian description of hydrodynamics a variable can change its value at a given point by advection, {\it i.e.} by transporting (with velocity $v_i$) material with a different value of that variable to this point, {\it e.g.} $(\partial/\partial t)^{(adv)} \rho = -\nabla_i (\rho v_i)$. For vectorial quantities also convection
(with vorticity $\bm{\omega} = (1/2)\textrm{curl} \,\bm{v}$) is possible. In particular, in a linearized description, where $\delta \Gamma_i$ is a vector of rotation angles, one gets $(\partial/\partial t)^{(trans)} \Gamma_i = - v_j \nabla_j \Gamma_i  + \omega_i$. 

All these transport contributions have to add up to zero in the entropy production. This is obtained by counter terms in the stress tensor, involving the isotropic pressure $p$ and a nonlinear stress $\sigma_{ij}^{nl}$, which exists in similar form also in nematic LC, where it is called Ericksen stress. For an exposition of the method and its application to tetrahedral phases cf. \cite{mbuch,BP-D2d}. The final, nonlinear result for the total reversible currents is
\begin{eqnarray}
 j_i^{\sigma, R} &=& \sigma v_i  + j_i^{\sigma, ph} \label{jsigmarev} \\
Y_i^R &=& v_j \nabla_j \Gamma_i - \frac12 \omega_i +\frac{1}{2\alpha} \epsilon_{ipq} \epsilon_{mjl} T_{pjk} T_{qlk} \,\omega_m \label{Yrev} \\
\sigma_{ij}^R &=& v_i g_j + \delta_{ij} p +\sigma_{ij}^{nl} + \sigma_{ij}^{ph} \label{sigmaijrev}
\end{eqnarray}
with $4 \sigma_{ij}^{nl} = 2 \Psi^\Gamma_{kj} \nabla_i \Gamma_k + 
2 \Psi^\Gamma_{ki} \nabla_j \Gamma_k - 3 \epsilon_{ijk} \nabla_l \Psi^\Gamma_{kl}$.
In Eqs. (\ref{sigmarev}) - (\ref{sigmaijrev}), 
as well as in the following, the superscript $ph$ refers to the
reversible parts of the currents that carry phenomenological coefficients the value of
which cannot be simply determined by invariance arguments.

The last term in Eq. (\ref{Yrev}) demonstrates that in the nonlinear domain $\delta \Gamma_i$ does not behave like an ordinary vector under finite rotations. There is no phenomenological reversible coupling to symmetrized shear flow, with the effect that the tetrahedral structure cannot be oriented in simple shear flow. In the dynamic momentum Eq. (\ref{gdyn}), the pressure term appears as $\nabla_i p$, which is given by the Gibbs-Duhem equation
\begin{equation} \label{GibbsDuhem}
\nabla_i p = \sigma \nabla_i T + \rho \nabla_i \mu + g_j \nabla_i v_j - h_j ^\Gamma \nabla_i \Gamma_j
\end{equation}
where $h_i^\Gamma$ is defined in Eq. (\ref{hGamma})
Note that the stress tensor is either symmetric or the divergence of an antisymmetric tensor, which is the requirement of angular momentum conservation \cite{MPP}. This form is obtained by applying a condition on the conjugate quantities that follows from the rotational invariance of the energy density $\varepsilon$.

Very often LC are mixtures of several different components. For a binary mixture, whose components are individually conserved, there are two mass conservation laws for $\rho_1$ and $\rho_2$. They can be replaced by the total mass conservation Eq. (\ref{rhodyn}) and a dynamic equation for the concentration $c= \rho_1/\rho$
\begin{equation}
\frac{\partial}{\partial t}c+ v_i \nabla_i c  + \frac{1}{\rho}\, j_i^c =0 \label{cdyn}
\end{equation}
whose conjugate quantity, the osmotic pressure, $\Pi = \rho \mu_c$, is related to the difference of the chemical potentials $\mu_c = \mu_1 - \mu_2$. It follows from an appropriately extended $\varepsilon_0$ in Eq. (\ref{vareps}). The concentration variable is rather similar to the temperature variable, and $j_i^c$ has the same structure as $j_i^\sigma$. In particular, the dissipative part contains diffusion and thermodiffusion, 
\begin{equation} \label{jcdiss}
j_i^{c,D} = - D \nabla_i  \mu_c - D_T \nabla_i T
\end{equation}
where the latter also occurs in Eq. (\ref{jsigmadiss}) as $-D_T \nabla_i \mu_c$. There is also the tetrahedral-specific reversible coupling to flow, $j_{i}^{c, ph}= 
 \Gamma_3 T_{ijk} A_{jk} $, with the counter term $-  \Gamma_3 T_{ijk} \nabla_k \mu_c $ in the stress tensor, Eq. (\ref{sigmarev}).

%%%%%%%%%%%%%%%%%%%%%%%%%%%%%%%%%%%%%%%%%%%%%%%%

\subsection{Orienting, flow- and order-inducing external fields} \label{Tdfields}

In nematic LC it is well known that an external static electric fields orients the director due to the dielectric anisotropy effect. In equilibrium the director is either along the field or perpendicular to it. 
Therefore, rotations away from the field direction in the former case, or out of the perpendicular plane in the latter case, are no longer Goldstone modes, but lead to a relaxation towards the equilibrium orientation. Since this is typically a weak coupling, it is customary to keep director rotations as variables to describe the macroscopic dynamics, and taking the symmetry unchanged as D$_{\infty h}$. In the same spirit we will treat the tetrahedral phase in external fields in this section.

The electrostatic degree of freedom is described by the electric field $E_i$ and the displacement field $D_i$. In the Gibbs relation (\ref{Gibbs}) this adds the electric energy change $d \varepsilon^E = E_i d D_i$. To make contact with the familiar description in nematic LC, we switch to the Legendre transformed energy $\tilde \varepsilon^E \equiv \varepsilon^E - E_i D_i$ giving rise to $d \tilde \varepsilon^E = - D_i d E_i $. The variable $D_i$ is thereby given by the thermodynamic conjugate (and generalized force) $E_i$ via $D_i = - (\partial \tilde \varepsilon^E /\partial E_i) $. The tetrahedral symmetry allows for a cubic electric field energy \cite{Fel}
\begin{equation} \label{cubicE}
\tilde \varepsilon^E  = - \epsilon_1 T_{ijk} E_i E_j E_k
\end{equation}
that orients the tetrahedral structure such that one of the tetrahedral vectors, say $n^1_i$,  is parallel or antiparallel to the field $E_i$ depending on the sign of the susceptibility $\epsilon_1$. 
Switching from the tetrahedral structure to its inverse ($T_{ijk} \to - T_{ijk}$ or $N \to -N$) the role of $\mathrm{sign}(\epsilon_1)$ is reversed.

In principle, other vectorial external fields, like temperature or concentration gradients can have a similar orienting effect via an energy like $\varepsilon^T  \sim T_{ijk} (\nabla_i T) (\nabla_j T) (\nabla_k T)$. Boundaries with a polar surface normal act in the same way orienting one tetrahedral vector perpendicular to the surface.
A magnetic field $H_i$ cannot orient the tetrahedral structure due to the odd time reversal behavior of magnetic fields, but there is an additional orienting effect, when both, electric and magnetic fields are present, since the energy $\tilde \varepsilon^{EH} = - (\epsilon_2 / 3) T_{ijk} (E_i H_j H_k + E_j H_i H_k + E_k H_i H_j)$ is possible \cite{BPC02}.
The electric field energies lead to quadratic contributions to the displacement field that come in addition to the linear, isotropic one
\begin{equation} \label{Dstat}
D_i = \chi E_i + \epsilon_1 T_{ijk} E_j E_k + \epsilon_2 T_{ijk} H_j H_k. 
\end{equation}

For the relaxation of the tetrahedral structure in an external field, the orienting energy Eq. (\ref{cubicE}) provides the driving force. Since it defines the equilibrium orientation, it is zero for linear deviations, since $E_i E_j E_k \delta T_{ijk} =0$ (with $\delta T_{ijk} = 2\,\epsilon_{mpi} T_{pjk}^{eq}\,\delta\Gamma_{m}$), while for quadratic ones one gets 
\begin{equation}\label{orientation}
\tilde \varepsilon^E_2 = 
\epsilon_1 E_i E_j E_k  \,\delta^{(2)} T_{ijk}  = 
\frac{32}{9} \vert \epsilon_1 N \vert  E_0^3 (\delta \bm{\Gamma}_\perp)^2
\end{equation}
with $\delta^{(2)} T_{ijk} \equiv 2\,\epsilon_{mpi} \,\delta T_{pjk}\,\delta\Gamma_{m}$. It describes the energy related to rotations of the tetrahedral structure perpendicular to an external field of strength ${E}_0$. As a result there is a non-vanishing restoring force, even in the homogeneous case, $h_i^\Gamma  = h_i^{\Gamma \prime} = \partial \tilde \varepsilon^E / \partial \delta \Gamma_i$ and transverse rotations relax with the inverse relaxation time
\begin{equation}\label{relaxation}
\lambda^{rel} = 16 \,\frac{\vert \epsilon_1  N \vert}{\gamma^\Gamma} \, E_0^3
\end{equation}
according to the dissipative dynamics, Eq. (\ref{Ydiss}). The relaxation is cubic in the field strength, in contrast to the nematic case, where it is quadratic.

There is also the analog of flexoelectricity
\begin{equation}\label{flexo}
\tilde \varepsilon^{flex}_2 = e_1E_j E_k  \nabla_i T_{ijk}  = \frac{16}{9} e_1 N   E_0^2 \,(\textrm{curl} \bm{\Gamma})_z
\end{equation}
involving $\nabla_x \Gamma_y - \nabla_y \Gamma_x$ when the field 
is in z-direction. Here, the effect is quadratic in the field amplitude, 
while in nematic LC it is linear.

The existence of $T_{ijk}$ also allows for piezoelectricity
\begin{equation}\label{piezo}
\tilde \varepsilon^{piezo}_2 = d_1 T_{ijk} E_i u_{jk}
\end{equation}
in solid systems, where a strain tensor $u_{ij}$ describes elasticity.

\begin{figure}      
\begin{center}
\includegraphics[width=8.3cm]{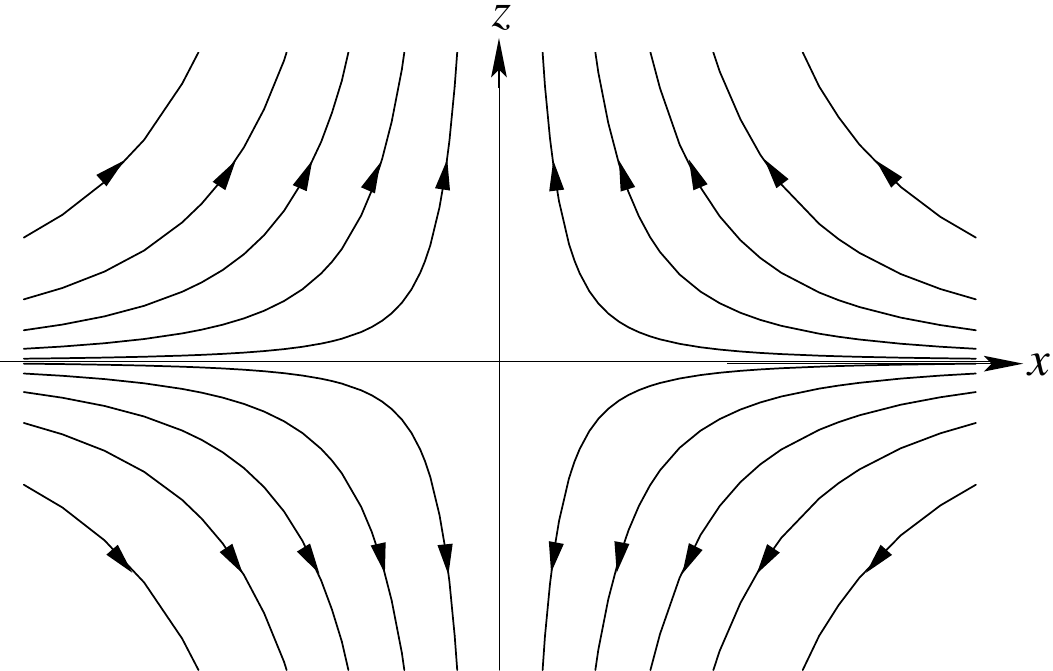}
\caption{Uniaxial 3-dimensional elongational flow in z-direction for $\Gamma_1 >0$; for $\Gamma_1<0$ the flow directions are reversed (biaxial elongational flow in the x/y plane) - from \cite{OMueller}.\label{fig:6}}
\end{center}
\end{figure}

The dynamics is also affected by an external field, in particular if there are
electric charges, $\rho_e$, as is often the case for LC. They are related to the displacement field by $\nabla_i D_i = \rho_e$ and the dynamic balance equation for $D_i$ is the charge conservation law 
\begin{equation} \label{chargedyn}
\frac{\partial}{\partial t} \rho_e + \nabla_i j_i^e = 0
\end{equation}
The reversible part of the electric current \cite{BPC02}
\begin{equation} \label{jerev}
 j_i^{e,R} =  v_i \nabla_i \rho_e + \Gamma_1 T_{ijk} A_{jk}
\end{equation}
contains a phenomenological coupling to flow, similar to the thermal and solutal currents in Eq. (\ref{heatrev}) and after (\ref{cdyn}). It is balanced to give zero entropy production by a contribution to the stress tensor
\begin{equation}\label{sigmarevel}
\sigma_{ij}^{e,ph} = -  \Gamma_1 T_{ijk} E_k 
\end{equation}
in analogy to Eq. (\ref{sigmarev}). If the electric fields (along the z-axis) orients the tetrahedral structure in the way described above, the stresses induced by Eq. (\ref{sigmarevel}) are $\sigma_{xx} = \sigma_{yy} = -(1/2) \sigma_{zz}$ giving rise (via viscous coupling) to 3-dimensional elongational flow, called uniaxial (biaxial) for $\Gamma_1 >0$ ($\Gamma_1<0$), cf. Fig. \ref{fig:6}. Note that a reversal of the field
direction also changes the velocity directions and interchanges uniaxial with biaxial elongational flow.
%%%%
%%%%
%%%%
%%%%

The dissipative part of the electric current \cite{BPC02}
\begin{eqnarray} \label{jedissE}
 j_i^{e, D} &=& \sigma^E E_i + D^E \nabla_i \mu_c + \kappa^E \nabla_i T 
 \end{eqnarray}
describes the coupled diffusions among the electric, thermal and concentration degrees of freedom (e.g. Ohm's law, Peltier effect, etc.) with appropriate counter terms in the heat and concentration current, Eq. (\ref{jsigmadiss}) and (\ref{jcdiss}). Here, the material parameters relating currents with forces are rank-2 material tensors that, however, are isotropic, since tetrahedral order cannot occur in such tensors. The
dissipative transport parameters have to fulfill 
certain positivity relations ({\it e.g.} $\sigma^E>0$ or $\kappa \sigma^E > [\kappa^E]^2$) to guarantee $R>0$.

In an isotropic liquid an external electric field can induce nematic order (Kerr effect \cite{PGdG}), which also leads to a shift of the thermodynamic isotropic to nematic phase transition. The effect is based on the dielectric anisotropy of the molecules and is quadratic in the field amplitude. In the \Td phase 
tetrahedral order provides an additional mechanism for inducing nematic order, as well as for shifting the nematic phase transition. However, both the strength of induced nematic order and the transition shift are {\it linear} in the field strength. This is due to the interaction energy \cite{N}
\begin{equation} \label{linEQ}
\tilde \varepsilon_{EQT} = \Gamma \, E_i Q_{jk} T_{ijk}
\end{equation}
which is linear in the field amplitude. This energy exists, since with Eq. (\ref{unitE}) $E_k T_{ijk}$ has the structure of uniaxial nematic order $(\delta_{iz} \delta_{jz} - (1/3) \delta_{ij})\sim Q_{ij}$. The coefficient $\Gamma$ is a true scalar quantity that can have either sign.

\begin{figure}      
\begin{center}
\includegraphics[width=8.3cm]{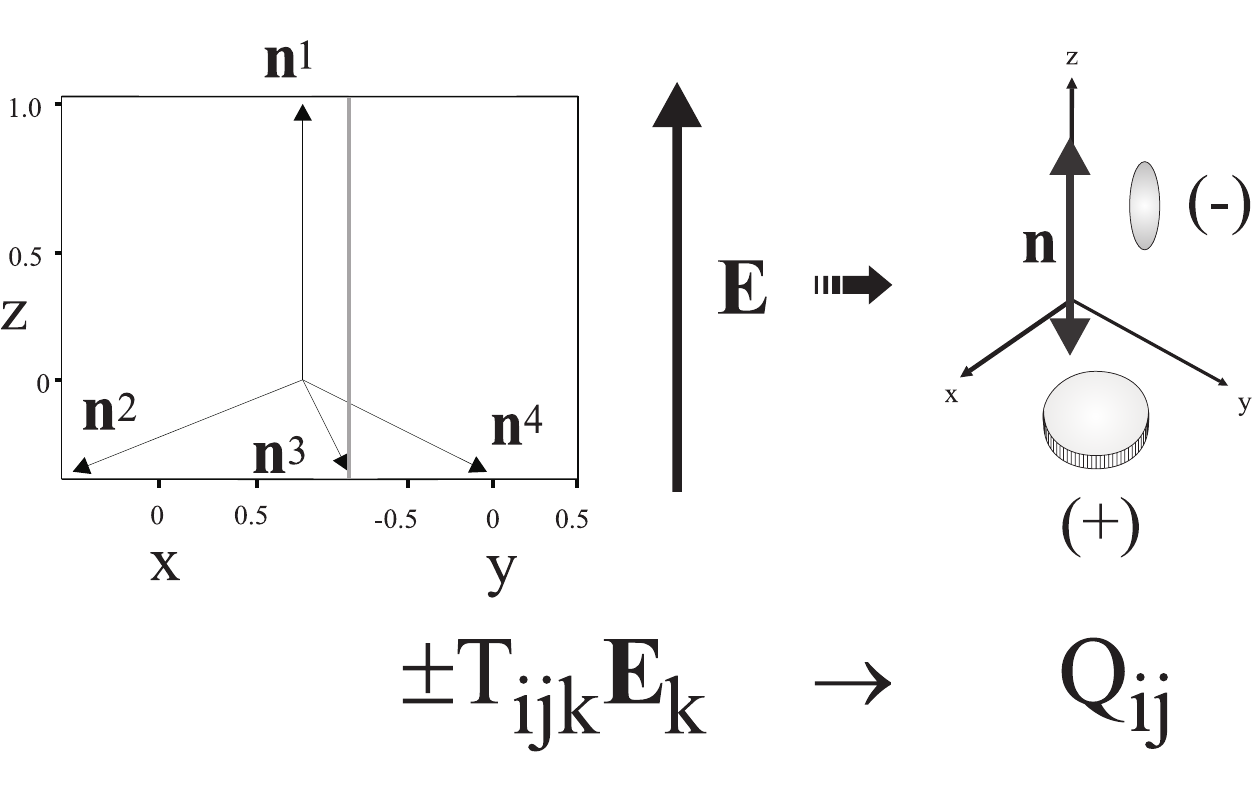}
\caption{Electric field induced nematic order, prolate (oblate) for $\Gamma >0$ ($<0$) - from \cite{N}. \label{fig:7}}
\end{center}
\end{figure}

Assuming in the \Td phase small unstable nematic fluctuations with energy \cite{pgdg2}
\begin{equation} \label{QQ}
\varepsilon_Q = (A/2) Q_{ij} Q_{ij} 
\end{equation}
with $A>0$, an electric field can stabilize them due to the coupling Eq. (\ref{linEQ}). Minimizing $\tilde \varepsilon_{EQT}  + \varepsilon_Q$ the stable induced nematic order is linear in the field strength
\begin{equation} \label{inducedQ}
Q_{ij}^{ind} = \frac{\Gamma}{A} \, E_k T_{ijk}
\end{equation}
and of prolate or oblate geometry depending on the sign of $\Gamma$, Fig. \ref{fig:7}.
%%%%
%%%%
%%%%
%%%%

The onset of the nematic phase transition is also shifted due to the interaction energy, Eq. (\ref{linEQ}). The Landau expansion for the onset of nematic order reads
\begin{equation} \label{LandauEQT}
\tilde \varepsilon = \frac49 \Gamma E_0 N S - \frac{\epsilon_a}{6} E_0^2 S + \frac{\alpha}{2} S^2 + \frac{\beta}{3} S^3 + \frac{\gamma}{4} S^4
\end{equation}
containing the linear (tetrahedral) and the quadratic (isotropic) electric contributions in addition to the field-free expansion of the nematic order parameter $S$; $N_0$ is the constant tetrahedral order parameter. We note that the prefactors of $S^2$, $S^3$ and $S^4$ in Eq. (\ref{LandauEQT}) 
are different from the prefactors one obtains 
when starting with $Q_{ij}$ instead of $S$.

Without the external field the first order transition takes place at $\alpha= \alpha_c \equiv (2 \beta^2 /9 \gamma) $, where a finite $S_c = - (2\beta)/(3 \gamma)$ occurs.
With the standard form $\alpha = a (T-T^*)$, where $T^*$ is the fictitious second order phase transition temperature, the transition temperature is $T_c = T^* + \alpha_c /a$. Assuming the changes due to the field to be small, the transition temperature and the order parameter jump are shifted
\begin{eqnarray} 
 T_c (E_0) &=& T_c + 3 \frac{\gamma}{\beta} \frac{L_E}{a}  \label{Tcshift} \\
S_c (E_0) &=& S_c + \frac92 \frac{\gamma}{\beta} L_E \label{Scshift} 
\end{eqnarray}
with $L_E = (4/3) \Gamma N E_0 - (1/6)\epsilon_a E_0^2$ demonstrating the linear field shift in the tetrahedral phase, in addition to the quadratic, isotropic one.

%%%%%%%%%%%%%%%%%%%%%%%%%%%%%%%%%%%%%%%%%%%%%%%%

\subsection{Strong, structure-changing external fields} \label{Tdstrong}

In this section we will deal with strong external electric 
fields that are able to distort the tetrahedral structure of the \Td phase. This might be possible, when the tetrahedral order is weak, in particular close to the phase transition or in the vicinity of defects. 

An external electric field not only provides the orientation of the (rigid) tetrahedral structure as described in the preceding section, it also invokes a torque on the individual tetrahedral vectors. Taking for definiteness $\bm{ E} = E_0 \bm{ n}^1$, with $E_0 \geq 0$, $\epsilon_1 > 0$,  and $N>0$ the electric free energy Eq. (\ref{cubicE}) 
\begin{equation} \label{cubicEn}
\tilde \varepsilon^E = - {\epsilon_1} N \sum^{4}_{\beta=1} (\bm{ E \cdot n}^{\beta})^3
\end{equation}
leads to the non-vanishing torques
\begin{eqnarray}\label{deltan2}
  \bm{ n}^2 \times \frac{\partial\tilde \varepsilon^E}{\partial \bm{ n}^2}  &\sim&
  (0,1,1) \\ \label{deltan3}
\bm{ n}^3  \times   \frac{\partial\tilde \varepsilon^E}{\partial \bm{ n}^3}  &\sim&
  (-1,0,-1) \\ \label{deltan4}
\bm{ n}^4 \times   \frac{\partial\tilde \varepsilon^E}{\partial \bm{ n}^4}
  &\sim&  (1,-1,0).
\end{eqnarray}
where we have used the orientation (\ref{unit}) for the tetrahedral vectors.
These torques are perpendicular to $\bm{ n}^1$ and tend to rotate $\bm{ n}^{2,3,4}$.
Assuming such a rotation with a (yet undetermined)
finite amplitude $b$, the rotated unit vectors are given by
\begin{eqnarray}\label{niEbeta}
\bm{ n}^{2 E} & = & \frac{1}{\sqrt{3}} \frac{1}{\sqrt{1 + 2 b^2}}
(1,- 1 + \sqrt{3}  b, 1 + \sqrt{3}  b) \\
\bm{ n}^{3 E} & = & \frac{1}{\sqrt{3}} \frac{1}{\sqrt{1 + 2 b^2}}
(- 1 - \sqrt{3} b, 1, 1 - \sqrt{3}  b) \\
\bm{ n}^{4E}  & = & \frac{1}{\sqrt{3}} \frac{1}{\sqrt{1 + 2 b^2}}
(- 1 + \sqrt{3} b, - 1 - \sqrt{3} b, - 1).
\end{eqnarray}
while $\bm{ n}^{1 E} = \frac{1}{\sqrt{3}} (1,1,-1) = \bm{ n}^1 
$ is undistorted. 

%%%%
%%%%
\begin{figure}      
\begin{center}
\includegraphics[width=7.5cm]{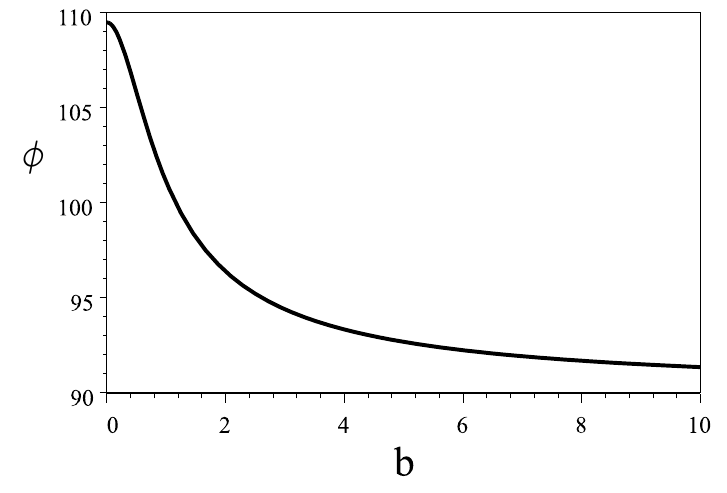}
\caption{The angle $\phi$ between $\bm{n}^1$ and $\bm{n}^{\bar \beta E}$ as a function of the deformation amplitude $b$. In the asymptotic limit $b \to \infty$ the latter are perpendicular to the field direction $\bm{n^1}$ - from \cite{L}. \label{fig:8}}
\end{center}
\end{figure}
%%%%
%%%%

These individual rotations distort the tetrahedral structure as is manifest in the relative angles, whose cosines are given by 
\begin{equation} \label{scalar1}
\bm{ n}^{1E}\bm{ \cdot n}^{\bar \beta E} =  - \frac{1}{3} \frac{1} {\sqrt{1 + 2 b^2}}
\end{equation}
for $\bar \beta=2,3,4$ and by
\begin{equation} \label{scalar2}
\bm{ n}^{2 E} \bm{\cdot n}^{3 E} =
\bm{ n}^{2 E} \bm{\cdot n}^{4 E}=
\bm{ n}^{3 E} \bm{\cdot n}^{4 E}=
- \frac{1}{3}\, \frac{1+3b^{2}}{1 + 2b^2}
\end{equation}
Equation (\ref{scalar1}) describes the rotation towards the field direction, where the tetrahedral angles $\phi (b=0) \equiv \phi_T \approx 109.5 ^o$ (or $\cos \phi_T = -1/3$) approach $\phi = 90^o$ in the asymptotic limit $b \to \infty$ (Fig. \ref{fig:8}). In parallel, the angles among the vectors $\bm{ n}^{\bar \beta E}$ change from $\phi = \phi_T$ in the tetrahedral case, $b=0$, towards $\phi = 120^o$ for $b \to \infty$, resulting in a regular pyramidal structure. For finite $b$ the distorted structure is of $C_{3v}$ symmetry with only one 3-fold symmetry axis left, $\bm{ n}^{1E}$, and three equivalent vertical symmetry planes containing this preferred axis and any of $\bm{n}^{\bar \beta E}$.

%%%%
%%%%
\begin{figure}      
\begin{center}
\includegraphics[width=7.5cm]{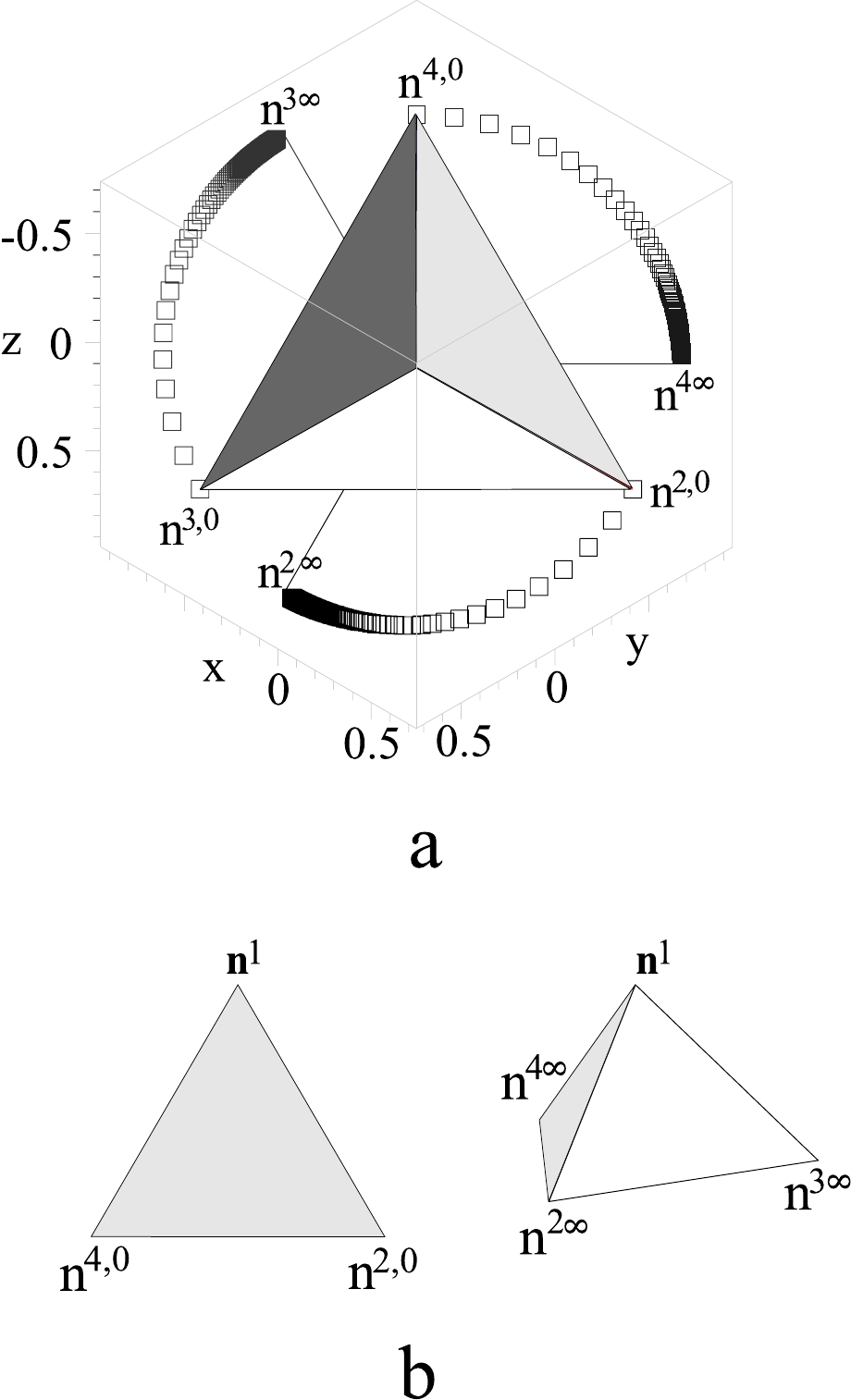}
\caption{a) The deformed tetrahedron viewed along the field direction for $b \in [0,\infty\}$ (200 values shown as $\Box$). Not only grow the angles among the $\bm{n}^{\bar \beta E}$ from $\phi_T$ to $120^o$ and shrink the angles between $\bm{ n}^{1E} $ and any $\bm{n}^{\bar \beta E}$ from $\phi_T$ to $90^o$, there is also an overall rotation of the structure about $\bm{ n}^{1E} $ of $90^o$ maximum. b) The transition from the tetrahedron to the pyramid and the accompanied rotation seen from the side - from \cite{L}. \label{fig:9}}
\end{center}
\end{figure}
%%%%
%%%%

In addition, there is an overall rotation of all $\bm{ n}^{\bar \beta E} $ about the field direction. This is obvious from 
\begin{equation} \label{scalar3}
\bm{ n}^{2E} \bm{\cdot n}^{2} =  \bm{ n}^{3E} \bm{\cdot n}^{3} = \bm{ n}^{4E} \bm{\cdot n}^{4} = \frac{1}{\sqrt{1 + 2 b^2}}
\end{equation}
demonstrating that for $b \to \infty$ all three $\bm{n}^{\bar \beta E}$ have been rotated by $90^o$, Fig. \ref{fig:9}. The rotation sense depends on the sign of $b$, but is irrelevant. What looks like a clockwise rotation when viewed from above, is a counter-clockwise one when viewed from below. The rotation sense is also changed, when the tetrahedral structure is replaced by its inverse. The equivalence of $b$ with $-b$ is also manifest in the energetics discussed below, where only $b^2$ occurs. 

If the system follows the torques provided by the external field, its electric free energy is certainly lowered. Indeed, the free energy of the distortion, $\Delta \tilde \varepsilon^E \equiv \tilde \varepsilon^E (\bm{n}^{\beta E}) - \tilde \varepsilon^E (\bm{n}^{\beta})$, is found using Eq. (\ref{cubicEn}) to be
\begin{equation}\label{statEgeneral}
\Delta \tilde \varepsilon^E =  - \frac{1} {18} \epsilon_1 E_0^3 \left( 1 - (1 + 2 b^2)^{-3/2}  \right) \leq 0.
\end{equation}
It vanishes by definition for $b=0$ and is a monotonically decreasing function with increasing $b^2$.

On the other hand, the thermodynamic ground state is that of the undistorted tetrahedral structure. Therefore, any deviation from the tetrahedral angle $\phi_T$ increases the energy, which can be written phenomenologically in a kind of harmonic approximation as
\begin{eqnarray}\label{statrelangle1}
\varepsilon_{def} &=&  \frac{B_1} {2} \sum_{
\gamma,
\beta > \gamma}
\Bigl(\bm{n}^{\gamma E} \bm{\cdot n}^{\beta E} + \frac{1}{3}\Bigr)^2 \\ \label{statrelangle2}
&=&  \frac{B_1} {6} \left[ \Bigl(1 - \frac{1}{\sqrt{1 + 2 b^2}}\Bigr)^2
+ \Bigl(\frac{b^2}{1 + 2 b^2}\Bigr)^2 \right] \geq 0.
\end{eqnarray}
It is zero for $b=0$ and positive for finite $b$, increasing monotonically with increasing $b^2$. 

%%%%
%%%%
\begin{figure}      
\begin{center}
\includegraphics[width=7.5cm]{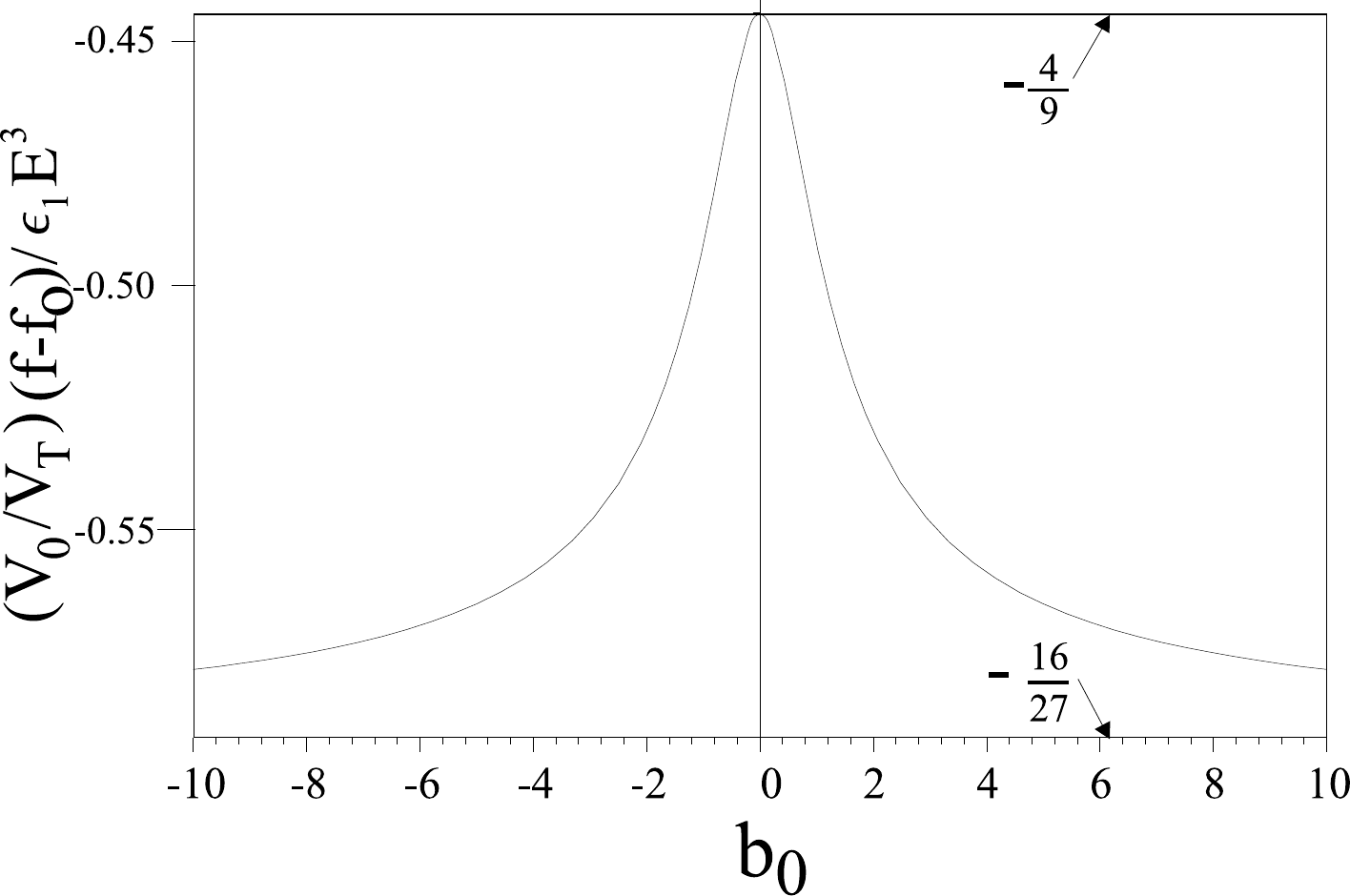}
\caption{The total deformation energy $\varepsilon_{def}^{tot}$ scaled by $\epsilon_1 E_0^3$ and by $V_0/V_E$, cf. Eq. (\ref{volume}), as a function of $b_0$ (with a vertical offset of $-4/9$) - from \cite{L}. \label{fig:10}}
\end{center}
\end{figure}
%%%%
%%%%
The equilibrium value of $b^2 = b_0^2$ is given by the minimum of the sum of the two energies related to tetrahedral distortions due to external electric fields, $\varepsilon_{def}^{tot} = \Delta \tilde \varepsilon^E + \varepsilon_{def}$. Minimization leads to the condition
\begin{equation}\label{Minimumb}
\sqrt{1 + 2 b_0^2}\, \bigl( 2 + 4 b_0^2 -A\bigr) = 2 + 2b_0^2
\end{equation}
with $A= \epsilon_1 E_0^3 / B_1$. Generally, in the physically relevant limit when the external field only slightly distorts the \Td phase, $A$ is small. In that case\footnote{\scriptsize This relation also applies to the opposite limit, when $A$ is large.}, $b_0^2 = A/4$, and the total distortion energy $\varepsilon_{def}^{tot} = - (3/16) A^2$ is negative. In Fig. \ref{fig:10} it is shown that $\varepsilon_{def}^{tot}$ is negative for all values of $b_0^2$. This means there is no threshold for the deformations due to an external field and even a very small field leads to a non-zero deformation. However, the deformation  amplitude $b_0$ will be very small and hardly measurable, in particular for large $B_1$ . Of course, in the rigid limit, $B_1 \to \infty$, there is $b_0^2 \to 0$ and $\varepsilon_{def}^{tot} \to 0$. This scenario is rather robust, {\it e.g.} the small amplitude behavior is the same, when $\varepsilon_{def}$ is replaced by the more complicated deformation energy $T_{ijk}^{E} T_{ijk}^{E} - T_{ijk} T_{ijk}$, where $T_{ijk}^{E}= N \sum_{\beta=1}^4  n_i^{\beta E} n_j^{\beta E} n_k^{\beta E}$.

Finally we remark that the deformations induced by the electric torques also change the volume of the structure given by the four vertices $\bm{ n}^{\beta E} $
\begin{equation}
\label{volume}
\frac{V_{E}}{V_{0}}=\frac{1}{16} \frac{4+ 9 b_0^2}{(1+2b_0^2)^{3/2}} 
\,\,\left( 1+3\sqrt{ 1+2b_0^{2}} \right)
\end{equation}
with $V_0 = 8 / (9 \sqrt{3})$ the volume of the undistorted tetrahedron. In the limit $b \to \infty$ the volume of the pyramid is $V_\infty = \sqrt{3}/4$, which is by 15,625\% smaller than without distortion.
Assuming that the volume of the structure is related to the volume of the molecules involved, it means that the electric torques lead to an increase of the density of the system.  

For a tetrahedral phase that is deformed by a strong external field rotations of aligned tetrahedral vector $\bm{n}^1$ are clamped, {\it e.g.} they relax on a time scale faster than the other hydrodynamic variables. Thus, compared to an isotropic fluid the only additional degree of freedom is the rotation of the (deformed) tetrahedron about the field. On the other hand, such a phase is polar and uniaxial, and the material properties become anisotropic due to the external field. In Sec. \ref{C3v} we will give a more detailed account of the full hydrodynamics of a thermodynamic phase with $C_{3v}$ symmetry, where the preferred direction exists spontaneously and its rotations are hydrodynamic excitations.

%%%%%%%%%%%%%%%%%%%%%%%%%%%%%%%%%%%%%%%%%%%%%%%%

\subsection{The chiral T phase} \label{chiralT}

\begin{figure}[t]
\includegraphics[width=7.3cm]{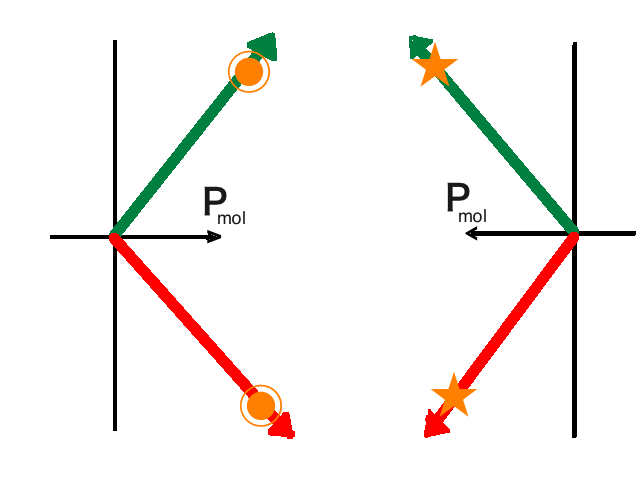}
\caption{Model of a symmetrically chiralized bent-core molecule (left) with its mirror image (right); filled (orange) circles mean e.g. positive chirality, while the (orange) stars indicate negative chirality - from \cite{PB14}.}
\label{fig:11}
\end{figure}

The \Td phase considered so far (made up of achiral molecules) is achiral. However, using chiral molecules one can get its chiral analog, the T phase \cite{Fel}. This is similar to conventional nematic LC that become cholesteric (chiral nematic), when the molecules are replaced by chiral ones or chiral molecules are added. Bent-core molecules can be chiralized in different ways. To get  the T phase, the simplest way is to assume that the two tails of such molecules are symmetrically chiralized, Fig. \ref{fig:11} (left). If such a molecules is mirrored at a plane or inverted, the chirality is changed, Fig. \ref{fig:11} (right), and the two forms cannot be brought into coincidence by mere rotations.  
 %
%%%%
%%%%
%%%%%
%%%%%

\begin{figure}[t]
%\begin{center}
\includegraphics[width=8cm]{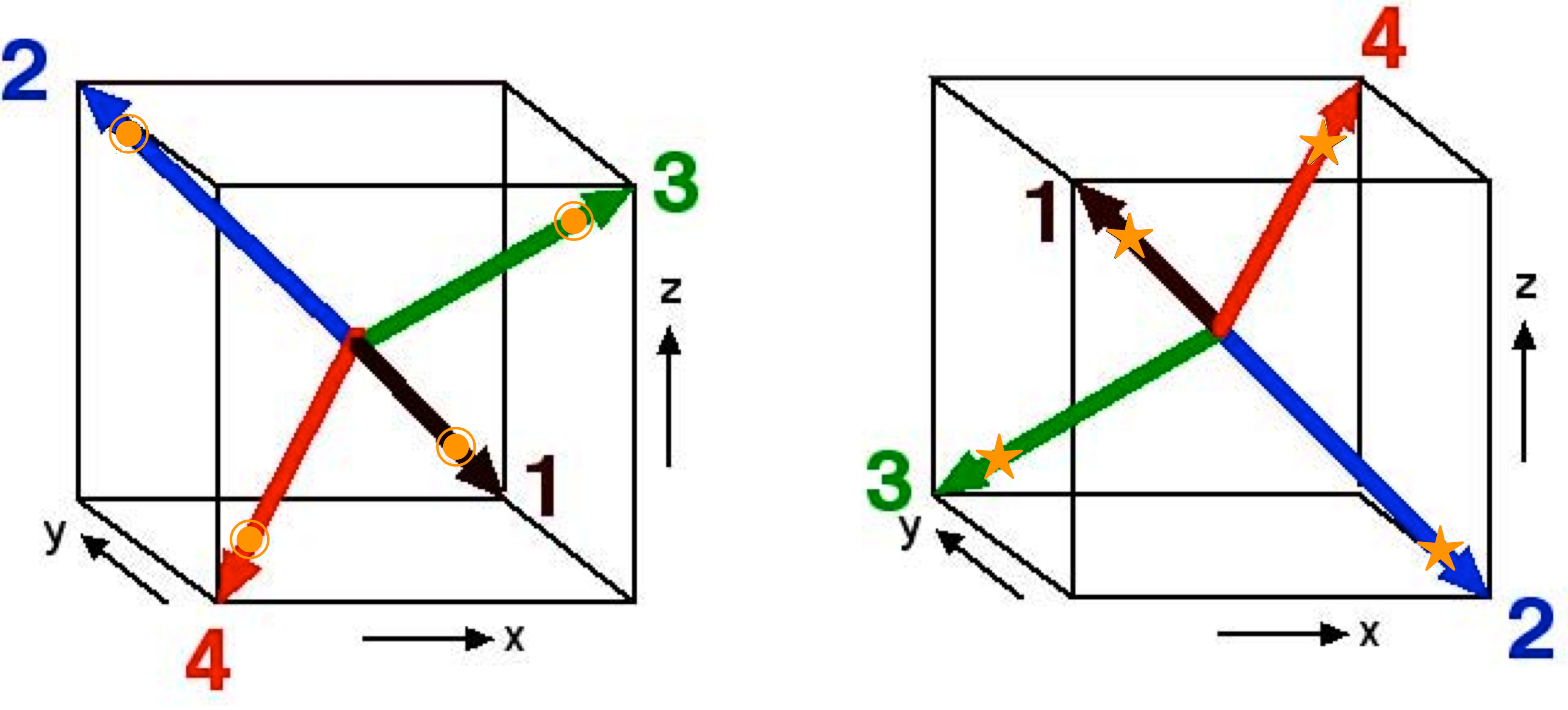}
\caption{Model of two bent-core molecules (1-4 and 2-3) with the same chirality, arranged to fit into the tetrahedral geometry (left) with its mirror image (right); not only is the tetrahedral geometry inverted, but also the molecular chirality has changed - from \cite{PB14}.}
\label{fig:12}
\end{figure}

\begin{figure}[ht]
\begin{center}
\includegraphics[width=7.9cm]{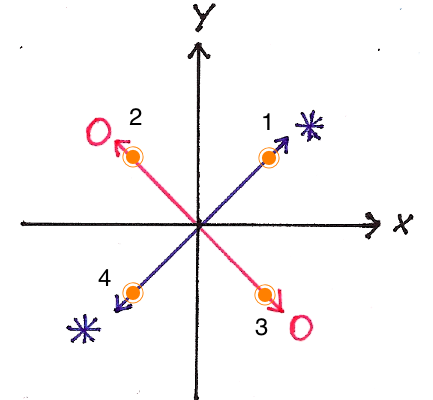}
\caption{Projection of the tetrahedral structure of Fig. \ref{fig:12} onto the $x/y$ plane. The $z$ axis is reduced to a 2-fold axis, since a $\pi/4$ rotation with an additional inversion preserves the structure, but changes the chirality. For the same reason, the planes spanned by vectors 1/4 or 2/3 are no mirror planes - from \cite{PB14}.}
\label{fig:13}
\end{center}
\end{figure}

To get a chiral phase, one has to employ a specific model, where two bent-core molecules of the same chirality are combined in a steric arrangement resembling the tetrahedral vectors 1-4 and 2-3 in Fig. \ref{fig:12},
What has been a $\bar 4$ axis in the T$_d$ phase is here reduced to a (proper) 2-fold symmetry axis, and the planes spanned by vectors 1/4 and 2/3 are no longer mirror planes, Fig. \ref{fig:13},
%%%%%
%%%%%
%%%%%
%%%%%
%%%%%
%%%%%
%%%%%
%%%%%
with the result that only three 2-fold and four 3-fold symmetry axes exist. The former are the $x,y,z$ directions, while the latter are the tetrahedral axes 1-4, which are equivalent since they have the same chirality. Such an arrangement of bent-core molecules ensures the compensation of the molecular polarity and results in the T phase being non-polar. 

The hydrodynamics of the achiral T$_d$ phase has been given above in Sec. \ref{Tdhydro}. We therefore concentrate on the differences between the hydrodynamics of the T compared to the T$_d$ phase. In both phases the same set of hydrodynamic equations is used. Differences occur in the static and dynamic couplings due to the chirality of the T phase, which is manifest by the existence of a pseudoscalar $q_0$. The rotational elastic gradient free energy (cf. Eq. (\ref{vareps})
\begin{equation}
\varepsilon_{g}=\frac{1}{2}K_{ijkl}^{\Gamma}(\nabla_{j}\Gamma_{i})(\nabla_{l}\Gamma_{k}) + q_0 K_1^{lin} \nabla_i \Gamma_i
\label{Tfrank}
\end{equation}
contains a chiral term $\sim q_0 K_1^{lin}$ linear in the gradients of the rotations $\delta \Gamma_i$. Generally, a linear gradient term favors a spatially inhomogeneous structure. In the present case, a helical rotation of $T_{ijk}$ about any of the 3-fold axes (the tetragonal vectors) reduces the free energy by $\Delta \varepsilon = -\tfrac12 (q_0 K_1^{lin})^2/(2K_1^\Gamma+K_2^\Gamma)$. What looks like a linear splay term is physically a linear twist contribution, quite similar to the familiar case of cholesteric LC.  The optimum helical pitch, $q_h = \frac32 q_0 K_1^{lin} /(2K_1^\Gamma+ K_2^\Gamma)$, is generally different from the chiral pseudoscalar of the phase, $q_0$, since there is no a priori reason that $K_1^{lin}$ is related to $K_{1,2}^\Gamma$. An analogous statement holds for ordinary cholesteric LC \cite{oswaldComm}. Helical rotations about the 2-fold axes do not lower the free energy, since the linear gradient term is zero in that case and the quadratic term, $\sim K_3^\Gamma$, increases the free energy.

The similarity to the cholesteric phase also holds for chiral Lehmann-type contributions, both static
(in the free energy), $\varepsilon_{c}=q_0 (\xi^{\rho}\delta\rho+\xi^{\sigma}\delta\sigma+\xi^{c}\delta c) \nabla_i \Gamma_i$,
and dynamic (in the dissipation function),
$R = q_0 h_k^\Gamma (\Psi^{E}E_{k}+\Psi^{T}\nabla_{k}T+\Psi^{c}\nabla_{k}c) $. They relate in the dissipative currents the scalar degrees of freedom (temperature, concentration, density etc.) with the rotations of the tetrahedron.  

The reversible part of the current of tetrahedral rotations, $Y_i^R$, Eq. (\ref{Yrev}), contains additionally a chiral coupling to the rate of strain tensor 
\begin{align} \label{YA}
Y_i^R=& q_0 \lambda T_{ijk} A_{jk} \\
\sigma_{ij} =& q_0 \lambda T_{ijk} h_k^\Gamma
\label{sigmaomega}
\end{align}
with the appropriate counter term in the stress tensor.

\begin{figure}[ht]
\begin{center}
\includegraphics[width=7.9cm]{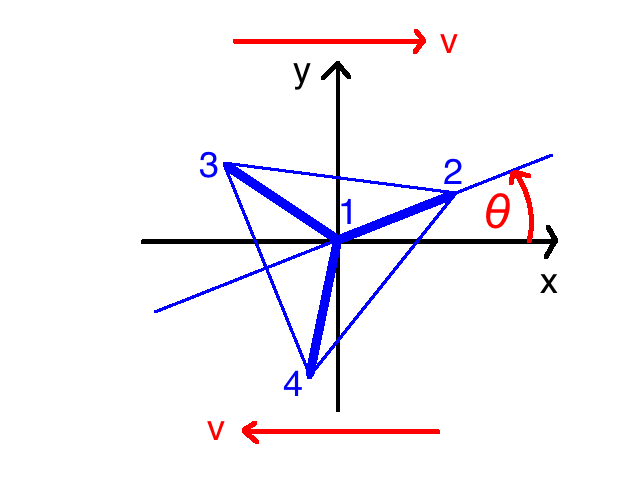}
\caption{Projection of the tetrahedral structure onto the $x/y$ plane with one of the tetrahedral axes ($\bm{n}^1$) along the z-axis (corners 2,3,4 lie below the x/y plane). This is also the vorticity direction of the simple shear $\nabla_y v_x = S$. The structure is rotated in the shear plane by an angle $\theta$ that is independent of the shear rate $S$ - from \cite{PB14}.}
\label{fig:14}
\end{center}
\end{figure}

Now that there are couplings to both, rotational and symmetric shear flows, a stationary  alignment of $T_{ijk}$ in simple shear is possible that is independent of the flow rate. This scenario is very much like the flow alignment in nematics, although there it is an {\it achiral} reversible effect. In particular, if one of the 3-fold tetrahedral axis is in the vorticity direction, the tetrahedron is rotated about this direction by an angle $\theta$, Fig. \ref{fig:14},
\begin{equation}
\frac{1}{\cos 2 \theta} = \frac{16}{27} \,q_0 \lambda
\label{Talign}
\end{equation}
%
%%%%%
%%%%%
%%%%%
%%%%%
that only depends on the material parameters $q_0 \lambda$. Only 3-axes, but no 2-axes, can be aligned. In the \Td phase there is no flow alignment, since there is no $q_0$. 
The remainder of the dynamics of the T phase is as in the achiral T$_d$ phase.

%%%%%%%%%%%%%%%%%%%%%%%%%%%%%%%%%%%%%%%%%%%%%%
%%%%%%%%%%%%%%%%%%%%%%%%%%%%%%%%%%%%%%%%%%%%%%

\section{Phases with Correlated Tetrahedral and Nematic Order} \label{TNcorr}

\subsection{The non-polar tetrahedral uniaxial nematic D2d phase} \label{D2d}

\subsubsection{Statics} \label{D2dstatics}

%%%%%%%sec%%%%%%%%%%%

\begin{figure}[ht]
\begin{center}
\includegraphics[width=7.9cm]{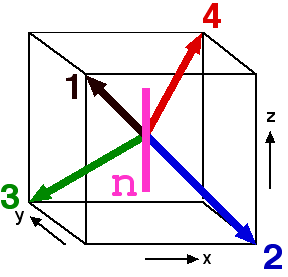}
\caption{The director $\bm{n}$ along the $\bar 4$ symmetry axis in the D2d phase. Vectors 1,2,3,4 denote the four tetrahedral vectors $\bm{n}^\beta$ with $\beta=1,2,3,4$ - from \cite{PB14}.}
\label{fig:15}
\end{center}
\end{figure}

When both, tetrahedral and nematic order are present, we have shown in Sec. \ref{tetranemaorder} that a possible ground state is the D2d phase. The nematic director $\bm{n}$ is fixed to be along one of the $\bar 4$ axes of the tetrahedron (the z axis in Fig. \ref{fig:15}). The angle between the director and the tetrahedral axes is half the tetrahedral angle $\phi_D =\phi_T/2$ or $\phi_D = \pi - \phi_T/2$ with $\bm{n \cdot n}^\beta = \pm 1/\sqrt{3}$.
The tetrahedral 3-fold axes ($\beta =1,2,3,4$ in Fig. \ref{fig:15}) are not symmetry axes any longer, but the z-axis is still a $\bar 4$ symmetry axis, while the x- and y-axis are reduced to be 2-fold.. The planes defined by vectors 1/4 and 2/3 are symmetry planes prohibiting chirality of the D2d phase. This phase is non-polar, because of the $\bm{n}$ to $-\bm{n}$ equivalence and the absence of a polar vector $T_{ijk}Q_{jk}=0=T_{ijk}n_{j}n_{k}$. It can be viewed as a uniaxial nematic LC with a transverse structure that resembles orthorhombic biaxial nematics, but without inversion symmetry.
%
%%%%%
%%%%%
%%%%%
%%%%%

Since the director and the tetrahedron are rigidly coupled, the number of hydrodynamic variables is the same as in the \Td phase. However, instead of using the three tetrahedral rotations $\delta \bm{\Gamma}$, Eq. (\ref{omega}), we will split them up into (two) rotations of the director $\delta \bm{n} = (\bm{n} \times \delta \bm{\Gamma})$ and one rotation
about the director
\begin{equation}\label{deltaOmega}
\delta\Omega\, \equiv \, \bm{n \cdot} \delta\bm{\Gamma}=(1/4\alpha)\,n_{i}\epsilon
_{ipq}T_{pjk}\, \delta T_{qjk}.
\end{equation}
By construction $\delta\Omega$ is even under
parity and time reversal and odd in $\bm{n}$, but is not a true
scalar (concerning its behavior under rotations - see below). There is no direct way of detecting this degree of freedom optically. Only through its (static) coupling to the director rotations (see below), it might be accessible to experiments.

The hydrodynamic description in terms of $\delta \bm{n}$ and $\delta\Omega$ facilitates the comparison with ordinary nematic LC and is closer to experiments, where the director and its rotations are accessible by optical means. We will assume the nematic as well as the tetrahedral order parameter strength, $S$ and $N$, respectively, to have relaxed on a very short time scale to their constant equilibrium values, $S_0$ and $N_0$, which we will take as unity.
We also restrict ourselves to a linear hydrodynamic
description, in particular we disregard the consequences of the nonlinear non-commutativity relation Eq. (\ref{noncommute}). A recipe how to deal with them and some results can be found in Sec. 4 of \cite{BP-D2d}. There, we also discuss other general aspects of nonlinear hydrodynamics, including the dependence of material parameters on state variables and material  tensors on orientations of the structure, as well as the occurrence of non-harmonic thermodynamic potentials, transport derivatives, and nonlinear pressure and stresses.

In this setting
the Gibbs relation Eq. (\ref{Gibbs}) takes the form
\begin{eqnarray} \label{GibbsD2d}
d \varepsilon &=& \mu d\rho + v_i d g_i + h_{i}^{\prime} d n_i + \Psi_{ij} d\nabla_j n_i \nonumber \\ &&+  h^{\Omega\prime} d \Omega + \Psi_{i}^\Omega d\nabla_j \Omega + T d\sigma
\end{eqnarray}
with the conjugates $h^{\Omega}=h^{\Omega\prime}-\nabla_i \Psi_i^\Omega$ and $h_i = h_i^\prime - \nabla_j \Psi_{ij}$ omitting the nonlinearities shown in Eq. (\ref{hGamma}).

The gradient free energy reads
\begin{eqnarray} \label{FrankD2d}
\varepsilon_{g}&=&\frac{1}{2}K_{ikjl}(\nabla_{i}n_{k})(\nabla_{j}n_{l})
+ K_{7}\delta_{ik}^{\perp}\epsilon_{lij}(\nabla_{l}\Omega)(\nabla_{j}n_{k})  \nonumber \\ &&
+( K_{5}n_{i}n_{j}+K_{6}\delta_{ij}^{\perp})(\nabla
_{i}\Omega)(\nabla_{j}\Omega)
\end{eqnarray}
with the transverse Kronecker $\delta_{ij}^{\perp} = \delta_{ij} - n_i n_j$ projecting onto the plane perpendicular to $n_i$.
There are four Frank-type orientational elastic coefficients related to distortions of the director
\begin{align}
K_{ijkl}  &  =K_{3}n_{i}n_{j}\delta_{kl}^{\perp}+(K_{1}-2K_{2})\delta_{ik}%
^{\perp}\delta_{jl}^{\perp}\nonumber\\
&  +K_{2}(\delta_{il}^{\perp}\delta_{jk}^{\perp}+\delta_{ij}^{\perp}\delta_{kl}%
^{\perp})+K_{4}n_{p}n_{q}T_{ijp}T_{klq} \label{Frank}%
\end{align}
which is one more than in the uniaxial nematic case: The $K_4$ term is related to $(\nabla_{x}n_{x})(\nabla_{y}n_{y})$, and vanishes in the transversely isotropic case.
In addition there are two coefficients related to distortions of $\Omega$ ($K_{5,6}$) and a mixed one ($K_7$). The latter links inhomogeneous $\Omega$ rotations,
$\nabla_{z}\Omega$, to director twist, $\nabla_{x}n_{y}-\nabla_{y}n_{x}$.
Assuming that the tetrahedral structure is clamped at solid surfaces with
$\mbox{\boldmath$n$}$ homeotropic, a circular Couette cell with a fixed plate at the bottom
and a rotating one at the top will create a finite $\nabla
_{z}\Omega$. By the $K_{7}$ coupling twist of the director is induced in
the x-y-plane.

The total number of seven Frank coefficients corresponds to the number of such coefficients in orthorhombic biaxial nematic LC. In particular, The $K_4$ term can be written as $ K_4  (m_i l_j + l_i m_j)(m_k l_l + l_k m_l)  $ using the transverse directors $\bm{m}$ and $\bm{l}$ of a orthorhombic biaxial nematic LC \cite{Nbiax}. This demonstrates that the lack of inversion symmetry does not influence the Frank-like (quadratic) gradient energy.  
On the other hand, however, a linear gradient energy is possible in the D2d phase 
\begin{equation} 
\varepsilon_{l}=\xi\,T_{ijk}\,n_{i}\nabla_{j}n_{k} \label{lingrad}%
\end{equation}
which is forbidden in ordinary nematic LC due to the inversion symmetry. It is related neither to linear splay, $\nabla_{x}n_{x}+\nabla_{y}n_{y}$
(present in polar nematics), nor to linear twist, $\nabla_{x}n_{y}-\nabla
_{y}n_{x}$ (present in chiral nematics), but involves the combination
$\nabla_{x}n_{y}+\nabla_{y}n_{x}$. As it is well known from cholesteric liquid
crystals \cite{PGdG} and polar nematics \cite{splay,polnema,polarchol}, the appearance of linear gradient terms in the
deformation energy of a director field allows for the possibility of an
inhomogeneous ground state. 

Before we start the discussion of the implications of the linear gradient energy, we mention related cross-couplings between director deformations of the $\xi$
-type (Eq.\@(\ref{lingrad})) and all the scalar
hydrodynamic variables
\begin{equation}
\varepsilon_{c}=T_{ijk}n_{i}\nabla_{j}n_{k}(\xi^{\rho}\delta\rho+\xi^{\sigma
}\delta\sigma) \label{cross}%
\end{equation}
unknown for usual nematics. Analogous terms for possible additional scalar
variables, like concentration variations $\delta c$ in mixtures, or variations
of the order parameters $\delta N$ or $\delta S$, can be written down
straightforwardly. These terms resemble the structure of static Lehmann terms in
cholesteric liquid crystals \cite{leh}, although here they do not involve a
(Lehmann) rotation of the director, but transverse bend deformations. 

The statics is completed by adding up all the energy contributions
$\varepsilon=\varepsilon
_{0}+\varepsilon_{g}+\varepsilon_{l}+\varepsilon_{c}+ \tilde \varepsilon_E$, where $\varepsilon_{0}$ is the part of isotropic liquids and $\tilde \varepsilon_E$ describes the influence of external fields (see below). The conjugate quantities then follow from $\varepsilon$ as partial derivatives according to Eq. (\ref{GibbsD2d}).

%%%%%%%sub%%%%%%%%%%%
\subsubsection{Ambidextrous Helicity \label{ambi}}

The linear gradient energy contribution Eq. (\ref{lingrad}) allows for a
inhomogeneous ground state. Indeed, it is straightforward to show that a
helical state has a lower free energy than the homogeneous state. In this
helical state the director and the tetrahedral structure
rotate together about one of the 2-fold axes.
\footnote{\scriptsize 
The picture of counter-propagating nematic and tetrahedral 
helices suggested in Ref. \cite{N} is based on a misinterpretation of the results  
and is not possible.
}
These are the $x$ and $y$ axis in the geometry of Eq.\@(\ref{unit}), where the
director is along the $z$ axis. Choosing the $x$ axis as helical axis for
definiteness, the director (and the $\bar4$ axis) is given by
\begin{equation}
n_{i} = \delta_{iz} C + \delta_{iy} S \label{nhelix}%
\end{equation}
with $C=\cos(q_{0} x)$ and $S= \sin(q_{0} x)$, while for the tetrahedral vectors one finds
\begin{equation}
\label{unithelix}\frac{\pm1}{\sqrt{3}}
\begin{pmatrix}
1 & 1 & -1 & -1\\
C-S & -C+S\ \ \ \  & C+S\ \  & -C-S\\
-C-S\ \ \ \  & C+S\ \ \  & C-S\ \ \  & -C+S
\end{pmatrix}
\end{equation}
where the minus sign refers to the inverted tetrahedral structure. It is this possibility to discriminate between original and inverted structure that allows for different helical rotation senses. 

\begin{figure}[ptb]
\begin{center}
\includegraphics[width=8.3cm]{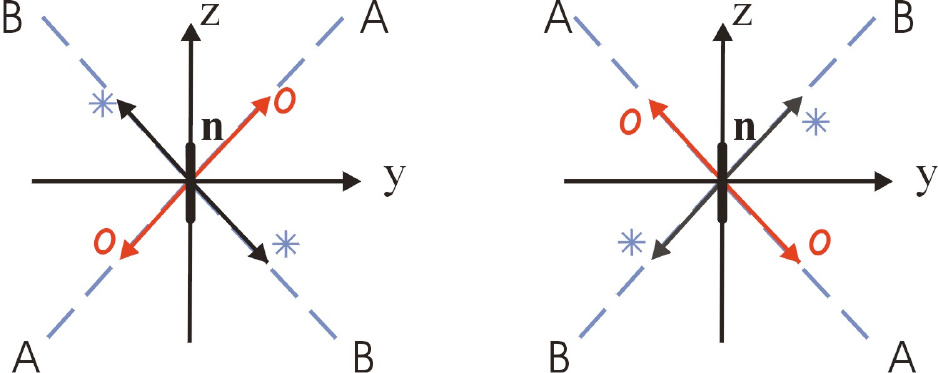}
\end{center}
\caption{ Projection of a D2d structure onto the y/z plane on the left, with the inverted structure on the right. The director
$\mbox{\boldmath$ n$}$ is along the z axis, the x axis sticks out of the
drawing plane, and circles and asters denote tetrahedral vectors that point out or into the drawing plane, respectively. Rotations about the x axis with the sequence $A\rightarrow z\rightarrow B$
constitute an opposite helical sense for the right and left structure. 
The sequence $B\rightarrow z\rightarrow
A$ reverses the rotating sense in both cases - from \cite{BP-D2d}.}
\label{fig:16}
\end{figure}

Indeed, the helical state has a free energy, which is by 
$\Delta\varepsilon=(1/2)(\xi^{2}/K_{2})$ smaller than that of the homogeneous state, independent of the sign of Eq. (\ref{unithelix}). On the contrary, the helical wave vector, $q_0 = \mp\, \xi/K_2$,  depends on that sign and the helical rotation sense is reversed for the inverted structure.
The two different possibilities are demonstrated in Fig. \ref{fig:16}, where the inverted structure on the right has the opposite rotation sense compared to the original structure on the left. We call this phenomenon 'ambidextrous {\it helicity}'. Ambidextrous means no energetic preference for the two possibilities, in analogy to 'ambidextrous {\it chirality}' {\it e.g.} in the C$_{B2}$ (B$_2$) phase, where the left and right handed helices are also energetically equivalent \cite{G1,G2}. In the latter case the term 'chirality' is appropriate, since this phase is structurally chiral, while the D2d phase is achiral and no pseudoscalar can be constructed from $\bm{n}$ and $T_{ijk}$. 

The helical wave vector $q_0$ is proportional to $\xi$, the material parameter of the linear gradient term, and changes sign with it. For materials with $\xi<0$ (there is no general principle that fixes this sign) the situation is as shown in Fig. \ref{fig:16}, while for the case $\xi>0$, the roles of 'original' and 'inverted' structure are interchanged. However, this is irrelevant, since what is called 'original' or 'inverted' is arbitrary.

%%%%%
%%%%%
%%%%%%
%%%%%%
The choice of one of the 2-fold axes (x- or y-direction) as the helical axis leads to the maximum energy gain, while for the rotation axis $\bm{b} = \bm{e}_x \pm \bm{e}_y$ no gain at all is found, since the linear gradient energy is zero in that case.

In a spontaneous formation of the D2d phase, helices of different rotation
sense and about different orthogonal axes might occur randomly at different places of
the sample, since all the possibilities discussed above are equally likely.
The D2d symmetry would be present only locally in the domains, and, when averaged, 
an almost isotropic behavior can be expected. 
Even, if in a D2d phase a pure helical state (with a single helicity and single
helix orientation) has been formed, averaging such a structure over
a length scale large compared to the pitch, the resulting 
symmetry of the system is 
isomorphic to the situation arising when a cholesteric phase is averaged over
length scales large compared to the cholesteric pitch.

Therefore, we will use the local description in the
following. This means, we assume locally D2d symmetry, but with the $\xi$-term in the free energy, Eq.\@(\ref{lingrad}), which reflects the lack of inversion symmetry. This procedure is
frequently used in cholesterics, which are locally described as nematics with
the additional linear twist energy term (reflecting chirality). If the D2d phase is in a homogeneous
state, the linear gradient free energy term always leads to the tendency of
forming localized helical domains.

%%%%%%%sub%%%%%%%%%%%
\subsubsection{Frustration by an External Electric Field \label{External}}

External electric
fields have an orienting effect on LC, in particular
the dielectric anisotropy orients the director of the nematic phase \cite{PGdG}, while the
tetrahedral structure is aligned by a cubic generalization of the dielectric
energy, Eq. (\ref{cubicE}), in the \Td phase. In the D2d phase both effects are present simultaneously and the (Legendre transformed) field-induced free
energy reads 
\begin{align} \label{epsED2d}
\tilde\varepsilon^{E}    =& -\frac{\epsilon_a}{2}n_{i}n_{j}E_{i}E_{j}%
-\epsilon_{10} T_{ijk}E_{i}E_{j}E_{k} \\ &- \nonumber
\frac{\epsilon_{11}}{3} T_{ijk} ( n_i E_j E_k + E_i E_j n_k + E_i n_j E_k)(\bm{n \cdot E})  
\end{align}
where, however, the two tetrahedral terms act the same way and can be 
combined via $\epsilon_{1}\equiv \epsilon_{10}+\epsilon_{11}$. 
We note that other possible terms one might think of such as 
$ \sim T_{ijk} n_j n_k$  vanish, since $n_i$ has only a $z-$ component in contrast to $E_i$. 
There is no need to
incorporate fourth order terms in $\tilde\varepsilon^{E}$ to guarantee convexity,
since we study external fields here. As in the \Td phase the cubic term gives rise to second harmonic generation in optical applications.\\

The nematic term is minimized for $\mbox{\boldmath$n$}$ parallel or
perpendicular to the field (for ${\epsilon}_{a}\gtrless 0$), while the
tetrahedral term forces one of the tetrahedral unit vectors to be parallel or
antiparallel to $\mbox{\boldmath$E$}$. However, in the D2d phase these two cases are incompatible, since
the director always makes an oblique angle (half of the tetrahedral angle, $\phi_{T}/2$, or $\pi-\phi_{T}/2$), with any of the tetrahedral
vectors, disproving the possibility for zero or 90 degrees. Therefore, the system is frustrated and the actual equilibrium orientation minimizes the sum of both terms, but not each term separately. As a result, the orientation depends on the relative strength of nematic vs. tetrahedral coupling ($\epsilon_a / \epsilon_1$). Since these couplings are of different powers in the field strength, $E_0$, the optimum orientation also depends on $E_0$. 

At small fields
the first term is dominant and the director orientation is the usual nematic
one. Above a
threshold field strength, ${E}_{c}$, it is energetically favorable to rotate the D2d structure rigidly such that
the director is tilted away from the dielectrically optimal orientation and, at the
same time, one of the tetrahedral vectors is tilted towards the
field by the same angle. Indeed, minimization of $\tilde\varepsilon^{E} $ with respect to the tilt angle of the director, $\theta_{E}$, leads to
\begin{equation}
\theta_{E}=0 \quad\quad  \mathrm{for} \quad\quad  E \leq E_c =\frac{\sqrt{3}}{8}\frac{\epsilon_a}{\epsilon_1} \label{threshold}
\end{equation}
and
\begin{equation}
\sqrt{3} \cos\theta_{E}=\gamma + \sqrt{\gamma^{2}+1} \quad\quad \mathrm{for} \quad\quad E \geq E_c \quad
\label{thetatilt}%
\end{equation}
with $\gamma
=(1/8) (\epsilon_{a} / \epsilon_1E_0)$ and $\gamma_{c}=1/\sqrt{3}$. Here, we have assumed positive dielectric coupling, $\epsilon_{a}>0$, and have chosen, without lack of generality, $\epsilon_1>0$. 

There is no jump of the tilt angle at $E_c$ and for very large fields ($\gamma\to 0$) $\theta_E \to \phi_T / 2$ meaning one of the tetrahedral vectors approaches asymptotically the field direction.

If $\epsilon_1/\epsilon_a$ is large enough for the threshold field to be within experimental reach,
 there is a unique way of identifying the D2d phase: Below the
threshold, the director is oriented parallel to $\bm{E}$. Increasing the
field beyond the threshold, the director turns away to a direction oblique to
the field - something that cannot happen in a conventional uniaxial nematic
phase. The presence of a helix further complicates the behavior. Any
homogeneous external field is incompatible with the combined helical structure
of director and tetrahedral vectors and tends to distort that structure.

For the dynamics discussed below we
assume $\epsilon_{a}$ to be positive and the nematic dielectric
anisotropy to be the dominant effect such that the system is below the
threshold for reasonable applied fields. In that case the symmetry of the D2d
phase is preserved and the hydrodynamic description is
valid. In this case
rotations of the structure away from the electric field direction cost energy
\begin{equation} \label{E2D2d}
\tilde\varepsilon^{E}_2=\frac{1}{2}\left(  {\epsilon}_{a}+\frac{32}{9}
\epsilon_1\,E_0\right)  E_0^{2}\, (\delta\mbox{\boldmath$n$})^{2}
\end{equation}
with an effective, field dependent susceptibility (in the big parentheses).
This provides the restoring force for the relaxation of the director.

%%%%%%%sub%%%%%%%%%%%

\subsubsection{Dynamic Properties \label{Dyn}}

While the dynamic laws for mass, momentum, and entropy density are the same as in the \Td phase, Eqs. (\ref{rhodyn}), (\ref{gdyn}), and (\ref{sigmadyn}), the dynamic equation for the symmetry variable, $(\partial/\partial t) \Gamma_i + Y_i=0$, is now written in terms of $\delta\bm{n}$ and $\delta \Omega$
\begin{eqnarray} 
 \frac{\partial}{\partial t} n_i +X_i &=& 0 \label{ndyn} \\
\frac{\partial}{\partial t} \Omega + Y^\Omega &=& 0 \label{Omegadyn} 
\end{eqnarray}
with $X_i = \epsilon_{ijk} n_j Y_k$ and $Y^\Omega=n_i Y_i$.

In the D2d phase the structure of the reversible director dynamics is the same as in
uniaxial nematics. There are advection and convection and a phenomenological coupling to symmetric shear flow \cite{forsternem,MPP,forsterann,forster}
\begin{equation} \label{alignment}
X_i^R = v_j \nabla_j n_i + (\bm{\omega} \times \bm{n})_i - \lambda_{ijk} A_{jk}
\end{equation}
with $2 \lambda_{ijk} = \lambda (\delta_{ij}^{\perp} n_k + \delta_{ik}^{\perp} n_j)$.
This allows for a steady alignment in simple shear at a tilt angle governed by $\lambda$, the sole reversible transport parameter. The appropriate back flow term in the stress tensor, $\sigma_{ij}^R = - \lambda_{kji} h_k$ guarantees zero entropy production.

For rotations about the $\bm{n}$ there is no phenomenological reversible coupling, only transport,
\begin{equation}
Y^{\Omega R}=v_{i}\nabla_{i}\Omega-n_{i}\omega_{i}\ \text{\,.} \label{Yrev3}
\end{equation}
and, therefore, no flow alignment in the plane perpendicular to $\bm{n}$. Obviously, $\delta\Omega$ is not constant under rotations (as a true scalar is), but behaves like the component of a rotation angle, {\it e.g.}, as $\delta
\bm{m\cdot}(\bm{n} \times \bm{m})$ in
biaxial nematics. 

In the \Td phase there are reversible phenomenological couplings between flow and the currents of conserved hydrodynamic variables, Eqs. (\ref{sigmarev}) and (\ref{heatrev}), that also exist in the D2d phase. Here they are of the uniaxial form and read for the heat current and the stress tensor
\begin{eqnarray} \label{Gamma2a}
j_{i}^{\sigma,ph}  & =& (\Gamma_{21}\delta_{li}^{\perp}+\Gamma_{22}
\,n_{l}n_{i})T_{ljk}A_{jk}\\
\sigma_{ij}^{ph}  & =&  -(\Gamma_{21}\delta_{lk}^{\perp}+\Gamma_{22}\,n_{l}n_{k})T_{ijk}\nabla
_{l}T \label{Gamma2b}
\end{eqnarray}
Appropriate anisotropic generalizations exists for the couplings involving the concentration current (cf. end of Sec. \ref{Tdhydro}) or the electric current, Eq. (\ref{jerev}) and (\ref{sigmarevel}), and the stress tensor, with parameters $\Gamma_{11}, \Gamma_{12}, \Gamma_{31}, \Gamma_{32}$ accordingly. The physical implications of those couplings has already been discussed in Sec. \ref{Tdhydro}.

In the dissipative dynamics there are couplings between scalar hydrodynamic variables and director rotations of the $\xi$-type, Eq. (\ref{lingrad}), {\it e.g.} described by the contribution to the entropy production
\begin{equation} \label{LehmannD2d}
R_{T,n} = \Psi^{T} T_{ijk}n_{j}h_{i} \nabla_{k}T.
\end{equation}
Appropriate couplings exist for the electric field $E_i$ or the concentration current $\nabla_i c$ replacing the temperature gradient. There is no such term for the mass current, since the mass density does not have a dissipative current. Similar terms would arise, when (gradients of the thermodynamic conjugates of) the order parameters, $S$ and $N$, are considered. These terms are the dissipative analog to the static couplings in $\varepsilon_c$, Eq. (\ref{cross}). They are not present in ordinary nematic LC, but they are similar in structure to the dynamic Lehmann effects in cholesteric LC \cite{leh}. There, inversion symmetry is broken by the helical wave vector $q_0$, while here it is the tetrahedral tensor $T_{ijk}$. As a difference, in the D2d phase transverse bend 
deformations of the director are involved, while in the cholesteric case director rotations occur.

Dissipation of the director is isotropic, $2R_{n,n}= (1/\gamma) h^2$ as in the nematic phase. The same is true for the dissipation of the transverse rotations
\begin{equation}
R_{\Omega,\Omega}=\frac{1}{2\gamma^{\Omega}}(h^\Omega)^{2}
\end{equation}
with $h^\Omega = \partial \varepsilon/ \partial \Omega$ that leads to the dissipative part of the quasi-current $Y^{\Omega D}=(1/\gamma^{\Omega})h^\Omega$. While the \Td phase has one rotational viscosity, $\gamma^\Gamma$, there are two in the D2d phase, the nematic one, $\gamma$, and a second one, $\gamma^{\Omega}$, which are generally different from from each other due to the anisotropy of the different rotations.

Other dissipative effects are anisotropic as in ordinary uniaxial nematic LC, {\it e.g.}, thermal conductivity $2R_{T,T} = \kappa_{ij} (\nabla_i T)(\nabla_j T)$ or electric conductivity $2 R_{E,E} = \sigma_{ij}^E E_i E_j$ etc.. Only the viscosity is slightly more complicated in the D2d phase, since for $2R_{v,v}   =\nu_{ikjl}(\nabla_{i}v_{k})(\nabla_{j}v_{l})$ the viscosity tensor
\begin{align}
\nu_{ikjl}  &  =(\nu_{1}-2\nu_{2})\delta_{ik}^{\perp}\delta_{jl}^{\perp}+\nu
_{2}(\delta_{ij}^{\perp}\delta_{kl}^{\perp}+\delta_{il}^{\perp}\delta_{jk}
^{\perp})\nonumber\label{viscosity}\\
&  +\nu_{3}(\delta_{ij}^{\perp}n_{k}n_{l}+\delta_{il}^{\perp}n_{j}n_{k}+\delta
_{kl}^{\perp}n_{i}n_{j}+\delta_{jk}^{\perp}n_{i}n_{l})\nonumber\\
&  +\nu_{4}n_{i}n_{j}n_{k}n_{l}+\nu_{5}(\delta_{ik}^{\perp}n_{j}n_{l}+\delta
_{jl}^{\perp}n_{i}n_{k})\nonumber\\
&  +\nu_{6}n_{p}n_{q}T_{pij}T_{qkl}
\end{align}
contains six viscosities as in the case of orthorhombic biaxial nematic LC \cite{mason}. The last term in Eq. (\ref{viscosity}) vanishes in the uniaxial nematic case.

%%%%%%%sub%%%%%%%%%%%

\subsubsection{Relative rotations \label{AddRem}}

In the D2d phase the orientation of the
director relative to the tetrahedral structure is fixed. In particular,
the director rotates always together with the tetrahedral structure, $\delta \bm{n} = \bm{n} \times \delta\bm{\Gamma}$. This means that any difference between $\delta \bm{n}$ and $\bm{n} \times \delta\bm{\Gamma}$ vanishes on time and length scales much shorter than the hydrodynamic ones. Under certain conditions, however, 
relative rotations 
\begin{equation} \label{relrot}
\bm{J}= \delta \bm{n} - \bm{n} \times \delta\bm{\Gamma}
\end{equation}
persist and are included as non-hydrodynamic variables in the macroscopic dynamics. 

This is similar to
the smectic A phase, where the layer normal and the director are locked to be
parallel. Under certain conditions, like the vicinity to the nematic phase
transition \cite{liu} or strong external shear \cite{auernh}, this coupling
can weaken allowing the two preferred directions to differ from each other for
some time before they have been relaxed back. Another example of relative rotations arises
for mixtures of a rod-like and a disk-like uniaxial nematic phase
\cite{mixture}. Such relative rotations play a prominent role in nematic
elastomers \cite{garmisch,physica}, where they are responsible for elastic
anomalies \cite{softmat,menzel}.

The relative rotations $\bm{J}$ are even under spatial
inversion, odd under the replacement of $\mbox{\boldmath$n$}$ by
$-\mbox{\boldmath$n$}$, and are invariant under rigid rotations. 
They do not involve rotations of the tetrahedral structure about the director, $\bm{n} \cdot \delta\bm{\Gamma}$, since $\bm{n} \cdot \bm{J}=0$. 

The free energy of relative rotations
\begin{equation}
\varepsilon_{J}=\frac{1}{2}D_{1}\bm{J}^2 \label{epsJ}
\end{equation}
diverges in the rigidly locked case, where the stiffness coefficient $D_{1}\to \infty$. In Eq. (\ref{epsJ}) we have neglected some bilinear couplings between relative rotations and gradients of the  director or gradients of $\Omega$, which are of the linearized form $J_x \nabla_z n_x - J_y \nabla_z n_y$ and $J_x \nabla_y \Omega + J_y \nabla_x \Omega$. Gradients of 
electric fields couple similarly. 

The dynamics is given by the balance equation
\begin{equation}
\frac{\partial}{\partial t}J_{i}+Y_{i}^{JR}  +Y_{i}^{JD}=0 \label{dJ}
\end{equation}
The reversible part
\begin{equation}
Y_{i}^{J\,R}=v_{k}\nabla_{k}J_{i}+\lambda_{ijk}^{J}A_{jk} \label{Yrevrel}
\end{equation}
contains the transport derivative and a phenomenological coupling to deformational flow
where $\lambda_{ijk}^{J}=\lambda_{J}(\epsilon_{ikp}n_{p}n_{j}+\epsilon
_{ijp}n_{p}n_{k})$ carries one phenomenological parameter. There is no
coupling to rotational flow, since $\delta \bm{n}$ transforms the same way as 
$\bm{n} \times \delta \bm{\Gamma}$ under rigid rotations (cf. Sec.\ref{Tdhydro}). Lacking the coupling to
rotational flow, shear flow does not lead to a (shear flow) alignment of $J_{i}$.

The dissipative dynamics can be derived from the appropriate part of the
dissipation function, $Y_{i}^{J\,D}=(\partial/\partial L_{i})R^{J}$ where
\begin{equation}
R^{J}=\frac{1}{2}\zeta^{\perp}\delta_{ij}^{\perp}L_{i}L_{j}+\zeta_{ij}%
^{n}L_{i}h_{j}+\zeta_{ijk}^{E}L_{i}\nabla_{j}E_{k} \label{RJ}%
\end{equation}
is expressed by 
$L_{i}=(\partial/\partial J_{i})\varepsilon_{J}$, the thermodynamic
conjugate of the relative rotations. 
The transport parameter $\zeta^{\perp}$
governs the relaxation of relative rotations with the relaxation time
$1/(\zeta^{\perp}D_{1})$, which is zero in the rigidly locked case. The material tensors $\zeta_{ij}^{n}=\zeta_{n}n_{k}\epsilon_{ijk}$ and $\zeta_{ijk}^{E}
=\zeta_{E}(\epsilon_{ikp}n_{p}n_{j}+\epsilon_{ijp}n_{p}n_{k})$ provide dissipative couplings between relative
rotations and director textures or inhomogeneous electric fields.

%%%%%%%%%%%%%%%%%%%%%%%%%%%%%%%%%%%%%%%%%%%%%%%%

\subsection{The non-polar, low symmetry tetrahedral biaxial nematic phases} \label{D2S4}

\subsubsection{The non-polar tetragonal S4 tetrahedral phase \label{S4}}

\begin{figure}[ht]
%\begin{center}
\includegraphics[width=7.4cm]{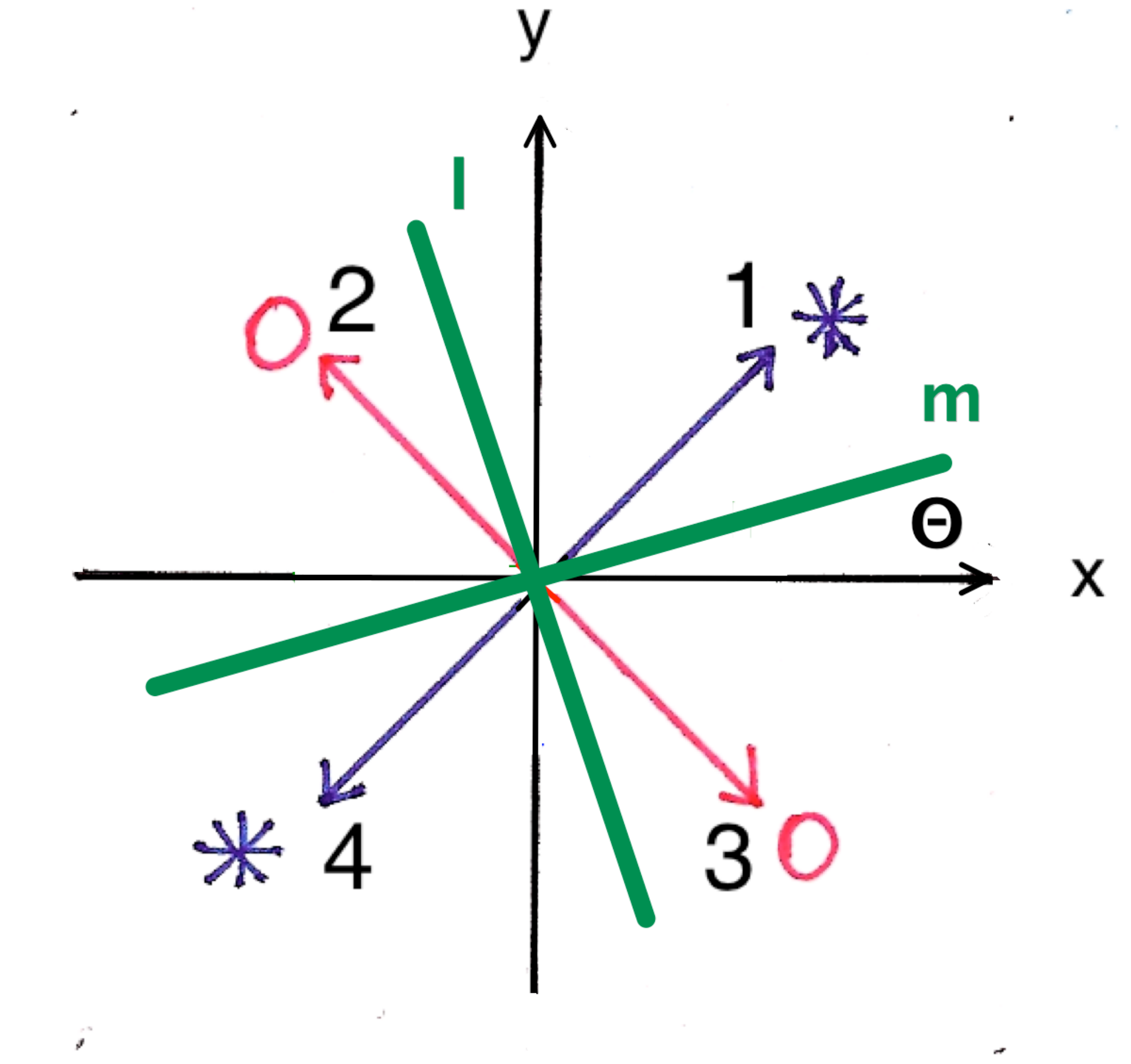}
\caption{The projection of the 4 tetrahedral vectors $\bm{n}^\alpha$ ($\alpha=1,2,3,4$) into the $x,y$ plane perpendicular to the director $\bm{n}$ in the S4 phase. A circle (asterisk) denotes  tetrahedral vectors that also have a component sticking out of (pointing into) the projection plane. The tetragonal nematic directors $\bm{m}$ and $\bm{l}$ are equivalent  and rotated by an angle $\Theta \neq 0,\pi/4$, thereby removing any mirror planes as well as the 2-fold axes. The $z$ axis remains to be an improper $\bar 4$ axis. For $\Theta=0$, the structure is equivalent to a D2d phase - from \cite{PB14}.}
\label{fig:17}
\end{figure}

If one adds two transverse biaxial directors $\bm{m}$ and $\bm{l}$ along the $\bar 4$ tetrahedral directions, the structure is equivalent to a D2d phase. In case the biaxial directors  are rotated in the transverse plane by a finite angle other than $\pi/4$ and $\pi/2$, as in Fig. \ref{fig:17}, an $S_4$ symmetric  S4 is obtained. It is obvious to see that due to this rotation the mirror planes are removed as well as both 2-fold rotation axes ($x,y$ axes). Only the (improper) 4-fold symmetry axis ($\bm{n}$ or $z$ axis) is left. Due to the existence of an improper rotation axis, there is no chirality.

%%%%%%%
%%%%%%%
%%%%%%%%
%%%%%%%%

The hydrodynamics of this phase is rather similar to the D2d phase, in particular the hydrodynamic variables are the same: Rotations of the preferred direction $\bm{n}$ (the tetragonal axis) and a rotation about this axis. The latter can again be described by appropriate rotations of the tetrahedral structure, 
$\delta \Omega \equiv (27/128) n_i \epsilon_{ipq} T_{pjk} \delta T_{qjk}$,
where we have used $N = 1$ in the normalization relation
$ 27 \alpha = 32 N^2$. 
The only difference is the reduced symmetry of S4 compared to D2d, which is manifest in more complicated structures of material tensors of fourth order (and higher) and, in addition, by a few more non-vanishing elements of the equilibrium tensor $T_{ijk}$ giving rise to some additional cross-couplings. The form of the hydrodynamic equations is the same as in the D2d phase and will not be repeated here. 

As an example for more complicated material tensors, the gradient free energy related to rotations of $\bm{n}$,
\begin{align} \label{Frank5S}
K_{ijkl}  &  =K_{3}n_{i}n_{j}\delta_{kl}^{\perp}+(K_{1}-2K_{2})\delta_{ik}
^{\perp}\delta_{jl}^{\perp}  \nonumber  \\
&  +K_{2}(\delta_{il}^{\perp}\delta_{jk}^{\perp}+\delta_{ij}^{\perp}\delta_{kl}
^{\perp})+K_{4}n_{p}n_{q}T_{ijp}T_{klq} \nonumber \\
&+ K_8 \delta^{\perp}_{rk} \delta^{\perp}_{tl}  T_{ijp}T_{rtp} .
\end{align}
contains five Frank bulk orientational elastic coefficients in the S4 phase (compared to four in D2d, Eq. (\ref{Frank})).
The $K_8$ term, which does not exist in the D2d phase, gives rise to new combinations of director variations, if linearized, of the form $\cos 2 \Theta \sin 2 \Theta (\nabla_x n_x - \nabla_y n_y)(\nabla_y n_x + \nabla_x  n_y)$ where $\Theta$ is the in-plane rotation angle as described in 
Fig. \ref{fig:17}.

On the other hand, the gradient energy involving $\delta \Omega$, Eq. (\ref{FrankD2d}), and the linear gradient term giving rise to ambidextrous helicity, Eq. (\ref{lingrad}), are as in the D2d phase. The same applies to the static Lehmann-type energy contributions, Eq. (\ref{cross}). There is still no linear gradient term w.r.t. $\nabla_i \Omega$, because of the invariance under $\bm{m} \Leftrightarrow \bm{l}$.

An example of a more complicated material tensor in the dynamics is the viscosity tensor, relating the stress tensor with the symmetric flow tensor, $\sigma_{ij} = -\nu_{ijkl} \nabla_l v_k$. It contains an additional 7th term (compared to Eq. (\ref{viscosity})), $ \nu_7 T_{rtp} (\delta^{\perp}_{rk} \delta^{\perp}_{tl}  T_{ijp} + \delta^{\perp}_{ri} \delta^{\perp}_{tj}  T_{klp})$ in the S4 phase, which has almost the same structure as the $K_8$ Frank term discussed above.
All other material tensors occurring in the static or dynamic part of the S4 hydrodynamics are of a rank less than four and have the same structure as in the D2d phase. 

Among the cases, where additional non-vanishing elements of the equilibrium tensor $T_{ijk}$
are relevant, is the linear gradient term, Eq. (\ref{lingrad}), $(\nabla_x n_y + \nabla_y n_x)\cos 2 \Theta + (\nabla_x n_x - \nabla_y n_y)\sin 2 \Theta$. It involves director structures not present in the D2d phase. 

In the reversible hydrodynamics there are couplings between shear flow and currents of temperature, concentration and charge, and {\it vice versa}, between shear stresses and gradients of temperature and concentration, and electric fields, Eq. (\ref{Gamma2a}) and (\ref{Gamma2b}), which in the S4 phase have the form
\begin{eqnarray} \label{shearflowS}
j_x^{\sigma,R} &=& \Gamma_{21} (\cos 2 \Theta A_{yz} + \sin 2 \Theta A_{xz} ) \nonumber \\
j_y^{\sigma,R} &=& \Gamma_{21} (\cos 2 \Theta A_{xz} - \sin 2 \Theta A_{yz} ) 
\nonumber \\
j_z^{\sigma,R} &=& \Gamma_{22} (\cos 2 \Theta A_{xy} + \sin 2 \Theta\, [A_{xx} - A_{yy}] ) 
\end{eqnarray}
and
\begin{eqnarray}\label{tempgradS}
\sigma_{xz} &=& - \Gamma_{21} (\cos 2 \Theta \,\nabla_y T + \sin 2 \Theta \,\nabla_x T) 
\nonumber \\
\sigma_{yz} &=& - \Gamma_{21} (\cos 2 \Theta \,\nabla_x T - \sin 2 \Theta \,\nabla_y T) 
\nonumber \\
\sigma_{xy} &=& - \Gamma_{22} \cos 2 \Theta \,\nabla_z T  
\nonumber \\
\sigma_{xx} &=& -\sigma_{yy}  = - \Gamma_{22} \sin 2 \Theta \,\nabla_z T  
\end{eqnarray}
with appropriate terms for the concentration and electric degrees of freedom involving $\Gamma_{31,32}$ and $\Gamma_{11,12}$. All terms $\sim \sin \Theta$ are new in the S4 phase and not present in D2d. They comprise hyperbolical flows and stresses and oblique currents.

The dissipative Lehmann-type couplings of the D2d phase, Eq. (\ref{LehmannD2d}), acquire more coupling elements in the S4 phase
\begin{eqnarray} \label{dissLehS1}
j_x^{\sigma,D} &=& -\psi^T (\cos 2 \Theta\,h_y + \sin 2 \Theta \,h_x) \nonumber \\
j_y^{\sigma,D} &=& -\psi^T  (\cos 2 \Theta \,h_x - \sin 2 \Theta \,h_y ) 
\end{eqnarray}
and
\begin{eqnarray}\label{dissLehS2}
\dot n_x &=& -\psi^T  (\cos 2 \Theta \, \nabla_y T + \sin 2 \Theta \,\nabla_x T) 
\nonumber \\
\dot n_y &=& -\psi^T (\cos 2 \Theta \,\nabla_x T - \sin 2 \Theta \,\nabla_y  T) 
\end{eqnarray}
with similar sets of equations involving the concentration and the electric degrees of freedom.

The orientation of the director (and the tetrahedral structure) in an external electric field is basically the same as in the D2d phase, Eq. (\ref{epsED2d}), including dielectric anisotropy and the cubic tetrahedral orientation leading to frustration. Assuming that the dielectric anisotropy effect is the dominant one orienting the director (and thus the $\bar 4$ axis) along the $z$ direction, a small oscillating electric transverse field will lead to a reorienting force on $\bm{n}$ of the transversely isotropic form $E_x^2 + E_y^2$ due to the dielectric anisotropy, while for the tetrahedral orientation the reorientation force is, in the S4 phase, of the form $\cos 2 \Theta E_x E_y + \sin 2 \Theta (E_x^2 - E_y^2)$.  Thus, this response to an external field can experimentally reveal the transverse anisotropy in the S4 phase different from the D2d case.

%%%%%%%%%%%%%%%%%%%%%%%%%%%%%%%%%%%%%%%%%%%%%%%%%%

\subsubsection{The non-polar orthorhombic D2 tetrahedral phase \label{D2}}

%%%%%%%%%
%%%%%%%%%
\begin{figure}[t]
\begin{center}
\includegraphics[width=7cm]{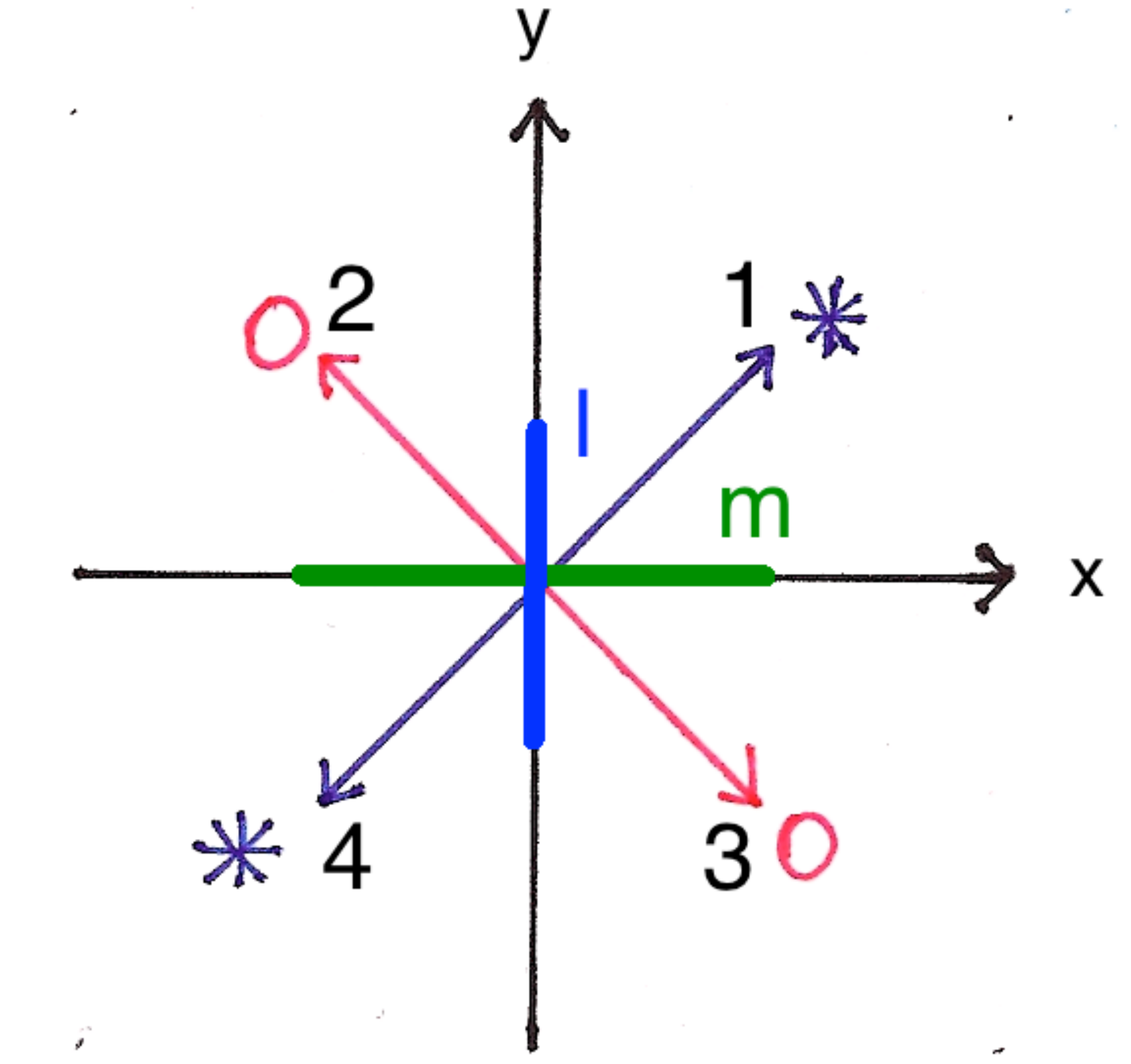}\\
\includegraphics[width=7cm]{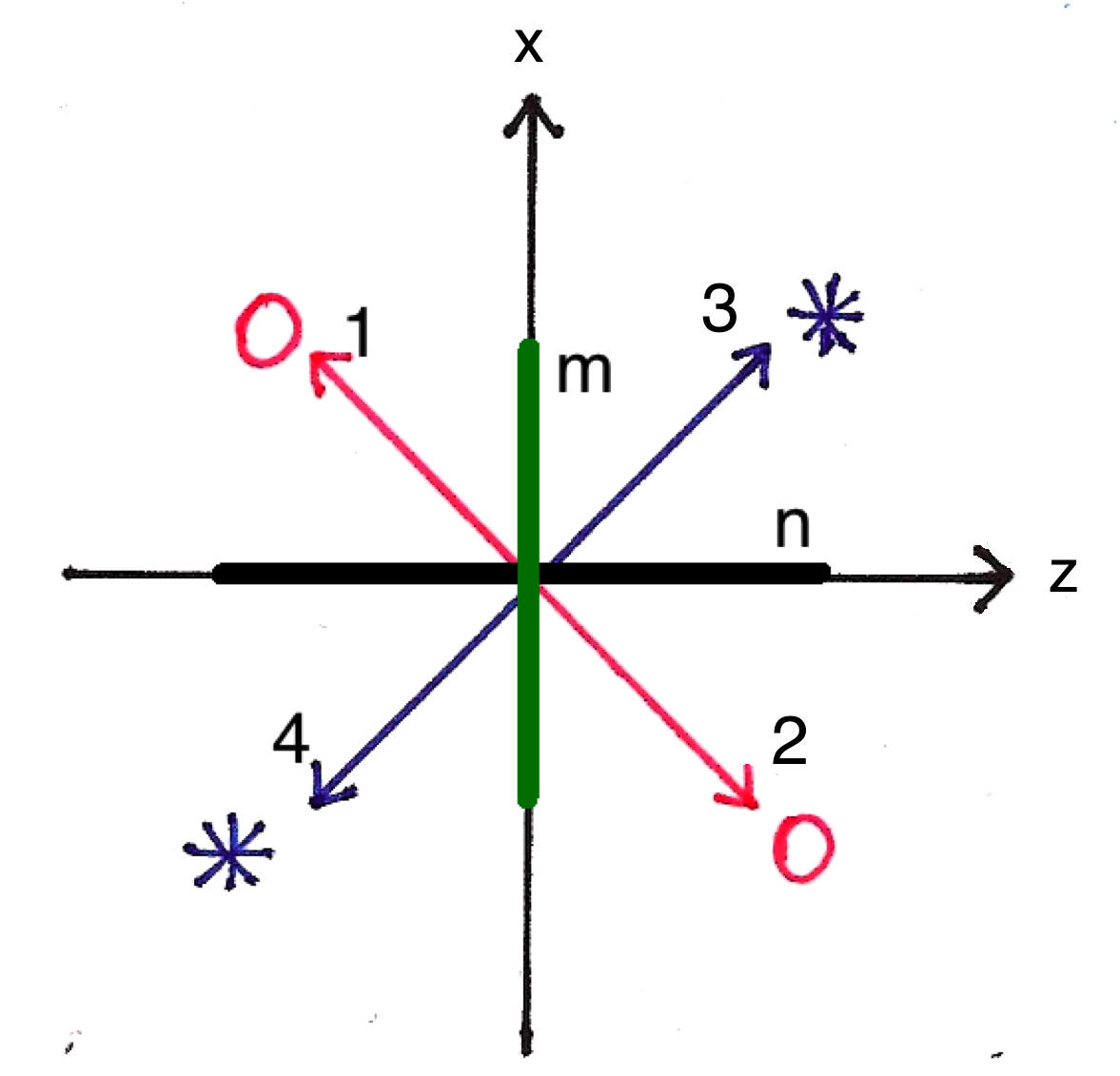}
\caption{Two projections of the 4 tetrahedral vectors $\bm{n}^\alpha$ ($\alpha=1,2,3,4$) and the orthorhombic directors $\bm{l}, \bm{m},\bm{n}$ in the D2 phase. Stars and circles as in Fig. \ref{fig:17}. Only three 2-fold symmetry axes (the $x,y,z$ axes) are left, but no mirror planes - from \cite{PB14}.}
\label{fig:18}
\end{center}
\end{figure}
%%%%%%%%%
%%%%%%%%%

A D2 phase can be viewed as an orthorhombic biaxial nematic of $D_{2h}$ symmetry (with mutually orthogonal directors $\bm{l}, \bm{m},\bm{n}$) that are rigidly attached to the three (improper) $\bar 4$ axes of the tetrahedral structure, cf. Fig. \ref{fig:18}. As a result, only proper (2-fold) symmetry axes are left, but no mirror planes rendering the phase chiral. The latter is expressed by the pseudoscalar quantity $q_0 \equiv \bm{n \cdot}(\bm{m} \times \bm{l}) n_i m_j l_k T_{ijk}$ or equivalently by $q_0^\prime \equiv \bm{n \cdot}(\bm{l} \times \bm{m}) n_i m_j l_k T_{ijk} = - q_0$ indicating that both kinds of handedness are present. This "ambidextrous chirality" \cite{G1} is of structural origin, in contrast to the molecule-based chirality of the chiral T phase, Section \ref{chiralT}.  

The hydrodynamics of the D2 phase is that of orthorhombic biaxial nematics \cite{Nbiax,biaxnemliu}, amended by effects of the broken inversion symmetry and chirality due to $T_{ijk}$. Like in the D2d and in the biaxial nematic phases, the hydrodynamic degrees of freedom (Goldstone modes) are the three independent rotations of the rigid structure. They can be realized by the three rotations $\delta \bm{n}$, $\delta \bm{m}$, and $\delta \bm{l}$ that preserve their mutual orientation due to the conditions $m_i \delta n_i + n_i \delta m_i =0$,  
$l_i \delta n_i + n_i \delta l_i =0$, and $m_i \delta l_i + l_i \delta m_i =0$. The tetrahedral structure follows those rotations rigidly. Alternatively, one could use rotations of the preferred axis $\delta \bm{n}$ and the rotation about this axis $\delta \Omega = (\bm{n}\times\bm{m})_i \delta m_i$ as variables. There are 12 bulk Frank-like quadratic rotation elastic coefficients, cf. \cite{BP26}.
  
In the following we will concentrate on the $T_{ijk}$- and $q_0$-induced effects. 
There are six linear gradient terms in the free energy, not present in ordinary biaxial nematics
\begin{eqnarray}
\label{lingradD2}   \nonumber
\varepsilon_{l} &=& T_{ijk}  (\xi_1 n_i  \nabla_j n_k + \xi_2  m_i \nabla_j m_k + \xi_3  l_i \nabla_j l_k) \\
&+&  q_0 \epsilon_{ijk} (k_1 n_i \nabla_j n_k + k_2 \,m_i\nabla_j m_k + k_3 \,l_i\nabla_j l_k) \quad
\end{eqnarray}
The first line is the generalization of Eq. (\ref{lingrad}) of the D2d phase describing ambidextrous helicity (cf. Sec.\ref{ambi}), since the inverted structure is different from the non-inverted one, but leads to the same energy reduction. The second line contains the linear twist terms of biaxial cholesterics \cite{PB90}. In the D2 phase, they describe ambidextrous chirality, since  $q_0$ and $-q_0$ are equally likely to occur. It is not possible to individually minimize each of the six terms for steric reasons. In that sense, the D2 phase is frustrated. 

If there are linear gradient terms, there are also static Lehmann-type energy contributions \cite{leh,Lehmann}, bilinear
in linear director gradients and variations of the scalar variables, $\gamma\in\{\sigma,\rho,c\}$ 
\begin{equation}
\label{statFredD2}   
\varepsilon_{c} =\sum_\gamma (\delta \gamma) \bigl(T_{ijk} g_{ijk}^\gamma+ q_0 \epsilon_{ijk} G_{ijk}^{\gamma}\bigr)
\end{equation}
with $g_{ijk}^\gamma= \xi_1^\gamma n_i  \nabla_j n_k + \xi_2^\gamma  m_i \nabla_j m_k + \xi_3^\gamma  l_i \nabla_j l_k$ and $G_{ijk}^{\gamma}$ as $g_{ijk}^\gamma$, but with the coefficients 
$\xi_{1,2,3}^\gamma$ replaced by $k_{1,2,3}^\gamma$.
Again, they are either a generalization of the D2d case, Eq. (\ref{cross}) or of the chiral nematic case.

The dissipative Lehmann-type terms also come in two classes, either due to the lack of inversion symmetry (as in the D2d phase, Eq. (\ref{LehmannD2d})) or due to chirality (as in cholesterics), for $\nabla_k Q\in\{\nabla_k T, \nabla_k \mu_c, E_k\}$,
\begin{equation}
\label{dissFred}   
2 R_L = \sum_Q (\nabla_k Q) \bigl(T_{ijk} h_{ij}^Q + q_0 \epsilon_{ijk} H_{ij}^{Q}\bigr)
\end{equation}
where $h_{ij}^Q=(\psi_1^Q m_j m_p + \psi_2^Q l_jl_p) n_i h_p^n + \psi_3^Q l_jl_p m_i  h_p^m$
and $H_{ij}^Q$ as $h_{ij}^Q$, but with coefficients $\psi_{1,2,3}^Q$ replaced by different ones $\Psi_{1,2,3}^Q$.  The molecular fields, $h_i^{n}$, $h_i^{m}$, follow from the Frank gradient energy in the standard way. The chirality-based static and dynamic Lehmann-type contributions lead to rotations of the directors due to applied thermodynamic forces \cite{leh,Lehmann} and the inverse effects \cite{inverse}, while the $T_{ijk}$-based ones do not have such a simple geometric interpretation. 
 
The phenomenological reversible couplings between {\it e.g.} the heat current and deformational flow, which is characteristic for tetrahedral systems
\begin{eqnarray}
\label{revAcurr}   
j_i^{\sigma,ph} = \Gamma_{ip}^\sigma \,T_{pjk} A_{jk} \\
\sigma_{ij}^{ph} =  -T_{pij}  \Gamma_{kp}^{\sigma} \nabla_k T
\label{revsigmagrad}   
\end{eqnarray}
contain one and two parameters for the \Td and the D2d phase, respectively, while there are 3 for the D2 phase, since  
\begin{equation} \label{rank2ortho}
\Gamma_{ij}^\sigma = \Gamma_1^\sigma n_i n_j + \Gamma_2^\sigma m_i m_j + \Gamma_3^\sigma l_i l_j
\end{equation}
is of the standard orthorhombic form. The same holds for the appropriate couplings to the concentration and the electric current.

Due to the chirality there is a phenomenological, reversible contribution to the tetrahedral rotation, Eq. (\ref{Omegadyn}), that couples to deformational flow  
\begin{align} \label{YAD2}
Y^{\Omega R} =& q_0 \lambda^\Omega n_i  T_{ijk} A_{jk} \\
\sigma_{ij}^{ph} =&  q_0 \lambda^\Omega n_k  T_{ijk} \, h^\Omega
\label{sigmaOmegaD2}
\end{align}
where $h^\Omega$ is the conjugate to $\Omega$, Eq. (\ref{GibbsD2d}). The stress tensor carries the appropriate counter terms that guarantee zero entropy production. There is one material parameter, $\lambda^\Omega$, involved, which is generally different from $\lambda$, the parameter that governs the flow alignment of the preferred direction $\bm{n}$, Eq. (\ref{alignment}).
Together with the (linearized) response to rotational flow, $Y^{\Omega R} = - \omega_i n_i$,Eq. (\ref{Yrev3})), there is now also a stationary alignment of the tetrahedral rotation in planar simple shear flow, in contrast to the D2d phase.
This is similar to the case of the chiral T phase, Eq. (\ref{YA}).

There are 4 orientational viscosities
\begin{align}
\label{dissrotviscD2}   
2 R_\gamma = &\bigl(\frac{1}{\gamma_1} m_i m_j + \frac{1}{\gamma_2} l_i l_j \bigr) h_i^n h_j^n + + \frac{1}{\gamma_3} l_i l_j  h_i^m h_j^m \nonumber \\ &+\frac{1}{\gamma^\Omega} (h^\Omega)^2
\end{align}

The orientation by external fields of the combined nematic and tetrahedral structure in the D2 phase has several different origins 
\begin{align} \label{epsED2}
\tilde\varepsilon^{E}    =&  -\frac{1}{2}\epsilon^a_{ij}E_{i}E_{j} -\frac{1}{2}\chi^a_{ij}H_{i}H_{j} \nonumber
-\epsilon_{10} T_{ijk}E_{i}E_{j}E_{k} \\ -& \nonumber
\frac{\epsilon_{11}}{3} T_{ijk} ( n_i E_j E_k + E_i E_j n_k + E_i n_j E_k)(\bm{n \cdot E})  
\\- & \nonumber
\frac{\epsilon_{12}}{3} T_{ijk} ( m_i E_j E_k + E_i E_j m_k + E_i m_j E_k)(\bm{m \cdot E})  
\\ -& \nonumber
\frac{\epsilon_{13}}{3} T_{ijk} ( l_i E_j E_k + E_i E_j l_k + E_i l_j E_k)(\bm{l \cdot E})  
\\ +& \nonumber    q_0\,\epsilon_{ipq}  T_{ijk} \left( \chi_{jp}^E E_k E_q  + \chi_{jp}^H H_j H_k \right)  \\
+&   q_0 \epsilon_{pik}  \zeta_{jp}^{EH} E_i H_j H_k  
\end{align}
where all second rank tensors in Eq. (\ref{epsED2}) are of the form $\epsilon^a_{ij}= \epsilon^a_1 n_i n_j + \epsilon^a_2 m_i m_j$. 
There is the electric and magnetic anisotropy, and the tetrahedral orientation, described by quadratic and cubic field-dependent energies, respectively. Compared to the energy expressions for the \Td and D2d phase, Eqs. (\ref{cubicE}) and (\ref{epsED2d}), respectively, there are more coefficients involved due to the biaxiality. In addition, there are terms quadratic in the field due to the chirality, anisotropy, and tetrahedral order in the D2 phase. Obviously, there is no orientation that minimizes those contributions individually, and orientational frustration has to be expected.

%%%%%%%%%%%%%%%%%%%%%%%%%%%%%%%%%%%%%%%%%%%%%%%%

\subsection{The polar, low symmetry tetrahedral biaxial nematic phases} \label{C32v}

\subsubsection{ The polar trigonal C3v tetrahedral phase} \label{C3v}

%%%%%%%
%%%%%%%
\begin{figure}[ht]
%\begin{center}
\includegraphics[width=7.4cm]{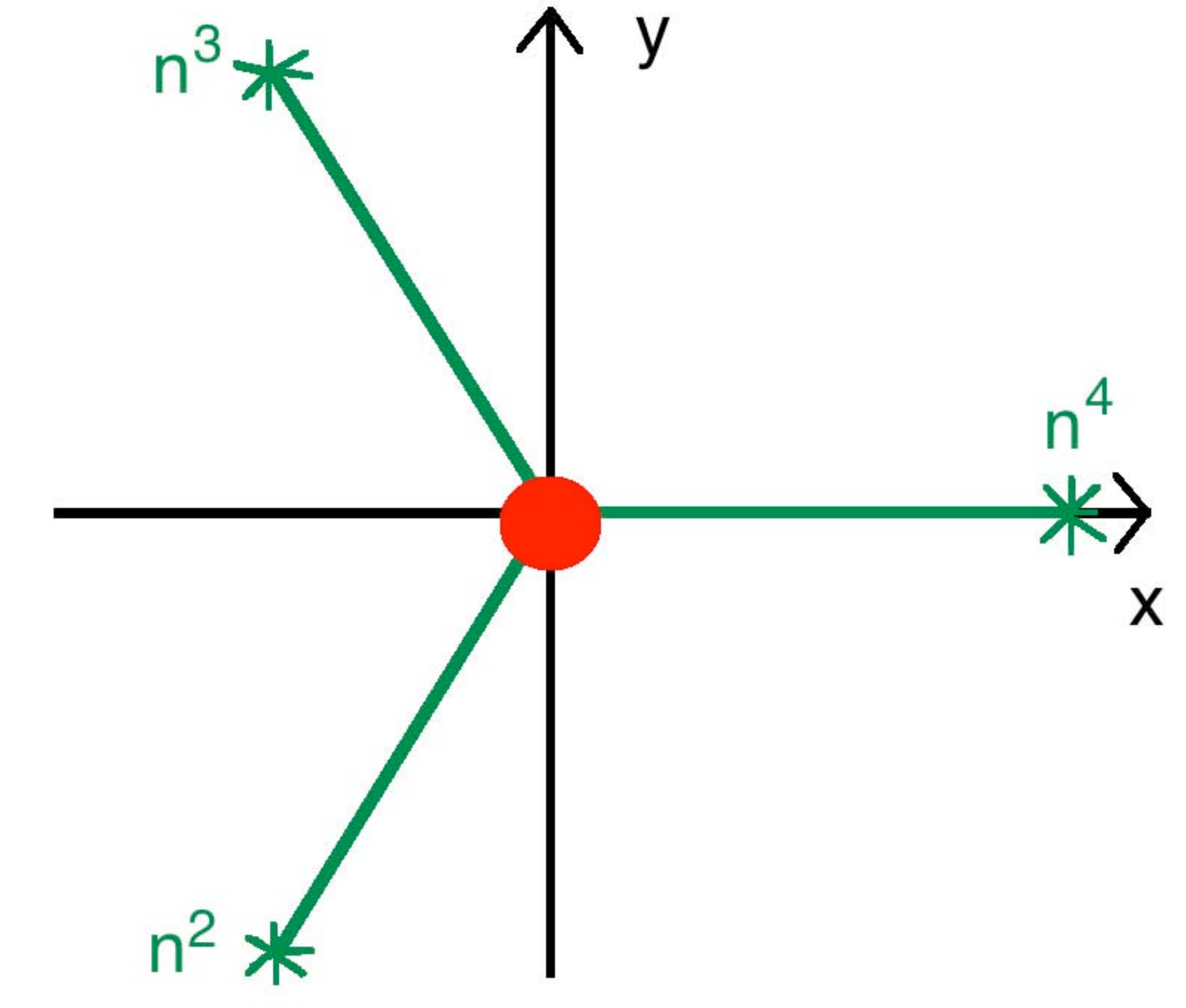}
\caption{The projection of the three tetrahedral vectors, $\bm{n}^{2,3,4}$, into the x/y plane. They all have a component pointing into that plane (asterisk). The tetrahedral vector $\bm{n}^1$ is sticking out of the plane along the z- axis, which is the polar, 3-fold symmetry axis of the C3v phase. The nematic director $\bm{n}$ is along the z-axis (red dot). This representation of the tetrahedral vectors corresponds to Eq. (\ref{unitE}).}
\label{fig:19}
\end{figure}
%%%%%%%%
%%%%%%%%

This phase is obtained, when the nematic director is along one of the tetrahedral vectors, $\bm{n} \parallel \bm{n}^1$ ($\parallel \bm{e}_z$ in Fig. \ref{fig:19}). The polarization
$P_i = T_{ijk} Q_{jk} = P_0 p_i$ is along that direction. We take the absolute value of the polarization
$P_0 = |T_{ijk} Q_{jk}| = (4/9)SN$ as a constant. The hydrodynamic variables are the rotations of the polarization $\delta p_i$ with $\bm{p \cdot}\delta \bm{p} = 0$ as in polar nematic LC \cite{polnema} and, in addition, rotations of the tetrahedral structure about the polar direction
$\delta\Omega\equiv \bm{p \cdot} \delta\bm{\Gamma}=(1/4\alpha)p_{i}\epsilon_{ipq}T_{pjk}\delta T_{qjk}$, similar to the D2d phase. However, here $\delta\Omega$ is odd under spatial inversion, in contrast to the D2d case, Eq. (\ref{deltaOmega}), and the rotation axis is a tetrahedral direction and not a $\bar 4$ axis as in the D2d case.

The structure of the hydrodynamics is therefore rather similar to that in the D2d phase, although there are subtle differences due to the polarity and the 3-fold rotational symmetry (about the polar axis) in the C3v phase. In particular, the linear gradient term, 
\begin{equation} \label{lingrad3Cv}
\varepsilon_{l} = \xi_p^\prime T_{ijk} p_i \nabla_j p_k = \xi_p \nabla_i p_i
\end{equation}
describes splay, as in polar nematic LC \cite{polnema}, allowing for spontaneous splay phases. In the C3v phase there is no ambidextrous helicity related to this linear gradient term. 
Similarly, the static Lehmann-type energy 
\begin{equation} \label{statLehmann3Cv}
\varepsilon_{c}=  ( \xi^\sigma \delta \sigma +  \xi^\rho \delta \rho+ \xi^c \delta c) \nabla_i p_i
\end{equation}
is as in the polar nematic case. The Frank orientational elasticity tensor carries 8 coefficients as in the S4 phase, Eqs. (\ref{Frank5S}).

The dissipative Lehmann-type terms have the form
\begin{equation}
\label{dissFredC3v}   
2 R_L =  \delta_{jl}^{\perp} p_i T_{ijk} ( \Psi^T \nabla_k T + \Psi^c \nabla_k \mu_c + \Psi^E E_k)h_l^p 
\end{equation}
where $h_l^p$ is the conjugate to $\delta p_i$.

Third rank material tensors, {\it e.g.} reversibly relating flow with gradients of temperature, which contain three material parameters  in the polar nematic case \cite{polarchol} and two in the D2d phase (cf. {\it e.g.} Eq. (\ref{Gamma2a})), have 5 coefficients in the C3v phase
\begin{eqnarray} \label{revPT3Cva}
j_{i}^{\sigma,ph}  & =&[ (\Gamma_{21}\delta_{li}^{\perp}+\Gamma_{22}%
\,p_{l}p_{i})T_{ljk}  \\ \nonumber &&+ \phi_1 p_i p_j p_k + \phi_2 p_i \delta_{jk}^{\perp} + \phi_3 (p_j \delta_{ik}^{\perp} + p_k \delta_{ij}^{\perp}) ]A_{jk}\\
\sigma_{ij}^{ph}  & =&  -[(\Gamma_{21}\delta_{lk}^{\perp}+\Gamma_{22}\,p_{l}p_{k})T_{ijk} \\ \nonumber && + \phi_1 p_i p_j p_l + \phi_2 p_l \delta_{ij}^{\perp} + \phi_3 (p_j \delta_{il}^{\perp} + p_i \delta_{jl}^{\perp}) ]\nabla_{l}T \label{revPT3Cvb}
\end{eqnarray}
with $\delta_{ij}^{\perp} = \delta_{ij} - p_i p_j$. Appropriate equations are obtained for concentration and electrical current.

The fourth rank viscosity tensor contains, as in the D2d phase, six coefficients, cf. (\ref{viscosity}) with $n_i$ replaced by $p_i$. The 7th term found in the S4 phase, $\sim \nu_7$, is zero due to the trigonal symmetry of $T_{ijk}$, Eq. (\ref{unitE}). On the other hand, second rank material tensors are of the standard uniaxial form, since the third rank tensor cannot influence second rank material properties.
 
There is flow alignment of the preferred direction, as in polar nematic LC and in the D2d phase, cf. Eq. (\ref{alignment}). The dynamics of rotations, $\Omega$, about the preferred direction (not present in polar nematic LC) is as in the D2d phase, with no alignment due to shear flow, and no alignment in an electric field. There is one rotational viscosity with respect to $\Omega$, which is different from that for $\delta \bm{p}$. 

The orientation of the preferred axis in an external electric fields is dominated by the polarization, which is along the field in equilibrium. There is no orientational frustration by an external field, since with the polarization also one of the tetrahedral vectors is along the field. Deviations of the polar direction from the field cost energy, $\tilde \varepsilon^E_2 = \tfrac12 P_0 E_0 (\delta p_i)^2$, and lead to a relaxation, which is linear in the field amplitude $E_0$, in contrast to the D2d phase, Eq. (\ref{E2D2d}).

%%%%%%%sub%%%%%%%%%%%

\subsubsection{The polar orthorhombic C2v tetrahedral phase} \label{C2v}

%%%%%%%
%%%%%%%
\begin{figure}[ht]
\includegraphics[width=7.4cm]{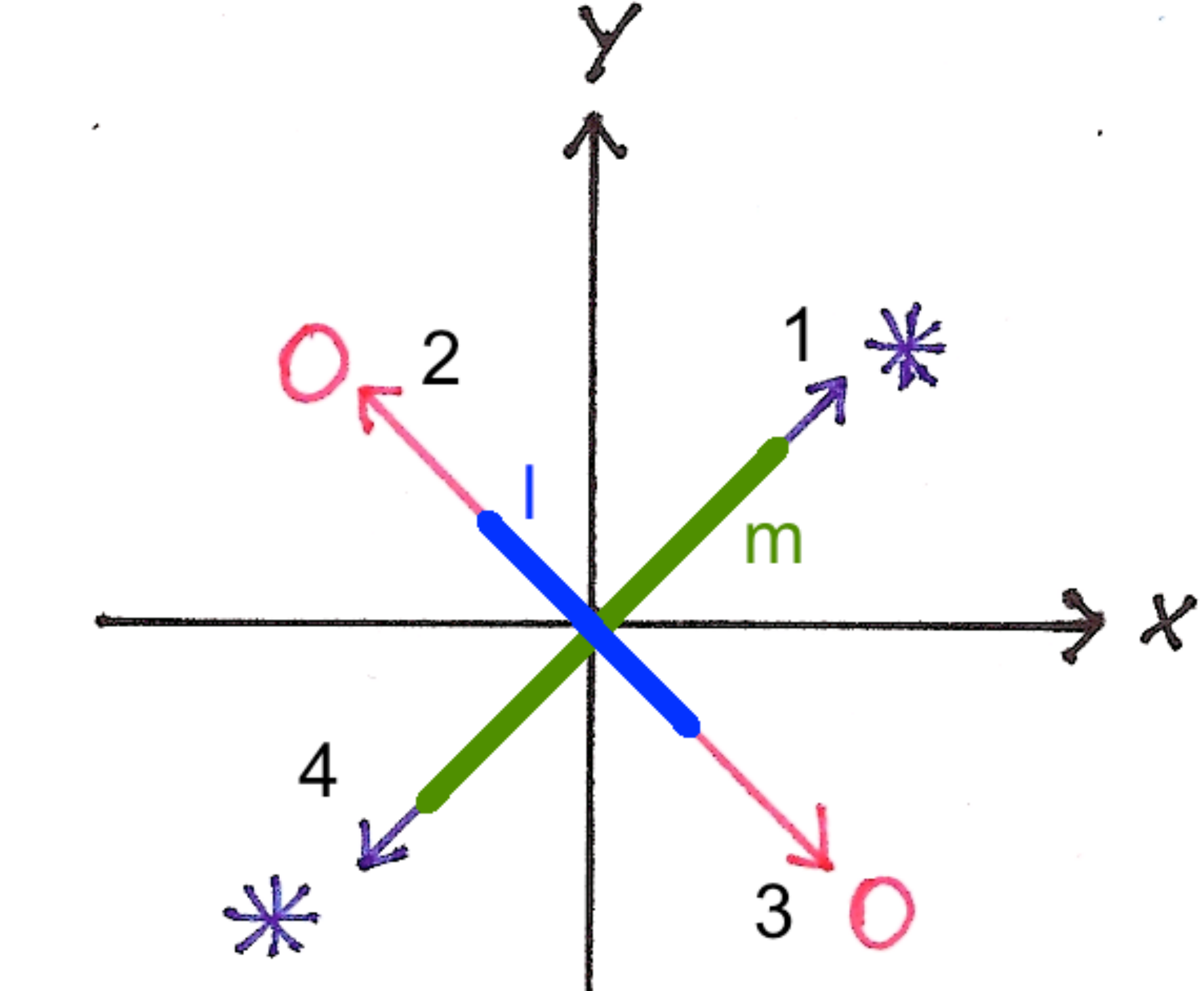}
\caption{Structure of the orthorhombic C2v phase: The tetrahedral vectors are as in Fig. \ref{fig:18} (upper graph), corresponding to Eq. (\ref{unit}). The biaxial nematic directors $\bm{m}$ and $\bm{l}$ are in the $\bm{n}^1/\bm{n}^4$ and $\bm{n}^2/\bm{n}^3$ planes, which are mirror planes. The perpendicular z-axis is polar, since $\bm{m}$ and $\bm{l}$ are not equivalent, and is the only (2-fold) symmetry axis left - from \cite{PB14}.}
\label{fig:20}
\end{figure}
%%%%%%%%
%%%%%%%%

This phase can be viewed as an orthorhombic biaxial nematic with the three mutually orthogonal, non-equivalent directors $\bm{n,m,l}$, where in addition inversion symmetry is broken due to the tetrahedral structure $T_{ijk}$. A possible spatial representation of the structure is shown in Fig. \ref{fig:20}. It is polar with the polar direction $p_i \sim T_{ijk} (m_j m_k - l_j l_k)$ along the director $\bm{n}$. The absolute value of the polarization is taken as constant. 

The hydrodynamic variables are the rotations of the polar direction, $\delta p_i$, as in a polar nematic LC, and the rotation, $\delta \Omega=p_i \delta \Gamma_i$, of the tetrahedral structure about the polar direction. Since the latter is a $\bar 4$ axis of the tetrahedral structure, this variable is somewhat similar to the appropriate one in the D2d phase and, in particular, there is no flow alignment of such rotations. Another way of setting up the hydrodynamics is the use of the rotations of the directors, $\delta \bm{m}$ and $\delta \bm{l}$ that preserve their mutual orientation, as in the D2 phase described in Sec. \ref{D2}. The tetrahedral structure follows those rotations rigidly. 

The system shows four linear gradient terms 
\begin{equation}
\label{lingradC2v}   
\varepsilon_{l} = T_{ijk}  (\xi_1 p_i  \nabla_j p_k + \xi_2  m_i \nabla_j m_k + \xi_3  l_i \nabla_j l_k) +  \xi_p  \nabla_i p_i  
\end{equation}
The first three terms are linear gradient terms similar to that of the D2d phase favoring spontaneous ambidextrous helical rotations of the tetrahedral structure about the different $\bar 4$ axes. The last one is the standard linear splay term of polar systems. 
All the spontaneous structures favored by the individual linear gradient terms are mutually incompatible, and a rather complicated, inhomogeneous ground state may occur.

The static Lehmann-type energy has 4 coefficients for the thermal, solutal, and electric degree of freedom, each ($\gamma \in \{\sigma,\rho,c\}$)
\begin{equation}
\label{statFred2Cv}   
\varepsilon_{c} =\sum_\gamma (\delta \gamma) \bigl(T_{ijk} g_{ijk}^\gamma+ \xi_p^\gamma \nabla_i p_i \bigr)
\end{equation}
with $g_{ijk}^\gamma= \xi_1^\gamma p_i  \nabla_j p_k + \xi_2^\gamma  m_i \nabla_j m_k + \xi_3^\gamma  l_i \nabla_j l_k$ combining the tetrahedral and the polar contributions. 

A similar combination of polar and tetrahedral  effects are found for the reversible flow/temperature gradient
\begin{eqnarray}
\label{revsigma2Cva}   
j_i^{\sigma,ph} &=& (\phi_1^\sigma  p_i p_j p_k + \phi_2^\sigma  p_i \delta_{jk}^{\perp} + \phi_3^\sigma  [p_j \delta_{ik}^{\perp} + p_k \delta_{ij}^{\perp}] \nonumber \\  &&+ 
\Gamma_{iq}^\sigma  \,T_{qjk} )A_{jk} \\
\sigma_{ij}^{ph} &=& -(\phi_1^\sigma  p_i p_j p_k + \phi_2^\sigma  p_k \delta_{ji}^{\perp} + \phi_3^\sigma  [p_j \delta_{ik}^{\perp} + p_i \delta_{kj}^{\perp}] \nonumber  \\ &&\,\,\,\,+ \Gamma_{kq}^\sigma  T_{qji} )\nabla_k T
\label{revsigma2Cvb}   
\end{eqnarray}
with $\Gamma_{ij}$ given by Eq. (\ref{rank2ortho}). There are analogous couplings of flow to concentration gradients and electric fields.

The dissipation of the rotation $\delta p_i$ and of $\delta \Omega$ is as in the D2 phase, Eq. (\ref{dissrotviscD2}). The dissipative Lehmann-type terms are the same as the achiral ($q_0=0$) part of those of the D2 phase, Eq. (\ref{dissFred}).

The forth-rank Frank tensor, $K_{ijkl}$, of the quadratic gradient energy contains 
12 coefficients as in orthorhombic biaxial nematic LC \cite{Nbiax,biaxnemliu}. 
Quadratic contributions constructed out of the linear gradient terms
given in Eq. (\ref{lingradC2v}) are already contained in this number.
The viscosity tensor $\nu_{ijkl}$ carries 9 ordinary viscosities.

In an external electric field the polar and tetrahedral orientation energies 
along with 
the dielectric anisotropies of the biaxial directors take the form
\begin{align} \label{epsE2Cv}
\tilde\varepsilon^{E}    =& - \bm{P \cdot E} -\frac{1}{2}\epsilon^a_{ij}E_{i}E_{j}%
-\epsilon_{10} T_{ijk}E_{i}E_{j}E_{k} +\ldots
\end{align}
with $\epsilon^a_{ij}= \epsilon^a_1 p_i p_j + \epsilon^a_2 m_i m_j$.
Since there is no orientation that minimizes simultaneously all 
contributions,
frustrated orientations must be expected. 
The $\dots$ refer to the anisotropic tetrahedral orientation energies $\sim \epsilon_{11}, \epsilon_{12}, \epsilon_{13}$ of the D2 phase, Eq. (\ref{epsED2}).

%%%%%%%sub%%%%%%%%%%%

\subsubsection{The polar monoclinic C2 tetrahedral phase} \label{C2}

%%%%%%%
%%%%%%%
\begin{figure}[ht]
\includegraphics[width=7.4cm]{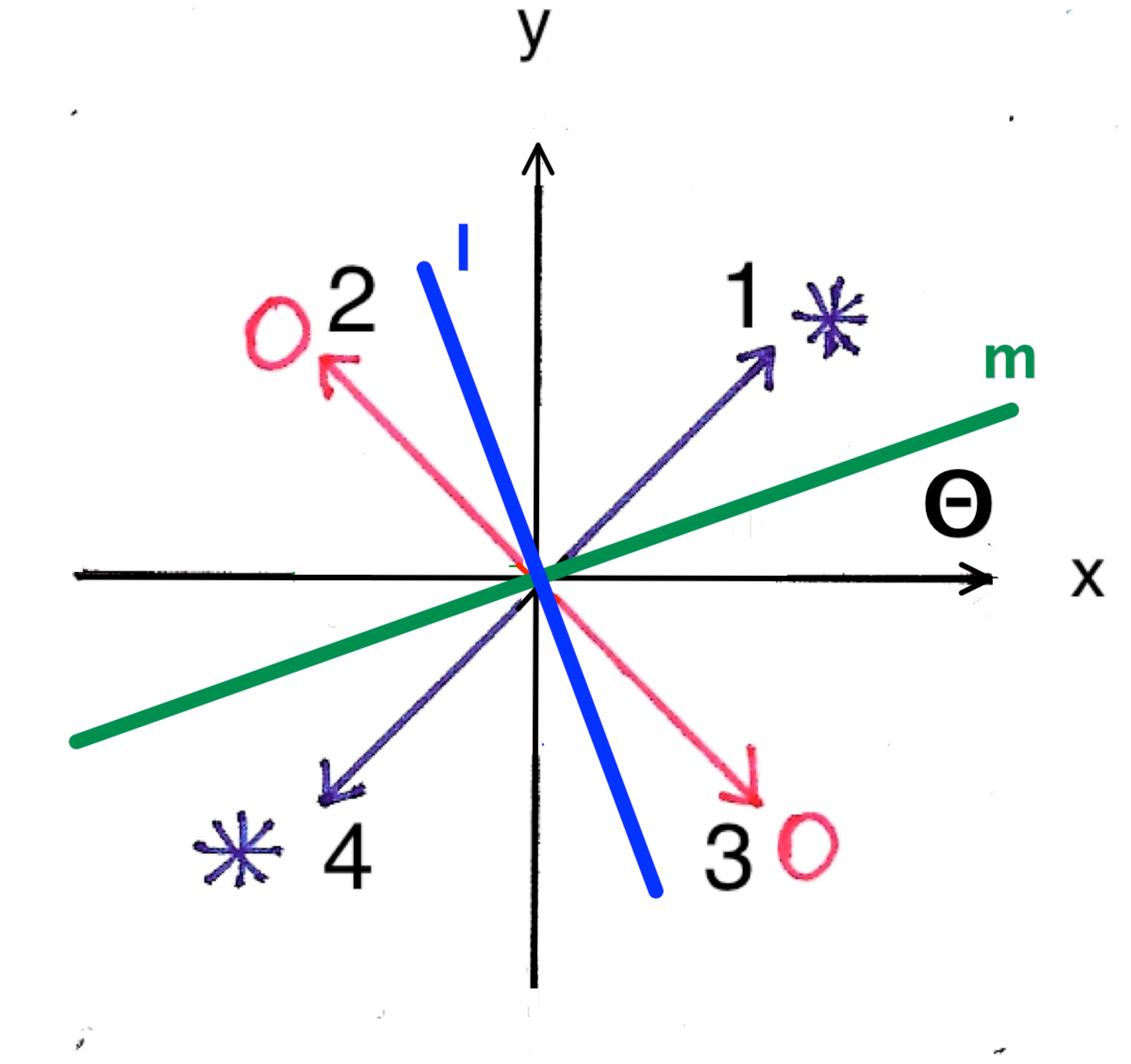}
\caption{The structure of the monoclinic C2 phase: Similar to Fig. \ref{fig:20}, but the biaxial nematic directors are rotated away from the 1/4 and 2/3 planes, which are therefore no longer mirror planes. The same structure is obtained, when in the S4 phase, Fig. \ref{fig:17}, the biaxial nematic directors $\bm{m}$ and $\bm{l}$ are made inequivalent - from \cite{PB14}.}
\label{fig:21}
\end{figure}
%%%%%%%%
%%%%%%%%

This phase is very similar to the C2v phase, but is in addition chiral and of the somewhat lower monoclinic symmetry, Fig. \ref{fig:21}. The chirality is manifest in the pseudoscalar quantity $q_0= n_i n_j m_k m_p l_q l_r \epsilon_{ikq} T_{jpr}$ that changes sign under spatial inversion. One can set up the hydrodynamics of the phase, basically, by adding chiral terms to those of the C2v phase. Alternatively, this phase is like the D2 phase, but is in addition polar with $p_i \sim T_{ijk} (m_j m_k - l_j l_k)$ the polar direction. Thus, its hydrodynamics is that of the D2 phase with the polar terms added.

In particular there are 7 linear energetic gradient terms, cf. Eqs. (\ref{lingrad}) and (\ref{lingradC2v}),
\begin{eqnarray}
\label{lingradC2}   
\varepsilon_{l} &=& \xi_p  \nabla_i p_i  
\nonumber + T_{ijk}  (\xi_1 p_i  \nabla_j p_k + \xi_2  m_i \nabla_j m_k + \xi_3  l_i \nabla_j l_k) \\
&+&  q_0 \epsilon_{ijk} (k_1 p_i \nabla_j p_k + k_2 \,m_i\nabla_j m_k + k_3 \,l_i\nabla_j l_k) \quad\end{eqnarray}
indicating spontaneous splay, ambidextrous helicity and ambidextrous chirality, all together. Of course, there is no spatial structure that minimizes all those terms individually resulting in frustrated textures. 
The lower (monoclinic) symmetry allows for 20 Frank-like coefficients and 13 ordinary viscosities \cite{mason}.

The static Lehmann-type energy $\varepsilon_c$ is the combination of that in the D2 and C3v phase, Eqs. (\ref{statFredD2}) and (\ref{statLehmann3Cv}) containing 7 coefficients for each $\gamma$.

The dissipative Lehmann-type terms are given by Eq. (\ref{dissFred}), where, however, the monoclinic symmetry allows for an additional term in the symmetric tensor $h_{ij}^Q$, since every symmetric 2-rank has the form
\begin{equation}
\label{2rankmono}   
a_{ij} = a_{11} \, p_i p_j + a_{22} \,m_i m_j + a_{33} \, l_i l_j + a_{23} \, p_k T_{ijk}
\end{equation}
bringing the number of coefficients in $R_L$ to 7 for each $Q$ in the C2 phase.

The phenomenological, reversible couplings between deformational flow and gradients of temperature, concentration and electric field are given by Eqs. (\ref{revsigma2Cva}) and (\ref{revsigma2Cvb}), if for the second rank tensors 
$\Gamma_{ij}$
the form Eq. (\ref{2rankmono}) is used.

The phenomenological, reversible couplings of the tetrahedral rotation with deformational flow in the chiral chiral tetrahedral phases read
\begin{align} \label{YAC2}
Y^{\Omega R} =& q_0 \lambda^\Omega_{ij} A_{ij} \\
\sigma_{ij}^{ph} =&  q_0 \lambda^\Omega_{ij}  \, h^\Omega
\label{sigmaOmegaC2}
\end{align}
with $ \lambda^\Omega_{ij}$ having the form Eq. (\ref{2rankmono}).

There are four orientational viscosities
\begin{equation}
\label{dissrotviscC2}   
2 R_\gamma = \bigl(\frac{1}{\gamma}\bigr)_{ij} h_i^p h_j^p + \frac{1}{\gamma^\Omega} (h^\Omega)^2
\end{equation}
where $\bigl(\frac{1}{\gamma}\bigr)_{ij}$ has the transverse structure
\begin{equation}
\label{2rankmonotrans}   
a_{ij}^{\perp} = a_{22} \,m_i m_j + a_{33} \, l_i l_j + a_{23} \, p_k T_{ijk}
\end{equation}

For the energy, $\tilde \varepsilon^E$, responsible for the orientation of the tetrahedral and director structure in an external field, one can take the expression of the D2 phase, Eq. (\ref{epsED2}), if the polar energy $- \bm{P \cdot E}$ is added and for all transverse tensors, {\it e.g.} $\epsilon^a_{ij}$,  the form Eq. (\ref{2rankmonotrans}) is used.

%%%%%%%%%%%%%%%%%%%%%%%%%%%%%%%%%%%%%%%%%%%%%%%%
%%%%%%%%%%%%%%%%%%%%%%%%%%%%%%%%%%%%%%%%%%%%%%%%

\section{Homogeneously Uncorrelated Tetrahedral and Nematic Order: Splay-bend Phase} \label{TNnotcorr}

Here we discuss the case that the coupling of the nematic the tetrahedral orientation is negligible in the Landau free energy, Eq. (\ref{Genergy}), and only gradient energies matter. In that case, the relevant coupling is given by the linear gradient energy density, ${\cal D} T_{ijk} \nabla_k Q_{ij}$, Eq. (\ref{GLenergy}). Clearly, such a term favors inhomogeneous structures. As examples, we construct two (slightly) different types of splay-bend textures and show that they are minimum states \cite{O}. 

%%%%%%%%%%
%%%%%%%%%%
\begin{figure}
%\begin{center}
\includegraphics[height=21cm]{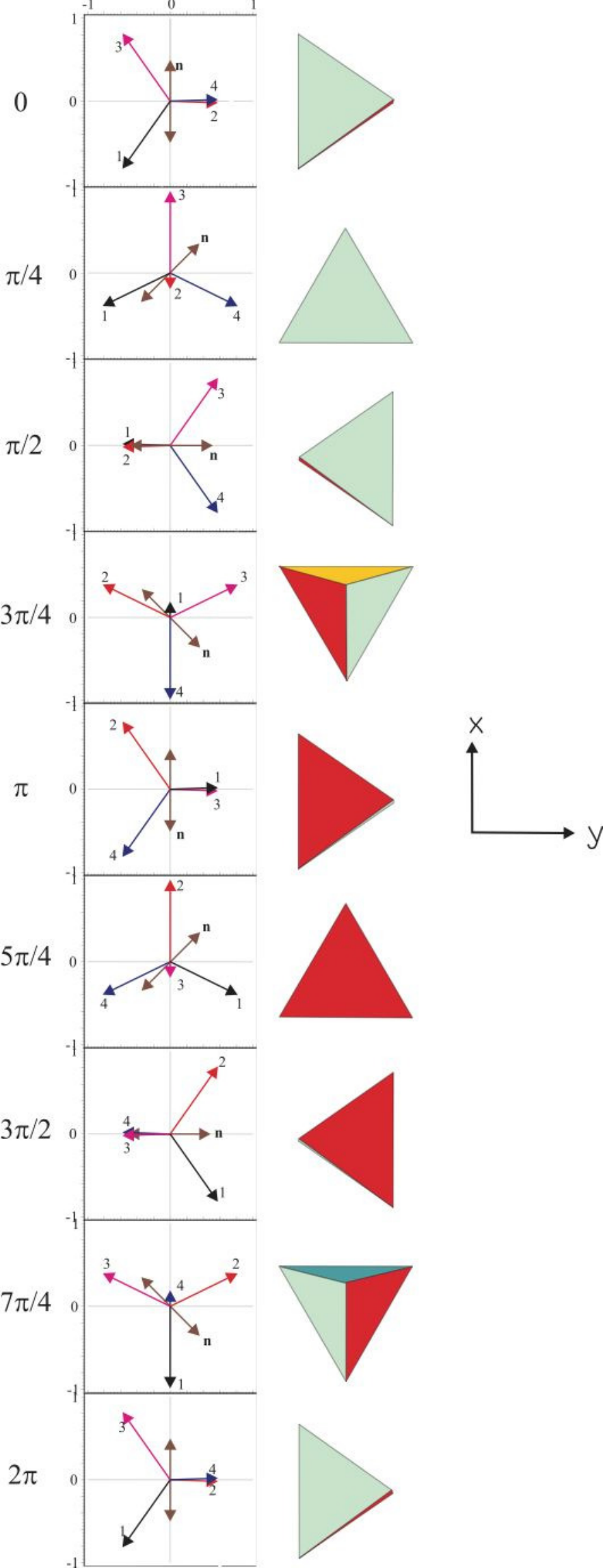}
\caption{ \label{xy} 
The orientation 
of the tetrahedra in the splay-bend texture 
viewed along the z direction (right) and the projection of 
the tetrahedral unit vectors (labeled 1 to 4), and of the 
director $\bm{ n}$ (with two arrows), in the (x,y)-plane (left)
for different values of $q_c x$ - from \cite{O}.} 
\label{fig:22}
\end{figure}
%%%%%%%%%%
%%%%%%%%%%

We start with a conformation similar to that of a D2d phase. The tetrahedral vectors $\bm{n}_0^\beta$ are given by Eq. (\ref{unit}), where the three $\bar 4$ improper rotation axes are given by the Cartesian x, y, z  direction. The uniaxial director $\bm{n}_0$ is along one of them, the x axis. We can now rotate the nematic and tetrahedral structure independently. For the director we assume a splay-bend texture applying a rotation of angle $qx$ about another $\bar 4$ axis, the z axis
\begin{equation} \label{2D}
{\cal R}_z(qx) \,\bm{n}_0 = {\bf n} = (\cos qx,  \sin qx, 0)
\end{equation}
with the rotation matrix
\begin{equation}\label{Rz}
\mathcal{R}_z (qx)= 
\begin{pmatrix}
\cos qx & \sin qx & 0 \\
-\sin qx & \cos qx & 0 \\
0&0& 1
\end{pmatrix}
\end{equation}
This is the standard 2-dimensional periodic splay-bend texture with wave vector q. Later we will also consider a 3-dimensional generalization of that.

A similar periodic splay-bend texture for the tetrahedral vectors is obtained by applying ${\cal R}_z(kx)$ to all of them, but in addition followed by a constant rotation by a fixed angle $\phi$ about the y axis
\begin{equation} \label{tetrarot}
{\cal R}_y(\phi) {\cal R}_z(kx) \,\bm{n}_0^\beta = {\bf n}^\beta 
\end{equation}
with
\begin{equation}\label{Ry}
\mathcal{R}_y(\phi) = 
\begin{pmatrix}
\cos \phi & 0 & -\sin \phi \\
0&1& 0\\
\sin \phi& 0 & \cos \phi
\end{pmatrix}
\end{equation}
Of course, since all tetrahedral vectors are rotated the same way, the tetrahedron is undeformed. The director is not rotated the same way and its orientation relative to the tetrahedral vectors varies periodically in space. The typical length scales involved are given by $1/k$ and $1/q$. Since the splay-bend structure has been obtained by rotations only, the structure is defect-free.

To investigate the energy gain due to the linear gradient energy term, one has to integrate over the whole space. Assuming the system size to be much larger than the length scales of the splay-bend texture, only in the commensurate case, $k^2 = q^2$ energies are obtained that do not vanish with the system size. The constant rotation angle $\phi$ is chosen, such that the energy gain from the linear gradient term is maximum. There are four different angles \cite{O} depending on ${\cal D} \gtrless 0$ and $k = \pm q$. However, all four cases are degenerate and give the same negative value of the linear gradient term.  
 
To calculate the energy change of the splay-bend texture relative to the
homogeneous state, one has to evaluate the total gradient energy of the texture
\begin{equation}\label{fgrad}
E_{GL} = \int dV \left( {\cal D} T_{ijk} \nabla_k Q_{ij}
+ \gamma (\nabla_k Q_{ij})^2
+ \delta (\nabla_k T_{ijl})^2 \right)
\end{equation}
which is still a function of $q$. The latter is determined by maximizing the energy gain, with the result \cite{O}
\begin{equation} \label{enred}
E_{GL}^{max}  = - c_1 \lambda  {\cal D}^2 \quad\quad  \textrm{and} \quad\quad  q_c  = c_2 \lambda | {\cal D}| 
\end{equation}
with $1/\lambda = (9/2) \gamma + (128/9) \delta$. The numerical factors $c_1\approx 1.09$ and $c_2 \approx 1.19$ are slightly larger than one. The nematic and tetrahedral order parameters are set to $S=1=N$. As expected, the linear gradient term leads to an inhomogeneous structure that is an energetic minimum despite the sign of the phenomenological parameter ${\cal D}$. 
An impression of the actual structure that leads to this energy gain is given in Fig. \ref{fig:22} for ${\cal D} >0$, $k=q$, and $\phi \approx 136^o$.

In contrast to the helical structure in chiral nematics and the (ambidextrous) helices in {\it e.g.} the D2d phase, Section \ref{ambi}, which show an energy density constant in space, for the splay-bend texture discussed above the energy density is space dependent. However, by a different choice of the rotation angle $\phi$, there is a homogeneous energy density also for the splay-bend case. This happens, in particular, for $\cos \phi = \mp 1/3$, corresponding to the tetrahedral and the dihedral angle. However, the energy gain is less than that for the optimized choice of $\phi$, since the $E_{GL}= - \lambda {\cal D}^2 $ and $q_c  = \lambda | {\cal D}|$.

The structure discussed above very likely is not the only one that leads to an energy minimum, and others might have an even lower free energy. An example is the slightly different splay-bend texture, where the nematic director is tilted into the third dimension, replacing Eq. (\ref{2D})
\begin{equation} \label{3D}
{\bf n} = (\beta  \cos qx, \beta \sin qx, \alpha)
\end{equation}
with $\alpha^2 + \beta^2 = 1$. This structure has an additional 
parameter that can be used to maximize the energy gain further. Indeed, for 
roughly $0.004 < \delta/\gamma < 2.86$ and an optimized $\beta_c$ the 
2-dimensional pattern are more favorable while outside this interval 3 D structures
are energetically preferred. 

There is the possibility that other, completely different inhomogeneous structures can rival the splay-bend textures considered here. However, for the latter there is experimental evidence that biaxial splay-bent structures are related to myelin textures observed in the original B7 phase of bent-core liquid crystals \cite{Halle}.

%%%%%%%%%%%%%%%%%%%%%%%%%%%%%%%%%%%%%%%%%%%%%
%%%%%%%%%%%%%%%%%%%%%%%%%%%%%%%%%%%%%%%%%%%%

\section{Review of the Experimental Situation} \label{exper}

It became clear early on in the study of liquid crystalline phases formed
by bent-core molecules \cite{shennem,weissflognem,pelzlnem,niorinem} that their nematic phases reveal unusual physical properties. 
The physical properties of bent-core nematic phases investigated up to about 2013 have been reviewed in Ref. \cite{jaklinemrev}. Here, we concentrate on those effects that can directly linked to the existence of tetrahedral order.

In Ref. \cite{gleeson2} flexoelectric effects in the nematic phase
were investigated and it was found that the values of the flexoelectric 
coefficients are about three orders of magnitude higher than for ordinary
nematics formed by rod-like molecules. To account for this unusual enhancement
it was suggested in \cite{gleeson2} that some tens of bent-core molecules  
form polar clusters. While it is quite intuitive that bent-core molecules
like to form clusters for reason of packing, no reason was presented 
why the clusters should be polar.
In \cite{gleeson3} the isotropic - nematic phase transition was studied
using magnetic birefringence and dynamic light scattering using moderate
magnetic fields. 
While the results found were qualitatively in accord with a classical
Landau picture, it was found that there is a density change at the isotropic -
nematic phase transition which is about an order of magnitude smaller
than for rod-like molecules. In addition, the relaxation rate of the 
isotropic phase fluctuations is slowed down and the viscosity 
of the orientational fluctuations are an order of magnitude higher.
The picture suggested \cite{gleeson3} is that of clusters of bent-core
molecules in the isotropic phase above the nematic phase, a picture which is also
consistent with the onset of tetrahedrally coordinated complexes. 
In \cite{ostapenko} the immediate vicinity of the nematic - isotropic transition
was investigated under high magnetic fields and it was demonstrated that a first order isotropic to nematic phase
transition could be induced, an observation unknown from compounds
made of rod-like molecules
near the isotropic - nematic phase transition in a magnetic field.
It is found that the measured change in phase
transition temperature to the nematic phase 
is found to be considerably 
larger than what is expected using a Landau picture for the orientational
oder parameter $Q_{ij}$ and the authors suggest that this
effect could be associated with the 
onset of tetrahedral order \cite{ostapenko}. 
In \cite{gleeson4} measurements of heat capacity, density,  
magnetic-field induced birefringence, line width, scattered light intensity,
and the viscosity associated with fluctuations of the quadrupolar
order parameter $Q_{ij}$ have been used to study a larger
temperature interval in the  
vicinity of the nematic - isotropic phase transition in a compound composed
of bent-core molecules.  
Two peaks are observed in the isotopic phase above the nematic phase in
heat capacity as well as in density measurements signaling the existence of two 
optically isotropic phases. The magnetic field induced birefringence is found 
to be no longer $\sim H^2$, but develops a curvature  
not compatible with a simple Landau picture for $Q_{ij}$.
The authors of Ref. \cite{gleeson4} also find that the viscosity associated 
with the quadrupolar orientational order parameter shows 
an unusual step change somewhat above the clearing point.
The authors suggest a Landau model comprising a quadrupolar orientational
order parameter $Q_{ij}$ as well as a tetrahedral order parameter $T_{ijk}$.
As a result of the Landau analysis involving two order parameters the
authors find that they can account for all their experimental results. This
provides the best experimental evidence available to date for the
presence of a tetrahedral phase with a completely isotropic phase at higher temperature.

Due to the broken inversion symmetry, electric field effects are prominent and rather special in tetrahedral phases.
First, for bent-core molecules, phase transitions isotropic - smectic B$_2$ 
\cite{bourny2002}
and isotropic - smectic C$_P$ \cite{Weissflog,schroeder2004} 
were studied in an external
electric field. It was found that an upward shift of the phase transition
temperature by up to about 10 K could be achieved, which was 
approximately linear in the applied electric field 
\cite{Weissflog}. At first sight both, the magnitude as well as 
the linearity of the shift in the applied voltage, come as a surprise
since for rod-like molecules large shifts of more 
than small fractions of a degree in the isotropic -
smectic phase transition temperature have never been observed.
Even more surprising is the linearity in the applied voltage.
Clearly an isotropic phase has no preferred polar direction, 
which could be oriented and could give rise to a linear response in the
electric field.
In addition, quadrupolar orientational order can only generate shifts
that are quadratic in the applied electric or magnetic fields
as already pointed out by de Gennes, when studying the isotropic -
nematic transition in low molecular weight liquid crystals
\cite{pgdg2}.  

As we have discussed in detail in Sec. \ref{Tdfields}, in a (optically isotropic) tetrahedral phase the situation is different. Here, transition shifts and induced nematic order occur, which are linear in the field strength, Eqs. (\ref{linEQ}), (\ref{inducedQ}), (\ref{Tcshift}), and (\ref{Scshift}). This also applies to tetrahedral - smectic transitions, since smectic order is always accompanied by
nematic order \cite{mukh}. Recently, isotropic  - nematic transitions in bent-core material have been studied directly. Electric field effects on this transition are described in Ref. \cite{ed2013}. They find a linear field dependence (their Fig. 4), which is
fully compatible with the assumption that the "isotropic" phase is rather a tetrahedral one. In addition, isotropic to isotropic transitions as well as a reentrant
isotropic phase has been described for another family of bent core molecules \cite{jacs}.
Again, the existence of two optically isotropic phases, and the linear response to electric fields in one of them, clearly points to the presence of tetrahedral order.

The electric properties of bent-core nematic phases 
were also shown to lead to some unusual spatio-temporal patterns 
in electroconvection \cite{gleeson1}.

The appearance of ambidextrous chiral domains in smectic phases formed of bent-core molecules has been reported quite early \cite{pelzlnem,niorinem}.
More recently \cite{tetratschierske} the issue of chirality shown
by nematic bent-core phases has been examined in detail 
experimentally, in planar as well as in homeotropic cells. It was suggested, using molecular modeling, that the nematic phase analyzed is of $D_2$ symmetry 
\cite{tetratschierske}.
We note that for a nematic phase with this symmetry the microscopic 
\cite{longa1,longa2} as well as the macroscopic (Sec. \ref{D2}) properties 
have been investigated. Tetrahedral order is crucial for the existence of, and the various chiral effects in $D_2$ symmetric nematic phases.
 
Several unusual smectic and nematic phases have been found in ferrocenomesogens
\cite{tetraferro2009,tetraferro2011} and linked to tetrahedral order. The latter is traced back 
to a tetrahedral association of the molecules. Among the macroscopic phenomena arising are large domains of opposite chirality (ambidextrous chiral domains) as well as helical superstructures \cite{tetraferro2009,tetraferro2011}. Both are compatible with tetrahedral order, Secs. \ref{ambi} and \ref{D2S4}.
For heterochiral areas in these compounds an optically isotropic appearance has been
found, again compatible with tetrahedral order.

%%%%%%%%%%%%%%%%%%%%%%%%%%%%%%%%%%%%%%%%%%%%%
%%%%%%%%%%%%%%%%%%%%%%%%%%%%%%%%%%%%%%%%%%%%

\section{Summary, Conclusions and Perspective  \label{Sum}}

Since the tetrahedral order parameter is of rank 3, most of the material relations are isotropic, in particular the dielectric tensor denoting the optical behavior (an exception is the viscosity tensor).  Therefore it is rather difficult to experimentally discriminate a tetrahedral phase from an isotropic one. However, there are two principal differences, one with respect to reversible (deformational) flows and the other regarding external electric fields. The former describes flow induced by gradients of temperature, concentration etc. (and {\it vice versa}), impossible in isotropic liquids. The latter comprise induced nematic order and transition shifts that are linear in the field amplitude, while they have to be quadratic in an isotropic phase. This only refers to electric fields and not to magnetic ones due to the different parity and time reversal properties of those fields. In the preceding Section we have discussed some experiments regarding the differences between isotropic and tetrahedral phases. Clearly, it would be highly desirable to perform measurements of 
the electric birefringence in addition to the magnetic birefringence.
The specific tetrahedral effects are due to the spontaneous broken inversion symmetry in these systems: The inverted structure is different from the original one, but energetically equivalent.

When combined with nematic order, depending on the geometric relation between the tetrahedral vectors and the nematic director(s), several different liquid crystal phases can arise that are achiral or (structurally) chiral, non-polar or polar, in any combination. Rather low symmetries can occur. Among those phases the D2d phase is of particular
interest, since, there, helical ground states of opposite helicity, but equal energy, are possible (ambidextrous helicity), despite the fact that the phase is achiral. The reason is the existence of a linear gradient term in the Frank-type energy allowed by the tetrahedral order. In a chiral system, on the other hand, there is a linear nematic twist allowed due to the existence of a pseudoscalar quantity, whose origin is either due to 
the chirality of the molecules or comes from the structure of the phase. In the latter case both 
types of handedness are energetically equal (ambidextrous chirality). As an example the D2 phase is discussed above.
Another specific feature of many tetrahedral nematic phases is the orientational frustration in an electric field, where the nematic dielectric anisotropy and the tetrahedral field orientation are incompatible with the relative orientation of the tetrahedral vectors and the nematic director(s).

We have shown in the bulk part of this review that tetrahedral 
or octupolar order and its consequences have been mainly analyzed
in soft matter physics, in particular in the field of liquid
crystals. It is worth pointing out, however, that octupolar order, 
in particular in two spatial dimensions, has also been used in the study 
of moving and deformable active particles as models for self-propelled
micro-organisms \cite{takao1,takao2,takao3}. 
In this case octupolar order comes into play  
when deformations of lower symmetry going beyond quadrupolar deformations
are considered.

From an applied mathematics point of view it has been demonstrated recently 
\cite{epifanio} for two spatial dimensions how the maxima of the associated
probability density are connected to a third rank totally symmetric
and traceless tensor.
It turns out that such a representation 
is equivalent to the diagonalization of this third rank tensor 
in two spatial dimensions. A suitable generalization of this analysis
to three dimensions is clearly desirable for the field of liquid crystals.

In this review we have restricted ourselves to phases with tetrahedral order, either alone or together with (uniaxial and biaxial) nematic order. Often, bent-core molecules form smectic phases and one can expect that tetrahedral order also plays a role there. However, many of those phases, {\it e.g.} those shown in Fig. \ref{fig:2}, and those relevant for applications, are polar. The existence of polarity has a strong influence on the structure of the macroscopic dynamics of such phases and an additional tetrahedral order does not lead to important new effects. On the other hand, for the non-polar smectic phases made of bent-core material, smectic C, C$_M$, and C$_T$ \cite{E2,CM}, the tetrahedral order provides new additional aspects \cite{tetraelast}. The same is true for gels and elastomers with tetrahedral order. An example is presented already in this review in Eq. (\ref{piezo}), where an electric field induces a strain field (linear in the field strength) in the presence of tetrahedral order. Tetrahedral elastic effects may also be able to explain some of the experimental findings in agglomerating bent-core systems \cite{dressel2015,tschierske16}.

The macroscopic theory presented here is based on the presence or lack of symmetries on the macroscopic level (in addition to certain general conservation laws and thermodynamic rules). The way molecules arrange on the microscopic level is not considered, since it is not necessary to do so for macroscopic properties. 
The only exception is Fig. \ref{fig:11} in Sec. \ref{chiralT}, which is an example of how to distribute chiral centers on the bent-core molecules in order to get chirality on the macroscopic level. It is 
meant to be a rather simple molecular picture. 
Often, {\it e.g.} in Refs. \cite{gleeson2,gleeson3}, tetrahedral order is related to the appearance of tetrahedral clusters or agglomerations of many bent-core molecules. Such a picture is in complete accordance with our description, where only the existence of tetrahedral order, but not its molecular realization is important.

%%%%%%%%%%%%%%%%%%%%%%%%%%%%%%%%%%%%%%%%%%%%%%%
%%%%%%%%%%%%%%%%%%%%%%%%%%%%%%%%%%%%%%%%%%%%%%%

\end{document}